\newif{\ifarxiv}
\newif{\ifdraft}
\newif{\ifremarks}

\arxivtrue  % for stuff that should only go into the arxiv preprint
\remarkstrue  % turn remarks on/off

\documentclass[11pt,a4paper]{article}
\ifdraft\overfullrule=5pt\fi
\pdfoutput=1
\usepackage[utf8]{inputenc}
\usepackage[left=2.5cm,right=2.5cm,top=2.8cm,bottom=2.8cm]{geometry}
\usepackage{graphicx}
\usepackage{amsmath,amssymb,mathtools}
\ifdraft\usepackage{showkeys}\fi % needs to be included before package "hyperref"
\usepackage[bookmarks=true,ocgcolorlinks=true]{hyperref} % ocgcolorlinks: color links on screen, but print them black. Requires pdf version 1.5
\usepackage{cite}
\usepackage{xcolor}
\usepackage{booktabs}
\usepackage{graphbox}
\usepackage[mathbold,autobold,greekcaps,greeklower]{mathfixs}
\renewcommand{\mathbf}{\mathbold}

\usepackage{lmodern}
\usepackage[T1]{fontenc}
\usepackage{microtype}
\usepackage{marvosym}

\newcommand{\colred}{\color[RGB]{255,0,0}}
\newcommand{\colgreen}{\color[rgb]{0,0.6,0}}

\ifremarks
\newcommand{\remarktb}[1]{{\renewcommand{\bfdefault}{b}{\color[RGB]{0,150,0}{\textbf{#1}}}}}
\newcommand{\remarkv}[1]{{\renewcommand{\bfdefault}{b}{\color[RGB]{150,0,0}{\textbf{#1}}}}}
\newcommand{\remarks}[1]{{\renewcommand{\bfdefault}{b}{\color[RGB]{0,0,150}{\textbf{#1}}}}}
\fi
\providecommand{\remarktb}[1]{\ignorespaces}
\providecommand{\remarkv}[1]{\ignorespaces}
\providecommand{\remarks}[1]{\ignorespaces}

\allowdisplaybreaks
\usepackage[font=small,labelfont=bf,width=0.85\textwidth]{caption}

\providecommand{\hypersetup}[1]{}
\providecommand{\hyperref}[2][]{#2}
\providecommand{\texorpdfstring}[2]{#1}
\providecommand{\pdfbookmark}[3][]{}

\hypersetup{plainpages=false}
\hypersetup{pdfpagemode=UseOutlines}
\hypersetup{bookmarksnumbered=true}
\hypersetup{bookmarksopen=true}
\hypersetup{pdfstartview=FitH}
\hypersetup{citecolor=[rgb]{0 .4 0}}
\hypersetup{urlcolor=[rgb]{.4 0 0}}
\hypersetup{linkcolor=[rgb]{0 0 .4}}
\hypersetup{citebordercolor={.6 .9 .6}}
\hypersetup{urlbordercolor={.7 .8 1}}
\hypersetup{linkbordercolor={1 .7 .7}}
\hypersetup{pdfborder={0 0 .5}}

\makeatletter
\renewcommand*\l@section[2]{%
  \ifnum \c@tocdepth >\z@
    \addpenalty\@secpenalty
    \addvspace{0.7em \@plus\p@}%
    \setlength\@tempdima{1.5em}%
    \begingroup
      \parindent \z@ \rightskip \@pnumwidth
      \parfillskip -\@pnumwidth
      \leavevmode \bfseries
      \advance\leftskip\@tempdima
      \hskip -\leftskip
      #1\nobreak\hfil \nobreak\hb@xt@\@pnumwidth{\hss #2}\par
    \endgroup
  \fi}
\makeatother

\newcommand{\namedref}[2]{\hyperref[#2]{#1~\ref*{#2}}}
\newcommand{\secref}[1]{\namedref{Section}{#1}}
\newcommand{\Secref}[1]{\namedref{Section}{#1}}
\newcommand{\appref}[1]{\namedref{Appendix}{#1}}
\newcommand{\Appref}[1]{\namedref{Appendix}{#1}}
\newcommand{\tabref}[1]{\namedref{Table}{#1}}
\newcommand{\Tabref}[1]{\namedref{Table}{#1}}
\newcommand{\figref}[1]{\namedref{Figure}{#1}}

\makeatletter
\def\mr@ignsp#1 {\ifx\:#1\@empty\else #1\expandafter\mr@ignsp\fi}%
\newcommand{\multiref}[1]{\begingroup%\let\protect\string%
\xdef\mr@no@sparg{\expandafter\mr@ignsp#1 \: }%
\def\mr@comma{}%
\@for\mr@refs:=\mr@no@sparg\do{\mr@comma\def\mr@comma{,\,}\ref{\mr@refs}}%
\endgroup}
\makeatother
\renewcommand{\eqref}[1]{(\multiref{#1})}

\makeatletter
\let\@myabstract\@empty
\let\@keywords\@empty
\let\@subject\@empty
\providecommand{\affiliation}[1]{\gdef\@affiliation{#1}}
\providecommand{\myabstract}[1]{\gdef\@myabstract{#1}}
\providecommand{\keywords}[1]{\gdef\@keywords{#1}}
\providecommand{\subject}[1]{\gdef\@subject{#1}}
\def\thetitle{\@title}
\def\theauthor{\@author}
\def\theaffiliation{\@affiliation}
\def\theabstract{\@myabstract}
\def\thesubject{\@subject}
\def\thedate{\@date}
\def\thekeywords{\@keywords}
\makeatother
\def\fillpdfdata{
\hypersetup{pdftitle={\thetitle}}%
\hypersetup{pdfsubject={\thesubject}}%
\hypersetup{pdfkeywords={\thekeywords}}%
}

\numberwithin{equation}{section}

\makeatletter\newcommand*{\etal}{%
    \@ifnextchar{.}%
        {et\penalty50\ al}%
        {et\penalty50\ al.\@\xspace}%
}\makeatother
\usepackage{xspace}
\newcommand*{\eg}{e.\,g.\@\xspace}
\newcommand*{\ie}{i.\,e.\@\xspace}

\makeatletter\newcommand*{\etc}{%
    \@ifnextchar{.}%
        {etc}%
        {etc.\@\xspace}%
}\makeatother

\newcommand{\nn}{\nonumber}

\newcommand{\brk}[1]{(#1)}
\newcommand{\lrbrk}[1]{\left(#1\right)}
\newcommand{\bigbrk}[1]{\bigl(#1\bigr)}
\newcommand{\Bigbrk}[1]{\Bigl(#1\Bigr)}
\newcommand{\biggbrk}[1]{\biggl(#1\biggr)}
\newcommand{\Biggbrk}[1]{\Biggl(#1\Biggr)}
\newcommand{\sbrk}[1]{[#1]}
\newcommand{\lrsbrk}[1]{\left[#1\right]}
\newcommand{\bigsbrk}[1]{\bigl[#1\bigr]}
\newcommand{\biggsbrk}[1]{\biggl[#1\biggr]}
\newcommand{\Bigsbrk}[1]{\Bigl[#1\Bigr]}
\newcommand{\Biggsbrk}[1]{\Biggl[#1\Biggr]}
\newcommand{\brc}[1]{\{#1\}}
\newcommand{\lrbrc}[1]{\left\{#1\right\}}
\newcommand{\bigbrc}[1]{\bigl\{#1\bigr\}}
\newcommand{\Bigbrc}[1]{\Bigl\{#1\Bigr\}}
\newcommand{\vev}[1]{\langle #1\rangle}
\newcommand{\smb}[1]{(#1)}

\newcommand{\floor}[1]{\lfloor#1\rfloor}
\newcommand{\abs}[1]{|#1|}

\newcommand{\lrabs}[1]{\left|#1\right|}

\newcommand{\comm}[2]{[#1,#2]}

\newcommand{\sfrac}[2]{{\textstyle\frac{#1}{#2}}}

\newcommand{\eps}{\varepsilon}
\newcommand{\p}{{+}}
\newcommand{\m}{{-}}
\newcommand{\superN}{\mathcal{N}}
\newcommand{\grp}[1]{\mathrm{#1}}

\newcommand{\Li}{\operatorname{Li}}
\newcommand{\smbop}[1]{\operatorname{S}\sbrk{#1}}

\newcommand{\Z}{\mathbb{Z}}

\newcommand{\order}[1]{\mathcal{O}(#1)}

\newcommand{\supMRL}{^{\mathrm{MRL}}}

\newcommand{\suprm}[1]{^{\text{#1}}}
\newcommand{\subrm}[1]{_{\text{#1}}}
\newcommand{\dd}{\mathrm{d}}

\newcommand{\gen}{\mathcal{C}}
\newcommand{\gcusp}{\Gamma\subrm{cusp}}
\newcommand{\Rsixope}{R_{6,(2)}\suprm{ope}}
\newcommand{\Rope}{R_{7,(2)}\suprm{ope}}

\newcommand{\Rnlope}{R_{n,(l)}\suprm{ope}}

\newcommand{\EulerGamma}{\gamma_{\mathrm{E}}}

\newcommand{\Gs}{G^{\mathbf{s}}}
\newcommand{\Gsshort}[1]{G^{\mathbf{s},#1}}
\newcommand{\lyndon}{\operatorname{Lyn}}
\newcommand{\id}{\operatorname{id}}

\newcommand{\cy}{\check y}
\newcommand{\Wl}{\mathcal{W}}
\newcommand{\Integers}{\mathbb{Z}}

\newcommand{\Rmmm}{R_{7,(2)}^{\m\m\m}}
\newcommand{\Rmpm}{R_{7,(2)}^{\m\p\m}}

\newcommand{\CL}{\mathrm{CL}}

\newcommand{\curve}{\gamma}
\newcommand{\cc}[1]{c_{#1}}
\newcommand{\cg}[1]{\kappa_{#1}}
\newcommand{\cgtilde}[1]{\tilde{\kappa}_{#1}}
\newcommand{\f}{\tilde{f}}

\newcommand{\defas}{:=}
\newcommand{\poly}{\mathcal{P}}
\newcommand{\polyQ}{\mathcal{Q}}
\newcommand{\wb}{\bar{w}}
\newcommand{\WBDS}{\mathcal{W}^{\text{\tiny$\grp{U}(1)$}}}

\makeatletter
\newcommand*\bigcdot{\mathpalette\bigcdot@{.6}}
\newcommand*\bigcdot@[2]{\mathbin{\vcenter{\hbox{\scalebox{#2}{$\m@th#1\bullet$}}}}}
\makeatother

\newcommand{\soft}[1]{\textsc{#1}}
\newcommand{\mathematica}{\soft{Mathematica}}
\newcommand{\filename}[1]{\texttt{#1}}
\newcommand{\code}[1]{\texttt{#1}}

\begin{document}

%%%%%%%%%%%%%%%%%%%%%%%%%%%%%%%%%%%%%%%%%%%%%%%%%%%%%%%%%%%%
% Title data
%%%%%%%%%%%%%%%%%%%%%%%%%%%%%%%%%%%%%%%%%%%%%%%%%%%%%%%%%%%%

\title{The Multi-Regge Limit from the Wilson Loop OPE}

\myabstract{The finite remainder function for planar, color-ordered, maximally helicity
violating scattering processes in $\mathcal{N}=4$ super Yang--Mills theory possesses a
non-vanishing multi-Regge limit that depends on the choice of a Mandelstam region. We
analyze the combined multi-Regge collinear limit in all Mandelstam regions through an
analytic continuation of the Wilson loop OPE.
At leading order, the
former is determined by the gluon excitation of the Gubser--Klebanov--Polyakov string.
We illustrate the general procedure at the example of the heptagon remainder function
at two loops. In this case, the continuation of the leading order
terms in the Wilson loop
OPE suffices to determine the two-loop multi-Regge heptagon functions in all Mandelstam
regions from their symbols. The expressions we obtain are fully consistent with recent
results by Del Duca \etal}

\subject{Mathematical Physics, Scattering Amplitudes}


%%%%%%%%%%%%%%%%%%%%%%%%%%%%%%%%%%%%%%%%%%%%%%%%%%%%%%%%%%%%
%%%%%%%%%%%%%%%%%%%%%%%%%%%%%%%%%%%%%%%%%%%%%%%%%%%%%%%%%%%%

\ifarxiv
\noindent
\mbox{}\hfill DESY 19-099
\fi

\author{%
Till Bargheer,\texorpdfstring{${}^{a,b}$}{}
Vsevolod Chestnov,\texorpdfstring{${}^b$}{}
Volker Schomerus\texorpdfstring{${}^b$}{}
}

\hypersetup{pdfauthor={\theauthor}}

\vfill

\begin{center}
{\Large\textbf{\mathversion{bold}\thetitle}\par}

\bigskip

\textsc{\theauthor}

\medskip

{\itshape
$^{a}$Institut f\"ur Theoretische Physik, Leibniz Universit\"at Hannover,\\
Appelstra{\ss}e 2, 30167 Hannover, Germany

\medskip

$^{b}$DESY Theory Group, DESY Hamburg,\\
Notkestra\ss e 85, D-22603 Hamburg, Germany
}

\medskip

{\ttfamily\brc{%
\href{mailto:till.bargheer@desy.de}{till.bargheer},%
\href{mailto:vsevolod.chestnov@desy.de}{vsevolod.chestnov},%
\href{mailto:volker.schomerus@desy.de}{volker.schomerus}%
}\href{mailto:till.bargheer@desy.de,vsevolod.chestnov@desy.de,volker.schomerus@desy.de}{@desy.de}}
\par\vspace{1cm}

\textbf{Abstract}\vspace{5mm}

\begin{minipage}{12.4cm}
\theabstract
\end{minipage}

\vfill

\end{center}

\fillpdfdata

\newpage

% \hrule
% \vspace{1ex}
\providecommand{\microtypesetup}[1]{}
\microtypesetup{protrusion=false}
\setcounter{tocdepth}{2}
\pdfbookmark[1]{\contentsname}{contents}
\tableofcontents
\microtypesetup{protrusion=true}
% \vspace{3ex}
% \hrule

% \vfill

% \newpage

%%%%%%%%%%%%%%%%%%%%%%%%%%%%%%%%%%%%%%%%%%%%%%%%%%%%%%%%%%%%
%%%%%%%%%%%%%%%%%%%%%%%%%%%%%%%%%%%%%%%%%%%%%%%%%%%%%%%%%%%%
\section{Introduction}
\label{sec:intro}

Constructing the scattering amplitudes of a four-dimensional quantum field theory
beyond a few orders of perturbation theory has been a long-term challenge for
theoretical physics. Back in the 1960s/70s, it gave rise to the analytic S-matrix
program, which aimed at reconstructing scattering amplitudes from their analytic
properties in the space of complexified kinematic invariants. At that time there
existed few tools to make precision predictions at high loop orders,
or even at finite
coupling. This has changed during the last two decades, in which our understanding
of gauge theories in general, and of the maximally supersymmetric $\mathcal{N}=4$
Yang--Mills (SYM) theory in particular, has advanced significantly. Many new
calculational techniques are now available, so that for the first time scattering
amplitudes in a 4d gauge theory are accessible even beyond perturbation theory.
In the case of $\mathcal{N}=4$ SYM theory, a concrete one-dimensional model could
be identified that describes the corresponding flux tube. This model turned out
to be integrable, and its solution allows us to compute scattering amplitudes at
any coupling~\cite{Basso:2013vsa,Basso:2013aha,Basso:2014koa,Basso:2014jfa,Basso:2014nra},
though far from the collinear limit many flux tube excitations
must be summed, which often restricts practical applications to near-collinear
kinematics.

A particularly interesting kinematic limit of scattering theory is the multi-Regge
limit. Roughly, it concerns regions in the space of kinematic variables that are
probed by particle colliders, in which the collision of two highly energetic
particles produces many new particles of much lower energies which escape
the scattering region, along with two particles that carry away most of initial
energy. The multi-Regge regime is not only experimentally relevant, but also
of significant theoretical interest. First of all, while the
limit entails remarkable simplifications, it still has a very rich structure
that has been studied extensively in the context of the analytic S-matrix program.
In Regge theory, whole classes of Feynman diagrams resum into the exchange of
effective particles called Reggeons, which give rise to the famous Regge
trajectories. Pole terms in the scattering amplitudes can be associated with
single-Reggeon exchange, while the exchange of multi-Reggeon states is seen
in cut terms.

The fundamental assumptions Regge theory makes about the analytic
structure of partial waves are rooted in fundamental properties of the theory, such
as confinement. Hence Regge theory applies to large classes of models, including
supersymmetric Yang--Mills theories. These possess the same multi-Regge behavior as
their bosonic cousins, at least in the leading logarithmic approximation (LLA). Indeed,
over the last decade, Regge theory has been pivotal in pushing the perturbative
expansion in $\superN=4$ supersymmetric Yang--Mills theory. When Bern, Dixon, and
Smirnov (BDS) proposed their celebrated all-loop formula for color-ordered,
maximally helicity violating (MHV) scattering amplitudes of $n$ gluons~\cite{Bern:2005iz},
the mismatch with the expected analytic structure of the
S-matrix was clearly demonstrated in the multi-Regge regime~\cite{Bartels:2008ce}.
Along with independent evidence from holography~\cite{Alday:2007he}, this showed
that the BDS formula was incomplete beyond five external gluons, and required a
finite correction term that is known as the remainder function. In constructing
the finite remainder for generic kinematics, predictions of Regge theory have
been used as essential boundary data for the perturbative amplitude bootstrap
to high loop orders, see~\cite{Caron-Huot:2019vjl} and references therein.

In spite of all the simplifications and integrability, constructing the finite
remainder function in multi-Regge kinematics remains a challenging problem. For
six external gluons, the problem was solved in~\cite{Basso:2014pla} to all orders,
completing a program that was initiated by Bartels \etal~\cite{Bartels:2008sc},
who determined the leading logarithmic approximation (LLA) to all orders in the
gauge coupling, with extensions to next-to-LLA (NLLA), and next-to-NLLA (NNLLA)
in~\cite{Fadin:2011we} and~\cite{Dixon:2014voa}.
The loop expansion of these
expressions was studied in~\cite{Dixon:2012yy}.
In the strong-coupling limit,
the multi-Regge limit of the hexagon remainder function was calculated
in~\cite{Bartels:2010ej,Bartels:2013dja}. For a larger number of external gluons,
results are sparse. The first expressions for the multi-Regge limit of the heptagon
remainder function in LLA and in all Mandelstam regions were proposed by
Bartels \etal in~\cite{Bartels:2013jna,Bartels:2014jya}. All other
results to date are restricted to two loops. The symbol of the two-loop remainder
function in the multi-Regge limit for any number of external gluons was
discussed in~\cite{Prygarin:2011gd,Bargheer:2015djt}, and the upgrade
to functions was completed in~\cite{DelDuca:2016lad,DelDuca:2018hrv,
DelDuca:2018raq}.

In comparison, the collinear limit that we mentioned in the introductory
paragraph is much better understood. In this case, the Wilson loop
operator product expansion
(OPE)~\cite{Basso:2013vsa,Basso:2013aha,Basso:2014koa,Basso:2014jfa,Basso:2014nra}
allows to obtain exact results that are based on the complete solution
of the relevant flux tube, which turns out be quantum integrable. Extracting
the analytic structure in general kinematics requires a resummation of all
flux-tube excitations, which is very difficult in general. Here, we pursue
a different, somewhat indirect route: We exploit the leading collinear
behavior of the flux-tube OPE to explore the weak-coupling expansion
of Regge theory. While this is certainly much less ambitious than the
beautiful all-order derivation of the hexagon remainder in multi-Regge
kinematics by Basso \etal~\cite{Basso:2014pla}, the structure of the
remainder function becomes richer for higher numbers of gluons. In
particular, there exist several disjoint Mandelstam regions, and more
cuts that can contribute to the multi-Regge limit of the remainder
function.

\medskip

The main goal here is to obtain constraints on the finite remainder function
of $\mathcal{N}=4$ supersymmetric Yang--Mills theory in multi-Regge kinematics
in all Mandelstam regions from the Wilson loop OPE.
The Wilson loop OPE allows to compute the remainder function in
collinear asymptotics for
completely spacelike polygon Wilson loops. Such completely spacelike
configurations define the main sheet of the remainder function.
On this sheet, the multi-Regge limit of the remainder
function is trivial, since the BDS amplitude is multi-Regge exact. In order to
go to non-trivial Mandelstam regions, in which the remainder function possesses
a non-vanishing multi-Regge limit, we need to analytically continue the OPE data.
Through these analytic continuations, we can obtain the remainder function in a
collinear limit of the multi-Regge regime. We shall refer to this limit as a
combined multi-Regge collinear limit. A priori, it may not be entirely clear that
the collinear limit sees enough of the branch cuts to reach the relevant
Mandelstam regions, but we shall provide evidence that this is the case. At the
example of the two-loop heptagon remainder function, which is known in all four
non-trivial Mandelstam regions, we shall show that our analysis fixes correctly
all parameters that cannot be determined from general consistency requirements
when we perform the lift from symbols to functions.

Let us briefly outline the content of the following sections. In \secref{sec:mrlfromope}, we
shall provide some of the kinematical background. In particular, we introduce
appropriate coordinates to describe and parametrize multi-Regge kinematics.
Then we turn to the collinear limit. We review the natural variables
of the Wilson loop OPE, and explain how to perform analytic
continuations in the collinear regime.
\Secref{sec:opeexpansion} contains a lightning review
of the flux tube and the Wilson loop OPE. The latter is used to compute leading terms
in the collinear limit of the heptagon remainder function. The remaining two
sections address the analytic continuation from the main sheet to the various Mandelstam
regions. For pedagogical reasons, we start with the hexagon case (\secref{sec:continuation6})
before turning to the more elaborate heptagon, which admits four Mandelstam
regions in which the remainder function does not vanish (\secref{sec:continuation7}). For all these
regions, we perform the analytic continuation. It turns out that the continuation of the
leading-order terms in the Wilson loop OPE, along with a few standard requirements,
suffice to lift the symbols of the two-loop multi-Regge heptagon to functions.
The expressions we obtain for the two-loop heptagon remainders in the multi-Regge
regime are fully consistent with recent results by Del Duca
\etal~\cite{DelDuca:2016lad,DelDuca:2018hrv,DelDuca:2018raq}.
We conclude with an
extensive outlook to further directions and open problems in \secref{sec:concl}. The paper also
includes several technical appendices, in particular on tessellations of the
Wilson loop and associated variables for the Wilson loop OPE and
multi-Regge limits for all multiplicities
(\appref{app:bsvtessdetail} and \appref{app:mrl-tessellation}).
\Appref{app:gfunction}
provides an extensive analysis of the multi-Regge limit of the two-loop
symbol and its lift to the two-loop heptagon function. We show that for the
two most interesting Mandelstam regions, single-valuedness, symmetry, and
collinear behavior determine the multi-Regge limit of the heptagon remainder
up to four parameters.
The latter are fixed by our continuation in \secref{sec:continuation7}.

%%%%%%%%%%%%%%%%%%%%%%%%%%%%%%%%%%%%%%%%%%%%%%%%%%%%%%%%%%%%
%%%%%%%%%%%%%%%%%%%%%%%%%%%%%%%%%%%%%%%%%%%%%%%%%%%%%%%%%%%%
\section{The Multi-Regge Limit}
\label{sec:mrlfromope}

Our ultimate goal is to compute the multi-Regge limit of the finite remainder function for
color ordered maximally helicity violating (MHV) amplitudes in the planar limit of
$\mathcal{N}=4$ super Yang--Mills theory from the Wilson loop operator product expansion (OPE).
This section contains some background material. After a short reminder on Mandelstam
regions and multi-Regge kinematics, we turn to collinear kinematics
and recall the OPE variables of~\cite{Basso:2013vsa}, which are
appropriate to discuss the combined multi-Regge collinear limit.
The final subsection
outlines how we calculate remainder functions in the combined multi-Regge collinear
limit from the Wilson loop OPE by summing relevant cut contributions,
see~\eqref{eq:MRLCLCLMRL} and~\eqref{eq:MRLCLCLMRL2} below.

%%%%%%%%%%%%%%%%%%%%%%%%%%%%%%%%%%%%%%%%%%%%%%%%%%%%%%%%%%%%
\subsection{Multi-Regge Regime and Mandelstam Regions}
\label{sec:variables}

We consider the scattering of $n$ gluons. With later kinematical limits in
mind we shall think of two incoming particles whose momenta we denote by
$p_1$, $p_2$ and $n-2$ outgoing particles with momenta $-p_3,...,-p_{n}$,
as shown in \figref{fig:configuration}. It will be convenient to label
momenta $p_i$ by arbitrary integers $i$ such that $p_{i+n}\equiv p_i$. In the
context of $\mathcal{N} = 4$ supersymmetric Yang--Mills theory, it is
advantageous to pass to a set of dual variables $x_i$ such that
\begin{equation} \label{eq:pfromx}
p_i=x_{i}-x_{i-1}.
\end{equation}
The variables $x_i$ inherit their periodicity $x_{i+n}=x_i$ from the periodicity
of the $p_i$ and momentum conservation. Let us also introduce the notation $x_{ij}
= x_i - x_j$. The $x^2_{ij}$ provide a large set of Lorentz invariants $x_{ij}^2 =
x_{ji}^2$. Throughout this note, we use a Lorentzian metric with signature
$(-,+,+,+)$. When expressed in terms of the momenta, the invariants read
\begin{equation} \label{eq:xij2}
x^2_{ij}=\lrbrk{p_{i+1}+\cdots+p_j}^2.
\end{equation}
Lorentz symmetry along with the mass-shell conditions $p_i^2=0$ imply that only
$3n-10$ of these variables are independent. We will not make any specific choice
here. On the main sheet, all the energies $-p_i^0$, $i=3,\dots,n$ are assumed to be
negative (\ie the particles with momenta $-p_3,\dots,-p_n$ are physical
outgoing particles).
We will refer to the Mandelstam invariants $x_{ij}^2$ that are negative on the
main sheet as $s$-like (\eg forward energies, see
\figref{fig:configuration}).
Those that obey $x_{ij}^2 \geq 0$ on the main sheet
are called $t$-like (\eg momentum transfers).

\begin{figure}
\centering
\includegraphics[align=c]{FigKinematics}
\quad
$\longleftrightarrow$
\quad
\includegraphics[align=c]{FigKinematicsDual}
\quad
\caption{Kinematics of the scattering process $2 \to n-2$. Forward
energies are labeled by~$s_i$, momentum transfers by~$t_i$. On the
right-hand side we show a graphical representation of the dual
variables $x_i$.}
\label{fig:configuration}
\end{figure}

The finite remainder function $R$ for an $n$-gluon scattering amplitude
is invariant under dual conformal symmetry~\cite{Drummond:2008vq},
and hence it can only depend on cross
ratios of the form
\begin{equation}
U_{ij}\equiv\frac{x_{i+1,j}^2x_{i,j+1}^2}{x_{ij}^2x_{i+1,j+1}^2}
\,,\qquad
3\leq\abs{i-j} < n-2\ .
\label{eq:Uij}
\end{equation}
Since the conformal group in four dimensions has $15$ generators, only $3n-15$ of these
cross ratios are independent. For the discussion of the multi-Regge limit, we adopt
the following choice~\cite{Bartels:2010ej,Bartels:2012gq}
\begin{equation}
u_{j,1}=U_{j+1,j+4}
\,,\qquad
u_{j,2}=U_{j+2,n}
\,,\qquad
u_{j,3}=U_{1,j+3}
\,,
\label{eq:uj}
\end{equation}
where $j=1,...,n-5$. Note that for $n < 6$ one cannot form any cross ratios,
and hence the remainder functions $R_n$ must be trivial for $n=4,5$. In the
case of $n=6$ external gluons, however, there exist $3$ independent cross
ratios, which we shall simply denote by $u_1,u_2,u_3$. And indeed it has been
argued in~\cite{Bartels:2008ce} that $R_6$ must be a non-vanishing function
of the cross ratios $u_i$ in order to correct for the unphysical analytical
structure of the Bern--Dixon--Smirnov (BDS) Ansatz~\cite{Bern:2005iz}.

In the multi-Regge limit, the absolute values of the $s$-like variables are much
larger than the $t$-like ones, which are kept finite. The precise characterization
of the limit in terms of Mandelstam invariants can be found
in~\cite{Bartels:2012gq}. Here, we shall mostly focus on the multi-Regge
limit of the remainder functions $R_n$, which depend on the Mandelstam
invariants only through the cross ratios $u$, of which a complete
independent set is given in~\eqref{eq:uj}. In the multi-Regge limit,
the so-called ``large'' cross ratios $u_{j,1}$ tend to $u_{j,1} \sim 1$ while the
remaining ``small'' ones tend to zero, \ie $u_{j,2}, u_{j,3} \sim 0$. Cross
ratios with the same index $j$ approach their limit values
such that the following ratios remain finite:
\begin{equation} \label{eq:uofw}
\biggsbrk{\frac{u_{j,2}}{1-u_{j,1}}}\supMRL
=:
\frac{1}{|1+w_j|^2}\quad  , \quad
\biggsbrk{\frac{u_{j,3}}{1-u_{j,1}}}\supMRL
=:
 \frac{|w_j|^2}{|1+w_j|^2}\, ,
\end{equation}
These expressions are parametrized by $n-5$ pairs of so-called
``anharmonic ratios'' $(w_j,\wb_j)$, $j=1,\dotsc,n-5$. In the above
formulas, $|f(w)|^2$ means $|f(w)|^2 = f(w) f (\wb)$, even if $w$
and $\wb$ are not complex conjugates of each other.
Our conventions concerning the
enumeration of gluons are shown in \figref{fig:configuration}. Here
and in the following, the superscript ``MRL'' instructs us to evaluate the expression
in square brackets in multi-Regge kinematics.

We are going to evaluate the multi-Regge limit for functions which
possess branch cuts, and so in order to make it well-defined, we need
to specify the sheet on which the limit is actually performed. There
exist $2^{n-4}$ different Mandelstam regions that are distinguished by
the signs of the energies $-p^0_i$ for $i= 4, \dots, n-1$. These regions
are reached by continuing the energies $-p^0_j$ of outgoing particles
with indices $j \in I \subset \brc{4,\dots, n-1}$ to negative values.
The choice of the subset $I$ labels the different Mandelstam regions.

To each such Mandelstam region $I$, we associate an $n$-component object
$\varrho^I = (\varrho^I_j)$ such that
\begin{equation}
\varrho^I_j =
\begin{cases}
-1 & \text{if } j \in I\,,
\\[1mm]
\phantom{-}0 & \text{if } j \in \brc{1 \equiv n+1,2}\,,
\\[1mm]
+1 & \text{otherwise}\,.
\end{cases}
\end{equation}
Since the first and last two entries of $\varrho$ are fixed to take the
values $\varrho_2 = 0 =\varrho_{n+1}$ and $\varrho_3 = 1 = \varrho_{n}$,
we will also use the $n-4$ component $\varrho = (\varrho_i, i = 4, \dots,
n-1)$ to label Mandelstam regions. When we go into a region $\varrho =
(\varrho_i)$, our curve in the space of kinematic invariants may wind
around the endpoints of some branch cuts of the remainder function.
Physical branch points are typically located at the points $U_{ij}=0$.
The winding numbers of the variables $U_{ij}$ around the points
$U_{ij}=0$ for the various Mandelstam regions are~\cite{Bartels:2014ppa}%
\footnote{Our
conventions here differ from those used in~\cite{Bartels:2014ppa} by an overall sign.}
\begin{equation} \label{eq:winding}
n_{ij}(\varrho) = \frac{1}{4} (\varrho_{i+2}-\varrho_{i+1})
(\varrho_{j+1}-\varrho_{j}) \ .
\end{equation}
Let us point out that the numbers $n_{ij}$ take values in the set of half-integers,
\ie $n_{ij} \in \mathbb{Z}/2$. The so-called large cross-ratios that become
$u_{ij} \rightarrow 1$ in multi-Regge kinematics, however, possess integer winding
numbers for all choices of $\varrho$.

If we perform the multi-Regge limit of the remainder functions $R_n$ in
the region in which all the energies are positive (\ie on the main
sheet, which is accessible to the Wilson loop OPE), the result turns out to
vanish,
\begin{equation} \label{eq:MRLMain}
[R_n(u,a)]\supMRL_{++\cdots+}  =  0 \, .
\end{equation}
In other words, in the region $\varrho_0=(+,+,\dots,+)$, the BDS formula is actually
multi-Regge exact. If it was only for this region, the multi-Regge limit would
not be able to see the difference between a vanishing and non-vanishing remainder
function.

As we have anticipated in the introduction, however, there exists
other regions in which the Regge limit of the remainder function
does not vanish. Of course, the non-vanishing terms must be associated
with the cut contributions that are picked up when we analytically
continue from the region $\varrho_0$ into a new region $\varrho$.
Hence, the multi-Regge limit is able to detect that the remainder
functions are non-zero, in spite of~\eqref{eq:MRLMain}.

%%%%%%%%%%%%%%%%%%%%%%%%%%%%%%%%%%%%%%%%%%%%%%%%%%%%%%%%%%%%
\subsection{Collinear Kinematics and Multi-Regge Limit}
\label{sec:collinear}

We want to study the multi-Regge limit using the Wilson loop OPE. The
latter is formulated as an expansion around a multi-collinear limit of
the external momenta. Luckily, the multi-collinear limit of the Wilson
loop OPE does have a non-trivial overlap with the multi-Regge limit,
at least up to $n=9$ external points.%
\footnote{In~\appref{app:mrl-tessellation}, we present a slightly different
parametrization that has an overlap with the multi-Regge limit for any
number of points.}
Hence we can zoom in on a combined multi-Regge collinear limit where
the Wilson loop OPE applies.

In order to study the combined multi-Regge collinear limit, we will
employ the natural variables that arose in the construction of the OPE
for null polygon Wilson
loops~\cite{Alday:2010ku,Gaiotto:2010fk,Gaiotto:2011dt,Sever:2011pc,Basso:2013vsa}.
The parametrization is based upon a tessellation of the polygon contour
with $n$ cusps into a sequence of $n-3$ null tetragons. Two of those
tetragons are boundary tetragons, the remaining $n-5$ tetragons are internal
tetragons. Compared to the hexagon and heptagon tessellations
of~\cite{Basso:2013vsa,Basso:2013aha}, our tessellation is cyclically
shifted.
% by one cusp, that is $x\suprm{here}_i\simeq x\suprm{there}_{i+1}$.
The tessellation is obtained by drawing unique null lines from
cusps $x_i$ to points $x'_i$ on edges $(x_j,x_j+1)$ of the polygon,
see \figref{fig:BSVtessellations}.
\begin{figure}
\centering
\includegraphics{FigBSVtessellations}
\caption{Polygon tessellations for up to $n=9$ points. The cases
$n=6,7$ are cyclic rotations of the parametrizations given in
Appendix~A of~\cite{Basso:2013aha}, the cases $n=8,9$ are
generalizations thereof, see~\protect\eqref{eq:ZBSV}. For $n=9$ we
found it convenient to
cyclically rotate again to find a non-trivial overlap between the
collinear and the multi-Regge limit.}
\label{fig:BSVtessellations}
\end{figure}
%
% cusps $x_i$ to points $x'_i$ on lines $(x_{n-i-1},x_{n-i-2})$, where
% $i=1,\dots,\floor{n/2}-2$, and from cusps $x_{n-i-1}$ to points
% $x'_{n-i-1}$ on lines $(x_{i-1},x_i)$, where
% $i=1,\dots,\floor{(n-1)/2}-1$. The internal tetragons are then formed by
% cusps $(x_i,x'_i,x_{n-i-1},x'_{n-i-1})$ or
% $(x'_{n-i-2},x_{n-i-2},x_i',x_i)$.
Each internal null tetragon is
preserved by three independent conformal transformations that are
parametrized by $\tau_i$, $\sigma_i$, and $\varphi_i$, where
$i=1,\dots,n-5$ now labels the internal tetragons.
% from ``bottom'' (near $x_n$) to ``top'' (near $x_{\floor{n/2}}$).
These variables are
conjugate to the energy, momentum, and helicity of flux tube
excitations in the frame defined by the respective tetragon~\cite{Alday:2010ku}.
In order to generate all conformally inequivalent configurations, we
start with a fixed ``reference'' polygon, and subsequently act
with the three conformal transformations that stabilize each
internal tetragon on all cusps $x_j$ that lie above that internal
tetragon.%
\footnote{Of course, one can alternatively act with the inverse
conformal transformations on the bottom part of the polygon. These two
choices are related by a global conformal transformation and hence
conformally equivalent.}
We will mostly use the exponentiated variables
\begin{equation}
T_i = e^{-\tau_i}
\,,\quad
S_i = e^{\sigma_i}
\,,
\quad
F_i = e^{i\varphi_i}
\,.
\label{eq:TSF}
\end{equation}
The Wilson loop OPE for the remainder function applies in
multi-collinear kinematics, which are attained when $T_i\ll 1$ with
$S_i$ and $F_i$ finite ($i=1,\dots,n-5$). Letting any $T_i\to0$ sends
all cusps that lie above the associated internal tetragon to points on
the top edge of that tetragon, thereby flattening the upper part of
the polygon. Further taking $S_i\to0$ sends all those cusps either to
the left end or to the right end of the top edge (depending on the
orientation of the tetragon). Conversely, taking $S_i\to\infty$ sends
all those cusps to the other end of the top edge.

Now, recalling the definition~\eqref{eq:uj} of the cross ratios
$u_{j,i}$, it is easy to see that the multi-Regge limit $u_{j,1}\to1$,
$u_{j,2},u_{j,3}\to0$ can be approached by letting all $T_i\to0$ as
well as all $S_i$ either to zero or to infinity. In fact, each triplet
$\brc{u_{j,1},u_{j,2},u_{j,3}}$ can be associated to one particular
internal tetragon, in the sense that the triplet approaches
$\brc{1,0,0}$ as the variables $\brc{T_i,S_i}$ of that tetragon are
sent either to $\brc{0,0}$ or to $\brc{0,\infty}$. For example,
consider the heptagon in \figref{fig:BSVtessellations}. The upper
internal tetragon is associated with the variables
$\brc{T_1,S_1,F_1}$, the lower tetragon with the variables
$\brc{T_2,S_2,F_2}$. Letting $T_1\to0$ and $S_1\to0$ sends cusps $3$
to cusp $2$ and cusp $4$ to the left end of line $\text{a}$. It is
clear that in this limit $u_{1,1}=U_{2,5}\to1$, $u_{1,2}=U_{3,7}\to0$,
and $u_{1,3}=U_{1,4}\to0$. Similarly, letting $T_2\to0$ and
$S_2\to\infty$ sends cusps $2$, $3$, and $4$ to the right end of line
$\text{b}$, upon which $u_{2,1}=U_{3,6}\to1$, $u_{2,2}=U_{4,7}\to0$,
and $u_{2,3}=U_{1,5}\to0$. Similar relations hold for all internal
tetragons and cross ratios $u_{j,i}$ of the four polygons
in \figref{fig:BSVtessellations}.

All in all,
the multi-Regge regime
corresponds to a double-scaling limit where $T_i\ll 1$
with either $r_i=S_i/T_i$ or $r_i=1/(S_iT_i)$ finite (depending on $n$ and $i$).
Explicit formulas for the momentum twistors parametrizing the $n$-gon
for $n=6,7,8,9$ are given in \appref{app:bsvtessdetail}.
For our choice of reference polygon (as given in~\appref{app:bsvtessdetail}),
we have to make the following identifications between the
multi-collinear parameters $r_i$, $F_i$ and the multi-Regge variables
$w_i$, $\wb_i$ to recover the multi-Regge
parametrization~\eqref{eq:uofw} in the limit $T_i\to0$:%
\footnote{The nonagon has permuted labels compared to the other
polygons because we chose to extend the octagon at the top instead of
the bottom (see \figref{fig:BSVtessellations}), and therefore have to
rotate the external labels in order to maintain a non-trivial overlap
between the multi-collinear and the multi-Regge limit. The benefit of
this choice is a simpler expression for the conformal transformations
that preserve the upper internal square and thus simpler expressions
for the cross ratios. See \appref{app:bsvtessdetail}, in
particular~\eqref{eq:M1234} there.}
\begin{align}
n&=6,7,8: & S_1&=r_1T_1           \,, & r_1&=\frac{1}{\sqrt{w_1\wb_1}} \,, & F_1&=\sqrt{\frac{w_1}{\wb_1}}  \,, \nn\\
 &        & S_2&=\frac{1}{r_2T_2} \,, & r_2&=\frac{1}{\sqrt{w_2\wb_2}} \,, & F_2&=\sqrt{\frac{w_2}{\wb_2}}  \,, \nn\\
 &        & S_3&=\frac{1}{r_3T_3} \,, & r_3&={\sqrt{w_3\wb_3}}         \,, & F_3&=-\sqrt{\frac{w_3}{\wb_3}} \,,
\label{eq:Frw678}
\\[2ex]
n&=9:     & S_1&=\frac{1}{r_1T_1} \,, & r_1&={\sqrt{w_3\wb_3}}         \,, & F_1&=\sqrt{\frac{w_3}{\wb_3}}  \,, \nn\\
 &        & S_2&=\frac{1}{r_2T_2} \,, & r_2&=\frac{1}{\sqrt{w_2\wb_2}} \,, & F_2&=\sqrt{\frac{w_2}{\wb_2}}  \,, \nn\\
 &        & S_3&={r_3T_3}         \,, & r_3&=\frac{1}{\sqrt{w_1\wb_1}} \,, & F_3&=-\sqrt{\frac{w_1}{\wb_1}} \,, \nn\\
 &        & S_4&={r_4T_4}         \,, & r_4&={\sqrt{w_4\wb_4}}         \,, & F_4&=\sqrt{\frac{w_4}{\wb_4}}  \,.
\label{eq:Frw9}
\end{align}
The arguments of large  logarithms in the multi-Regge limit are given by
\begin{equation}
\eps_j\equiv u_{j,2}u_{j,3}\,,
\qquad
j=1,\dots,n-5\,.
\label{eq:epsjdef}
\end{equation}
With the above parametrization, we have in the limit $T_i\to0$:
\begin{align}
n&=6,7,8: & \eps_1&=r_1^2T_1^4 \,, & \eps_2&=r_2^2T_2^4 \,, & \eps_3&=r_3^2T_3^4 \,,
\label{eq:reps678}
\\[1ex]
n&=9:     & \eps_1&=r_3^2T_3^4 \,, & \eps_2&=r_2^2T_2^4 \,, & \eps_3&=r_1^2T_1^4 \,, & \eps_4&=r_4^2T_4^4 \,,
\end{align}
and
\begin{align}
\frac{u_{j,2}}{1-u_{j,1}}&=\frac{1}{\abs{1+w_{j}}^2}+\sum_i\order{T_i^2} \,, &
\frac{u_{j,3}}{1-u_{j,1}}&=\frac{\abs{w_{j}}^2}{\abs{1+w_{j}}^2}+\sum_i\order{T_i^2} \,.
\label{eq:ujaofwwb}
\end{align}
Formulas for the cross ratios $u_{j,a}$ in terms of the tetragon
variables $T_j,S_j,F_j$  for the hexagon and the heptagon are given in
the respective sections below. The respective formulas for the octagon
and nonagon can be found in \appref{app:bsvtessdetail}.

%%%%%%%%%%%%%%%%%%%%%%%%%%%%%%%%%%%%%%%%%%%%%%%%%%%%%%%%%%%%
\subsection{Analytic Continuation in the Collinear Limit}

Our goal is to recover the combined multi-Regge collinear limit of the remainder
function in all Mandelstam or multi-Regge regions from the Wilson loop OPE.
The latter is computed on the main sheet $\varrho = (+, \dots,+)$
and must be continued into the non-trivial Mandelstam regions along some curve
$\curve$. The main goal of this subsection is to set up some notation
that allows to evaluate the result of such analytical continuations in collinear
kinematics. Since we are only interested in the combined multi-Regge collinear
limit, we shall implement the multi-Regge limit throughout.

As we will see in the next section, it is not too difficult to construct the
finite remainder function to leading order in the $T_i$ at low number of loops
from the Wilson loop OPE. The explicit expressions $\Rnlope$ turn out to contain several
functions of the variables $S_i$ which possess branch cuts ending at $r$
hypersurfaces $\sigma_\nu, \nu = 1, \dots, r$ of co-dimension two in the space
of complexified $S$-variables. We shall choose a set of generators $p_\nu$ for
the fundamental group $\pi_1$ of the complement, \ie $p_\nu \in
\pi_1(\mathbb{C}^{n-5}\setminus \{\sigma_\nu\,|\, \nu = 1 , \dots, r\})$. If
we continue the remainder function $\Rnlope$ along a curve $\gen_\nu$ associated to the
generator $p_\nu$, we may pick up a cut contribution $\Delta_\nu \Rnlope$ whose multi-Regge limit
may or may not vanish:
\begin{equation}
    \label{eq:Cicont}
    \bigsbrk{\gen_\nu  \Rnlope}\supMRL
    \equiv \bigbrk{1+{2\pi i}\Delta_\nu} \Rnlope\ .
\end{equation}
Let us now pick an arbitrary element $\curve \in \pi_1$ in our fundamental group.
By construction, $\curve$ can be written as a product of generators $p_\nu$, \ie it is a
finite product of the form $\curve= \prod_k p_{\nu_k}^{n_k}= p_{\nu_1}^{n_1}
p_{\nu_2}^{n_2} \cdots $ with $\nu_k \in \{1,\dots,r\}$ and $n_k \in \{\pm 1\}$.
Note that the fundamental group is not abelian so that the order of factors matters.
Continuation along a curve $\gen_\curve$ that is associated to the element
$\curve$ gives
\begin{equation}\label{eq:Cgcont}
\bigsbrk{\gen_\curve \Rnlope}\supMRL = \prod_k (1+ 2\pi i\Delta_{\nu_k})^{n_k} \Rnlope =
(1+2\pi i\Delta_{\nu_1})^{n_1} (1+2\pi i\Delta_{\nu_2})^{n_2} \cdots \Rnlope\ .
\end{equation}
Here, $(1+2\pi i\Delta)^{-1}$ is defined as a formal geometric series expansion in
$2\pi i\Delta$. Let us make a few comments. First of all, our symbols $(1+2\pi i\Delta_\nu)$
are a bit formal. One should first expand all the terms with $n_k = -1$, and then write the
right hand side as a sum of `products' of the $\Delta_\nu$. Each of the terms in this sum then
stands for \emph{the multi-Regge limit} of a particular multiple cut
contribution. We drop the leading term $1\cdot\Rnlope$, since it
vanishes in the multi-Regge limit.
We stress that
taking the multi-Regge limit does not commute with evaluating cut contributions, \ie even
if the multi-Regge limit $\Delta_\nu R$  of a cut contribution vanishes, the
multi-Regge limit $\Delta_\mu \Delta_\nu R$ of a double-cut contribution
may not vanish. We shall see examples later on.

The right hand side of eq.~\eqref{eq:Cgcont} can now be expanded in the number of cut
contributions, starting with those terms that contain a single cut $\Delta_\nu$,
\begin{equation}
\bigsbrk{\gen_\curve \Rnlope}\supMRL = \biggbrk{
 2 \pi i\sum_\nu\cc{\nu}\Delta_\nu
 +(2 \pi i)^2\sum_{\mu,\nu}\cc{\mu,\nu}\Delta_\mu\Delta_\nu
 + \dots}
\Rnlope
\label{eq:discRnlope}
\end{equation}
Let us point out that the sum on the right hand side is finite at any given loop order
$l$ since the maximal number of non-vanishing discontinuities at $l$ loops is $2l-1$.

In principle, there is a unique curve $\curve_\varrho$ associated to each Mandelstam
region, and if this curve was known, we could simply compute the combined multi-Regge
collinear limit in all regions
\begin{equation}  \label{eq:MRLCLCLMRL}
\bigsbrk{R_{n,(l)}^{\varrho}}^\CL = \bigsbrk{\gen_{\curve_\varrho} \Rnlope}\supMRL
\end{equation}
and obtain strong constraints on the remainder function in multi-Regge kinematics.
In practice, however, the curve $\curve_\varrho$ is not known, and the equality~\eqref{eq:MRLCLCLMRL},
\begin{equation} \label{eq:MRLCLCLMRL2}
\bigsbrk{R_{n,(l)}^{\varrho}}^\CL = \biggbrk{
 2 \pi i\sum_\nu\cc{\nu}\Delta_\nu
 +(2 \pi i)^2\sum_{\mu,\nu}\cc{\mu,\nu}\Delta_\mu\Delta_\nu
 + \dots}
\Rnlope
\end{equation}
imposes constraints on both sides of the equation, \ie on the remainder function
in multi-Regge kinematics and on the curve $\curve_\varrho$ (through
the coefficients $\cc{\mu}$, $\cc{\mu,\nu}$ \etc).
Eqs.~\eqref{eq:MRLCLCLMRL}
or~\eqref{eq:MRLCLCLMRL2} are our key to constraining the multi-Regge limits of the
remainder function in subsequent sections.

Even though the curve $\curve_\varrho$ is not known in general, the coefficients
$\cc{\bigcdot}$ that characterize the discontinuity expansion are not entirely free. The
first set of constraints comes from the winding numbers $n_{ij}(\varrho)$ we
defined in eq.~\eqref{eq:winding}. Let us recall that there exists a famous
projection from the fundamental or first homotopy group $\pi_1$ to the first
homology group $H_1$ of our space $\mathbb{C}^{n-5} \setminus \{ \sigma_\nu\,
|\, \nu = 1, \dots, r\} $. Elements of the latter are characterized by the
winding numbers of the former around the endpoints $\sigma_\nu$ of our branch
points. Recall that $H_1$ is an abelian group that is obtained from the non-abelian
fundamental group $\pi_1$ by equating all commutators to the unit element. This
set of winding numbers is clearly not sufficient to determine the associated
curve, but it allows us to compute the leading coefficients $\cc{\nu}$. In
fact, for each generator $p_\nu$ of the fundamental group, one can compute the
winding numbers $n_{ij}^\nu$ of the cross ratios $U_{ij}$. These allow to
constrain the coefficients $\cc{\nu} = \cc{\nu}(\varrho)$ as
\begin{equation} \label{eq:sdisc}
\sum_\nu n_{ij}^\nu  \cc{\nu}(\varrho) = n_{ij}(\varrho)
\,,
\end{equation}
where $n_{ij}(\varrho)$ are the known winding numbers~\eqref{eq:winding} of the cross
ratios $U_{ij}$ as one continues into the Mandelstam region $\varrho$.
As we will see below, there are as many independent cross ratios
$U_{ij}$ as there are generators $p_\nu$, and hence eq.~\eqref{eq:sdisc}
completely fixes the coefficients $\cc{\nu}$.
In addition to these constraints on the coefficients of single discontinuities,
we can also constrain the coefficients of multiple discontinuities. As an example,
let us consider the following equality
\begin{equation}
    \label{eq:acommut}
    d_{\mu, \nu}\brk{\gamma}
    := \cc{\mu, \nu}\brk{\gamma} + \cc{\nu, \mu}\brk{\gamma}
    = \cc{\mu}\brk{\gamma} \cc{\nu}\brk{\gamma} \equiv \cc{\mu} \cc{\nu}
    \,,
\end{equation}
which allows to determine the coefficient $\cc{\nu,\mu}$ from $\cc{\mu,\nu}$
along with the coefficients $\cc{\nu}$ of the single discontinuities.

To prove the middle identity in eq.~\eqref{eq:acommut}, we may ignore all
generators $\gen_{k}$ with  $k \ne \mu, \nu$ that appear in the expression for the curve $\gamma$.
Hence, for the purpose of computing $\cc{\mu,\nu}$ and
$\cc{\nu,\mu}$ we will think of the curve as a product of generators $p_\mu$, $p_\nu$ only \brk{and their
inverses}. Of course, the order in which these factors appear in the curve does
matter for the individual coefficients, but not for the anti-commutator
$d_{\mu,\nu}$ we defined in eq.~\eqref{eq:acommut}. In fact, since $d_{\mu, \nu}$ is
computed from the coefficients of double discontinuities, we obtain
\begin{align}
    &\brk{1 + 2 \pi i \Delta_\mu}^{n} \brk{1 + 2 \pi i  \Delta_\nu}^m -
    \brk{1 + 2 \pi i  \Delta_\nu}^{m} \brk{1 + 2 \pi i  \Delta_\mu}^n
    \nn\\
    &\quad\sim
      \brk{1 + 2 \pi i \mskip4mu n \Delta_\mu} \brk{1 + 2 \pi i \mskip4mu m \Delta_\nu}
    - \brk{1 + 2 \pi i \mskip4mu m \Delta_\nu} \brk{1 + 2 \pi i \mskip4mu n \Delta_\mu}
    = (2 \pi i)^2 n m \, \brk{\Delta_\mu \Delta_\nu - \Delta_\nu \Delta_\mu}.
    \nn
\end{align}
Consequently, in calculation of the anti-commutator $d_{\mu, \nu}$ we do not have to worry about
the order of generators $p_\mu$ and $p_\nu$. It is therefore straightforward to relate
the anti-commutator $d_{\mu, \nu}$ and the coefficients of the single discontinuities $\cc{\nu}$:
\begin{equation}
    d_{\mu, \nu}\brk{\gamma}
    = d_{\mu, \nu}\brk{\brk{1 + \Delta_\mu}^{\cc{\mu}} \brk{1 + \Delta_\nu}^{\cc{\nu}} + \dots}
    = d_{\mu, \nu}\brk{\cc{\mu} \cc{\nu} \Delta_\mu \Delta_\nu + \dots}
    = \cc{\mu} \cc{\nu}.
\end{equation}
Here, the ellipses stand for terms with $d_{\mu, \nu}\brk{\dots} = 0$, and $\cc{\mu}$,
$\cc{\nu}$ count the overall amount of $\Delta_\mu$, $\Delta_\nu$ respectively. This completes
the proof of eq.~\eqref{eq:acommut}.

In all of this discussion, we have ignored one detail that will start to appear from
$n=7$ external gluons. As we pointed out in \secref{sec:variables}, some
of the winding numbers of small cross ratios may fail to be integer. The meaning
of half-integer winding numbers is simple: We need to allow cross ratios to become
negative by going above \brk{winding number $+1/2$} or below
\brk{winding number $-1/2$} the origin. Without
loss of generality, we choose to append such continuations to the very end of
our curves. We shall explain this in more detail below when be discuss the
heptagon.

%%%%%%%%%%%%%%%%%%%%%%%%%%%%%%%%%%%%%%%%%%%%%%%%%%%%%%%%%%%%
%%%%%%%%%%%%%%%%%%%%%%%%%%%%%%%%%%%%%%%%%%%%%%%%%%%%%%%%%%%%
\section{OPE Expansion to Two Loops}
\label{sec:opeexpansion}
In this section we want to construct the collinear limit of the two-loop heptagon
remainder function from the Wilson loop OPE. The result is spelled out in the third
subsection. Its derivation needs some background about the flux tube in four-dimensional
$\mathcal{N}=4$ SYM theory and the Wilson loop OPE, which we provide in the first two
subsections in order to keep our discussion self-contained. While some parts of the
leading collinear terms in the two-loop heptagon remainder function had been computed
before, the complete result also contains new terms which we derive in the final
subsection, after a short warm-up with a one-loop calculation.

%%%%%%%%%%%%%%%%%%%%%%%%%%%%%%%%%%%%%%%%%%%%%%%%%%%%%%%%%%%
\subsection{The Flux Tube or GKP String}

The formula for the remainder function that we spell out in the next
subsection realizes an old idea in quantum field theory, namely to
construct the four-dimensional amplitudes in terms of a one-dimensional
quantum system that describes the famous flux tube and its
excitations. In the case of $\mathcal{N}=4$ super Yang--Mills theory,
this flux tube is also known as Gubser--Klebanov--Polyakov (GKP) string,
referring to the incarnation of the flux tube in the dual $AdS_5$
geometry~\cite{Gubser:2002tv}. Let us describe some facts about this
one-dimensional quantum systems that will become relevant below.

The excitations of the GKP string can be considered as multi-particle
states that are built up from a set of single-particle excitations.
The set of single-particle excitations is known~\cite{Basso:2010in} to
consist of six `scalars' $\varphi$, eight `fermions' $\psi, \bar \psi$,
two `gluons' $F, \bar F$, and so-called `gluon bound states' $D^kF$
and $D^k\bar F$, where $k$ can be any positive integer $k=1,2,3, \dots$. The
names of these excitations refer to their four-dimensional origin. They
carry an action of the four-dimensional R-symmetry group SO(6) under
which the scalars transform in the vector representations, while
fermions $\psi$ and $\bar \psi$ are spinors. All gluon bound states,
finally, transform as scalars under SO(6).
In the one-dimensional system, all the elementary particles can
move with some rapidity $u$. As a consequence of integrability,
their dispersion law is known for any value of the coupling parameter $g$. In
fact, these quantities are determined by the famous
Beisert--Eden--Staudacher (BES) equation~\cite{Beisert:2006ez}, which can be
solved to any desired order, both at weak and strong coupling. For us,
only the gluon and gluon bound state excitations are relevant. For these,
the leading order terms of the energy $E=E(u)$ and the momentum $p(u)$
take the form~\cite{Basso:2010in}
\begin{equation}
E_{D^kF}(u) = E_{D^k\bar F}(u) =
1+ k + 2g^2 \lrbrk{\psi\Bigbrk{\sfrac{k+3}{2}+iu} +
\psi\Bigbrk{\sfrac{k+3}{2}-iu} -
2 \psi(1)} + O(g^4)\ ,
    \label{eq:GluonEnergy}
\end{equation}
and
\begin{equation}
 \quad p_{D^kF}(u) = p_{D^k\bar F}(u)= 2u +
 2ig^2 \lrbrk{\psi\Bigbrk{\sfrac{k+1}{2}+iu} -
 \psi\Bigbrk{\sfrac{k+1}{2}-iu}} + O(g^4) \ .
\end{equation}
Here, we allow for $k$ to be $k=0$, in which case the formulas
give the dispersion law of the gluon excitations. Let us also mention
that the one-particle excitations possess a conserved $\grp{U}(1)$ charge $m$,
which we shall refer to as helicity. For the gluon bound states, this
is simply given by
\begin{equation}
m_{D^kF} = 1 + k
\,,\qquad
\ m_{D^k\bar F} = -1 - k
\,.
\end{equation}
Being a discrete quantum number, $m$ neither depends on the coupling
$g$ nor on the rapidity $u$. To complete the description of
single-particle excitations, let us note that the one-particle wave
functions
$\Psi(u)$ are integrated with a measure that depends on the rapidity. Once
again, this measure is known for all one-particle excitations and any
coupling. In the case of gluons and gluon bound states, it reads
\begin{equation}
\mu_{D^kF}(u) = \mu_{D^k\bar F}(u) = (-1)^{k+1} g^2
\frac{\Gamma(\frac{k+1}{2}+iu) \Gamma(\frac{k+1}{2}-iu)}
{\Gamma(k+1)(u^2+(k+1)^2/4)} + O(g^4)\ .
\end{equation}
Similar formulas also exist for the other one-particle excitations, \ie
the scalars and fermions. Since we won't need them below, we refrain from
spelling them out here. Let us only mention that all one-particle excitations
satisfy $E^{g=0}_X(u) \geq 1$. Equality holds only for scalars, fermions and
gluons, but obviously not for non-trivial gluon bound states.

From the one-particle states, one can now build up multi-particle
excitations. These can contain any number $N$ of single-particle
excitations, each with its own rapidity $u_a, a=1, \dots, N$. Since the
GKP string is integrable, the energy $E$, momentum $p$ and helicity $m$
of such multi-particle states can be computed as the sum of their
single-particle constituents. The interaction between the single-particle
excitations is described by a factorizable S-matrix, \ie it can be
built from the two-particle S-matrix. The latter is also known
explicitly, but since we will not need it here, at least not
directly, we will not give explicit formulas.

%%%%%%%%%%%%%%%%%%%%%%%%%%%%%%%%%%%%%%%%%%%%%%%%%%%%%%%%%%%%
\subsection{Wilson Loop OPE and Finite Remainder}

After this preparation, we are now able to state the main result
from~\cite{Basso:2013vsa, Basso:2013aha, Basso:2014koa, Basso:2014nra}.
According to Basso \etal, the finite remainder function of a color
ordered planar MHV $n$-gluon amplitude in $\mathcal{N}=4$ super
Yang--Mills theory is given by
\begin{equation}\label{eq:Remainder}
R_g(\tau_i,\sigma_i,\varphi_i) =
\log \mathcal{W}_g(\tau_i,\sigma_i,\varphi_i)
- \log \WBDS_g(\tau_i,\sigma_i,\varphi_i)
\end{equation}
where the first term takes the form
\begin{equation}\label{eq:WilsonloopOPE}
\mathcal{W}_g(\tau_i,\sigma_i,\varphi_i) =
\sum_{\Psi_i} \lrsbrk{\prod_{i=1}^{n-5} e^{-E^g_i\tau_i+
i p^g_i\sigma_i +i m_i\varphi_i}} P_g(0|\Psi_1)
P_g(\Psi_1|\Psi_2) \dots P_g(\Psi_{n-5}|0)\ .
\end{equation}
Let us explain the individual pieces of this formula. First of all, we
need to discuss the summation. The $n-5$ objects $\Psi_i$, $i = 1, \dots n-5$
that we sum over are $n-5$ multi-particle excitations of the GKP string.
Hence the sum consists of a discrete (but infinite) summation over the
single-particle content of the multi-particle states, along with an
integration over the rapidity variables. The rapidity integration must
be performed with the appropriate measure $\mu^g(u)$. To be quite precise,
it should also contain appropriate symmetry factors that depend on the
exact multi-particle content, but we will not need these below.

$E_i$, $p_i$, and $m_i$ denote the energy, momentum, and helicity of these multi-particle
states. Recall that these are simply obtained by summing the energy, momentum, and
helicity of the single-particle constituents. The kinematic invariants $\tau_i$,
$\sigma_i$, and $\varphi_i$ of our scattering process multiply the energies, momenta,
and helicities. Finally, the factors $P$ are known as pentagon transitions. One
should think of them as being determined uniquely by the S-matrix of the GKP
string. For gluon excitations, the pentagon transitions are
\begin{align}
P_{FF}(u|v) = P_{\bar F\bar F} (u|v)& =  - \frac{1}{g^2}\frac{\Gamma(iu-iv)}
{\Gamma(-\frac12+iu) \Gamma(-\frac12-iv)} + O(g^0)
\,, \\[2mm]
P_{F\bar F}(u|v) = P_{\bar F F}(u|v) & = \frac{\Gamma(2+iu-iv)}{\Gamma(\frac32+iu)
\Gamma(\frac32-iv)} + O(g^2)
\,,
\end{align}
and $P_F(0|u) = 1 = P_{\bar F}(0|u)$. The formulas can be extended to any pair of
multi-particle excitations, and in particular to gluon bound states and their
multi-particle composites. We will not need these formulas below. Let us only
mention that $P(\Psi_1|\Psi_2)$ satisfies an important selection rule. As we
pointed out above, all single-particle excitations transform under the
space-time R-symmetry SO(6). The action on single-particle states induces
an action on the multi-particle states~$\Psi_1$ and~$\Psi_2$. The pentagon
transition $P(\Psi_1|\Psi_2)$ intertwines this action. Since the vacuum $0$
of the GKP string is SO(6) invariant, the transition $P(0|\Psi)$ can only be
non-zero if the action of SO(6) on $\Psi$ contains a trivial subrepresentation.
If $\Psi$ is a single-particle state, it must be a gluon or gluon bound state
in order for $P(0|\Psi)$ to be non-trivial.

It remains to describe the second term in eq.~\eqref{eq:Remainder}.%
\footnote{See the discussion in Section~3.3 of~\cite{Alday:2010ku}.}
%
% namely the function $\WBDS_g$.
The function $\mathcal{W}_g$ in the first term
represents a certain ratio of polygon Wilson loops.
$\WBDS_g$ represents the same ratio, but in a free abelian
$\grp{U}(1)$ theory with coupling $\gcusp$. Both ratios
$\mathcal{W}_g$ and $\WBDS_g$ obey the same anomalous Ward identities,
such that their ratio $\mathcal{W}_g/\WBDS_g$ is UV finite, and its
logarithm equals the remainder function, as stated in
eq.~\eqref{eq:Remainder}. Concretely, the second term is given by
\begin{equation}
\log \WBDS_g(\tau_i,\sigma_i,\varphi_i) \equiv
\frac{\gcusp(g)}{4g^2}
\brk{\log\mathcal{W}_g}^{(1)}(\tau_i,\sigma_i,\varphi_i)\ ,
\end{equation}
where $\brk{\log\mathcal{W}_g}^{(1)}$ denotes the one-loop part of the
function $\log\mathcal{W}_g$ defined in eq.~\eqref{eq:WilsonloopOPE}. We
divide by $4g^2$ in order to remove the dependence on the coupling from
the one-loop result, and then multiply with the so-called cusp anomalous
dimension $\gcusp$. The latter describes the vacuum energy
of the GKP string, and it is also known for any value of the coupling. Its
weak coupling expansion reads
\begin{equation}
\gcusp(g) = 4 g^2 - \frac{4\pi^2}{3}g^4 + O(g^6) \ .
\end{equation}
It is not too difficult to work out explicit formulas for the function
$\WBDS$ for any number of external gluons.

%%%%%%%%%%%%%%%%%%%%%%%%%%%%%%%%%%%%%%%%%%%%%%%%%%%%%%%%%%%%
\subsection{The Heptagon Remainder Function}

We now describe the hexagon and heptagon remainder functions in the collinear limit
at two loops. We first discuss the general structure of the collinear expansion in
a bit more detail, mainly to fix our notation. Then, we state our results on the
two-loop heptagon remainder function. We restrict ourselves to the first non-trivial
terms in the collinear expansion of eq.~\eqref{eq:WilsonloopOPE}: For $n \in \brc{6,
7}$ we consider the two-loop term proportional to $g^4$ with only the lowest corrections
to the asymptotic expansion in $T_i \to 0$. As it was shown in~\cite{Basso:2013aha},
only the one-gluon excitation can contribute to these terms, which makes the problem
much more manageable.

Let us start with putting more structure to eqs.~\eqref{eq:Remainder} and~\eqref{eq:WilsonloopOPE},
following~\cite{Basso:2013aha}. For the hexagon, the
collinear ($T_1 \to 0$) expansion at weak coupling takes the form (see eq.\ (35)
in~\cite{Basso:2013aha})
\begin{equation}
    \Wl_{\mathrm{hex}}
    = 1
    + 2 T_1 \cos\brk{\varphi_1} \, \f_1\brk{T_1, S_1}
    + \order{T^2}
    \,,
    \label{eq:R62ope}
\end{equation}
where the second term proportional to $\f_1$ comes form the propagation
of a one-gluon excitation through the only internal tetragon of hexagon
\brk{see also \secref{sec:variables}}. Similarly, the leading term in
the collinear ($T_i \to 0$) expansion of the heptagon is (see eq.\ (38)
in~\cite{Basso:2013aha}):
\begin{align}
    \Wl_{\mathrm{hep}}
    = 1
    &+2T_1\cos\brk{\varphi_1}\,\f_1\brk{T_1,S_1}
     +2T_2\cos\brk{\varphi_2}\,\f_2\brk{T_2,S_2}
    \nn\\
    &+2T_1T_2\cos\brk{\varphi_1+\varphi_2}\,h_{12}\brk{T_1,T_2,S_1,S_2}
    \nn\\
    &+2T_1T_2\cos\brk{\varphi_1-\varphi_2}\,\bar h_{12}\brk{T_1,T_2,S_1,S_2}
     +\order{T^2}\,.
    \label{eq:R72ope}
\end{align}
Here, the first line comes from the one-gluon excitation in either of the
internal tetragons of the heptagon, and the second and third lines from
excitations in both of the tetragons.
The hexagon functions $\f_1$ and $\f_2$ are identical:
\begin{equation}
\f_1(T,S)\equiv \f(T,S)
\,,\quad
\f_2(T,S)\equiv \f(T,S)
\,,
\end{equation}
where $\f(T,S)$ is the usual hexagon function of~\cite{Basso:2013aha}.
The heptagon functions $h_{12}\equiv h$ and $\bar h_{12}\equiv\bar h$ are
also defined in~\cite{Basso:2013aha}.
The hexagon and heptagon functions are graded in powers of $\log\brk{T_i}$:
\begin{align}
    \f\brk{T, S} &= \sum_{L \ge 1} \sum_{p = 0}^{L - 1} \f^{\brk{p}}_L \bigbrk{\log\brk{T}}^p,
    \\
    h\brk{T_1, T_2, S_1, S_2} &=
    \sum_{L \ge 1} \sum_{p_1, p_2} h^{\brk{p_1, p_2}}_L \bigbrk{\log\brk{T_1}}^{p_1} \bigbrk{\log\brk{T_2}}^{p_2},
    \\
    \bar{h}\brk{T_1, T_2, S_1, S_2} &=
    \sum_{L \ge 1} \sum_{p_1, p_2} \bar{h}^{\brk{p_1, p_2}}_L \bigbrk{\log\brk{T_1}}^{p_1} \bigbrk{\log\brk{T_2}}^{p_2},
\end{align}
where the inner sum on the second line covers the $0 \le p_1 + p_2 \le
L - 1$ domain, whereas the inner sum in the last line covers $0\le
p_1+p_2\le L-2$.
The components $\f^{\brk{p}}_L$, $h_L^{\brk{p_1,
p_2}}$, and $\bar h_L^{\brk{p_1, p_2}}$
are proportional to $g^{2 L}$. The relevant terms in this grading up to two-loop order are
\begin{align}
    \f\brk{T,S}
    &= \f^{\brk{0}}_1\brk{S}+\f^{\brk{0}}_2\brk{S}+\log\brk{T} \f^{\brk{1}}_2\brk{S}+\dots\,,
    \label{eq:f2}
    \\[2mm]
    h\brk{T_1,T_2,S_1,S_2}
    &= h^{\brk{0,0}}_1\brk{S_1,S_2}
    + h^{\brk{0,0}}_2\brk{S_1,S_2}
    \nn\\
    & \quad + \log\brk{T_1}\,h^{\brk{1,0}}_2\brk{S_1,S_2}
    + \log\brk{T_2}\,h^{\brk{0,1}}_2\brk{S_1,S_2}
    + \dots\,,
    \label{eq:h2}
    \\[2mm]
    \bar h\brk{S_1,S_2}
    &= \bar h^{\brk{0,0}}_2\brk{S_1,S_2}+\dots\,.
    \label{eq:h2bar}
\end{align}
The component functions $\f^{\brk{0}}_L$, $\f^{(1)}_L$, $h^{\brk{0,0}}_L$, and
$\bar h^{\brk{0,0}}_L$ to two loop order can be found in the literature,
see the discussion around eqs.\ (118), (125) and (126) in~\cite{Basso:2013aha}
as well as the file \filename{Functionshf.nb} accompanying that paper, and
eq.\ (62) in~\cite{Papathanasiou:2013uoa} together with the file
\filename{MHV\_full.m} there. For $\tilde{f}_2^{(0)}$ and
$\tilde{f}_2^{(1)}$, see also~\eqref{eq:f20} and~\eqref{eq:f21} below.
The only missing pieces that we need to compute
ourselves are the functions $h^{\brk{1,0}}_2$ and $h^{\brk{0,1}}_2$ that appear
at two loops. As we shall show below, the first of these functions is given
by
\begin{align}
h^{(1,0)}_2(S_1,S_2)&=
\frac{2g^4}{S_1S_2}
\Bigsbrk{
(1+S_1^2)S_2^2\log\brk{1+S_1^2}\Bigbrk{\log\brk{1+S_1^2}-\log\brk{S_1^2}-2}
\nn\\&\mspace{70mu}
-2 S_1^2 \log\brk{S_1^2}
-2S_2^2\log\brk{S_2^2}
-2S_1^2(1+S_2^2)\log\brk{1+S_2^2}
\nn\\&\mspace{70mu}
-(S_1^2+S_2^2+S_1^2S_2^2)\Bigbrc{
    \log\brk{S_1^2}\log\brk{S_2^2}
    +\log\brk{S_2^2}\log\brk{1+S_2^2}
    \nn\\&\mspace{70mu}
    -\log\brk{S_1^2+S_2^2+S_1^2S_2^2}\Bigbrk{
        2+\log\brk{S_1^2}+\log\brk{S_2^2}+\log\brk{1+S_2^2}
    }
    \nn\\&\mspace{70mu}
    +\log\brk{S_1^2+S_2^2+S_1^2S_2^2}^2
}}
\label{eq:h2result}
\end{align}
The second function is then obtained by swapping the S-variables, \ie
\begin{equation}
h^{(0,1)}_2(S_1,S_2)=
h^{(1,0)}_2(S_2,S_1)\,.
\label{eq:h2resultother}
\end{equation}
The general one-gluon contribution to the OPE of the heptagon remainder function is
given in eq.\ (39) of~\cite{Basso:2013aha}
\begin{equation}
    h=\int \frac{du_1 du_2}{4\pi^2} \mu(u_1) P(-u_1 | u_2) \mu(u_2)
    e^{-\tau_1 E(u_1) + i p(u_1) \sigma_1 - \tau_2 E(u_2) + i p(u_2) \sigma_2}\ .
    \label{eq:hint}
\end{equation}
The helicity-breaking transition $\bar{h}$ has the same form with $P \to \bar{P}$.
In the following two subsections, we will describe how one can obtain explicit
results for all of the building blocks in eqs.~\eqref{eq:f2, eq:h2, eq:h2bar}
from the OPE integral~\eqref{eq:hint}.

%%%%%%%%%%%%%%%%%%%%%%%%%%%%%%%%%%%%%%%%%%%%%%%%%%%%%%%%%%%%
\subsection{Evaluation of the Collinear Remainder Function}

The goal of this final subsection is to provide a brief sketch of the methods
that allow to evaluate the remainder function in the collinear limit. Before
we dive into the full complexity of the two-loop integral~\eqref{eq:h2}, we
first calculate the function $h^{\brk{0,0}}_1$ at one loop in order to
illustrate the main steps in a more pedagogical setting.

%%%%%%%%%%%%%%%%%%%%%%%%%%%%%%
\subsubsection{One-loop Remainder Function}
\label{sec:1loopRem}

The one-loop contribution $h_1^{\brk{0,0}}$ comes from the first term in the
$g^2$ expansion of the integral~\eqref{eq:hint}
\begin{equation}
    h_1^{\brk{0,0}} = g^2 \int \frac{\dd u_1 \, \dd u_2}{4\pi^2}
    e^{2 i \brk{\sigma_1 u_1 + \sigma_2 u_2}}
    \frac{
        \Gamma\brk{\tfrac32 + i u_1} \Gamma\brk{-i u_1 - i u_2} \Gamma\brk{\tfrac32 + i u_2}
    }{
        \brk{u_1^2 + \tfrac14} \brk{u_2^2 + \tfrac14}
    }
    \,.
    \label{eq:h1int}
\end{equation}
We compute this integral via the residue theorem. The $u_1$--contour of integration
can be closed in the upper half-plane so that only the poles at $u_{1,\mathrm{pole}} =
i \brk{\tfrac12 + k_1},\ k_1 \in \Z_{\ge 0}$ contribute. Following~\cite{Papathanasiou:2013uoa}
we take $u_1 = u_{1, \mathrm{pole}} + \eps$ and pick the $\eps^{-1}$--term in the series
expansion $\eps \to 0$ to compute the residue. We then repeat the same procedure for $u_2$
and arrive at the following double sum over poles,
\begin{align}
    h_1^{\brk{0,0}} = \frac{g^2}{S_1 S_2} \Biggsbrk{
        &\sum\limits_{k_{1, 2} \in \Z_{\ge 1}}
        \frac{
            \brk{-S_1^{-2}}^{k_1} \brk{-S_2^{-2}}^{k_2}
        }{
            \brk{k_1 + 1} \brk{k_2 + 1} k_1 k_2
        }
        \times
        \frac{
            \Gamma\brk{1 + k_1 +k_2}
        }{
            \Gamma\brk{k_1} \Gamma\brk{k_2}
        }
        \nn\\
        &+
        \sum\limits_{k_1 \in \Z_{\ge 1}}
        \frac{\brk{-S_1^{-2}}^{k_1} + \brk{-S_2^{-2}}^{k_1}}{k_1 + 1}
        + 1
    }
    \,,
    \label{eq:h1sum}
\end{align}
where in the second line we combined the $k_1 = 0$ and $k_2 = 0$ residues
together into one separate term.
These sums turn out to be very easy to perform, and in \appref{app:sums} we
explain how one can express them in terms of simple logarithms, so
that the helicity-preserving function $h_1^{\brk{0,0}}$ takes the form
\begin{equation}
h_1^{(0,0)} =
g^2\biggbrk{
\frac{S_1}{S_2} \log\frac{S_1^2(1+S_2^2)}{S_1^2+S_2^2+S_1^2 S_2^2}
+\frac{S_1S_2}{2}\log\frac{(1+S_1^2) (1+S_2^2)}{S_1^2+S_2^2+S_1^2 S_2^2}
+(S_1 \leftrightarrow S_2)
}\,.
\label{eq:h1intev}
\end{equation}
Let us furthermore note that the helicity-breaking transition starts only at $g^4$,
\ie the one-loop contribution $\bar{h}_1^{(0,0)} = 0$ vanishes, see eq.\ (124)
of~\cite{Basso:2013aha}. After this preparation, we can proceed to higher loops.
%

%%%%%%%%%%%%%%%%%%%%%%%%%%%%%%
\subsubsection{Higher-Loop Remainder Function}

Here we describe the procedure to compute
higher terms in the $g^2$-expansion of the one-gluon contribution to the
remainder function in terms of multiple polylogarithms, which mostly
follows~\cite{Papathanasiou:2013uoa} and~\cite{Drummond:2015jea}. The
idea is exactly the same as in the previous \secref{sec:1loopRem}: First we need to extract the
desired loop contribution from the integral~\eqref{eq:hint}, then we apply
the residue theorem to convert integrals into sums over poles. These
sums can finally be expressed in terms of multiple polylogarithms.

When expanded to higher loops, the schematic structure of the integrand in
eq.~\eqref{eq:hint} becomes:
\begin{align}
    \frac{
        e^{2 i \brk{\sigma_1 u_1 + \sigma_2 u_2}}
    }{
        \prod\limits_{j = 1, 2}\brk{u_j + \tfrac{i}{2}}^{r_j} \brk{u_j - \tfrac{i}{2}}^{p_j}
    }
    \cdot
    &\Gamma\brk{\tfrac32 + i u_1} \Gamma\brk{-i u_1 - i u_2} \Gamma\brk{\tfrac32 + i u_2}
    \times
    \nn\\
    \times
    &\polyQ\Bigbrk{
        \Bigbrc{
            u_j,
            %\psi^{\brk{m_j}}\brk{\tfrac12 + i u_j},
            %\psi^{\brk{m_j}}\brk{\tfrac12 - i u_j},
            \psi^{\brk{m_j}}\brk{\tfrac12 \pm i u_j},
            \psi^{\brk{n_j}}\brk{\tfrac32 \pm i u_j}
        }_{j = 1, 2}
    }
    \, .
    \label{eq:hintGen}
\end{align}
Compared to our discussion in the previous \secref{sec:1loopRem}, the
$\Gamma$-functions do not change, the denominators acquire integer exponents $r_j$ and $p_j$,
and a polynomial $\polyQ$ of the rapidities $u_j$ and polygamma $\psi$-functions of various weight appear.
The location of relevant poles stays the same, only the residues
become more complicated. Therefore, we use the same procedure as above to
convert the integral~\eqref{eq:hintGen} into a sum over residues. As
in~\cite{Papathanasiou:2013uoa}, a few basic relations for $\Gamma$- and
$\psi$-functions turn out to be very useful in this step,
\begin{align}
    \Gamma\brk{z} \Gamma\brk{1 - z} &= \frac{\pi}{\sin{\pi z}}
    \,,
    \nn
    \\[2mm]
    \psi^{(n)}\brk{z + 1} &= \psi^{(n)}\brk{z} + \brk{-1}^n n! z^{-n - 1}
    \,,
    \nn
    \\[2mm]
    \psi^{(n)}\brk{z} &= \brk{-1}^n \psi^{(n)}\brk{1 - z} - \pi \partial_z^n
    \cot\brk{\pi z}
    \nn
    \,.
\end{align}
After some processing, we arrive at the following $L$-loop analogue of eq.~\eqref{eq:h1sum}:
\begin{align}
    h_L^{\brk{p_1, p_2}} & =
    \sum\limits_{k_{1,2} \in \Z_{\ge 1}}
    \frac{
        \brk{-S_1^{-2}}^{k_1} \brk{-S_2^{-2}}^{k_2}
    }{
        %\brk{k_1 + 1}^{r_1} \brk{k_2 + 1}^{r_2} k_1^{r_3} k_2^{r_4}
        \prod\limits_{j = 1, 2} \brk{k_j}^{p_j} \brk{1 + k_j}^{r_j}
    }
    \cdot
    \frac{
        \Gamma\brk{1 + k_1 + k_2}
    }{
        \Gamma\brk{k_1} \Gamma\brk{k_2}
    }
    \times \poly\Bigbrk{
        \Bigbrc{
            k_j,
            \psi^{\brk{m_j}}_{k_j},
            \psi^{\brk{n_j}}_{1 + k_j}
        }_{j = 1, 2}
    }\,,
    \label{eq:opehL}
\end{align}
where we have introduced shorthands $\psi_k^{\brk{m}} \defas \psi^{\brk{m}}\brk{k}$ for
polygamma functions of integer arguments. On the right hand side of eq.~\eqref{eq:opehL}, the polynomial $\poly$
depends on both summation indices $k_1, k_2$ and $\psi$-functions
of weight $m_j, n_j < 2 L$, the powers of common denominators $p_j$,
$r_j$ are bounded by the number of loops as well.
This double sum cannot be done in full generality for arbitrary loop number,
but for every fixed number of loops, we can make use of algorithms developed
in~\cite{Moch:2001zr}. In order to do so, we replace the ratio of $\Gamma$-functions
with a binomial $\binom{k_1 + k_2}{k_1}$, introduce a new variable $j = k_1 +
k_2$, and express $\psi$-functions in terms of the so-called $S$-sums (see eq.\ (3)
in~\cite{Moch:2001zr})
\begin{equation}
    S\brk{n; m_1, \dots, m_k; x_1, \dots, x_k}
    =
    \sum\limits_{i=1}^n
    \frac{x_1^i}{i^m_1}
    S\brk{i; m_2, \dots, m_k; x_2, \dots, x_k} \,.
\end{equation}
The $\psi$-functions of integer arguments are related to these objects via
\begin{align}
    \psi_{k}
    &= -\EulerGamma + S\brk{k - 1, 1, 1}
    \\[2mm]
    \psi_{k}^{(m)}
    &= \brk{-1}^{m + 1} m! \, \brk{\zeta_{m + 1} - S\brk{k - 1, m + 1, 1}}
    \,.
    \label{eq:psitos}
\end{align}
After all these conversions, we make use of the \soft{XSummer} package~\cite{Moch:2005uc} for
\soft{FORM}~\cite{Ruijl:2017dtg} that rewrites these sums over residues in terms of multiple
polylogarithms (see also~\cite{Moch:2001zr} for a description of the algorithm). Since the
only missing pieces at two loops are the functions $h_2^{(1,0)}$ and
$h_2^{(0, 1)}$ in eq.~\eqref{eq:h2result}, we can just filter the
corresponding terms with $\tau_i \equiv \log\brk{T_i}$
that come from the~$g^2$ corrections to the gluon energy~\eqref{eq:GluonEnergy}
in the two-loop expansion of eq.~\eqref{eq:hint}, perform the residue
resummation, and convert the produced
answer to classical polylogarithms using the
package~\cite{Frellesvig:2016ske}, which finally
yields~\eqref{eq:h2result} and~\eqref{eq:h2resultother}.

With all this, we are finally able to write the full expression~\eqref{eq:R72ope} for the
two-loop remainder function in collinear kinematics in terms of only classical polylogarithms
$\brc{\log, \Li_2,\Li_3}$. This form of the answer makes it easy to understand the analytical
structure, and hence it provides a good starting point for our discussion in the next
sections.

%%%%%%%%%%%%%%%%%%%%%%%%%%%%%%%%%%%%%%%%%%%%%%%%%%%%%%%%%%%%
%%%%%%%%%%%%%%%%%%%%%%%%%%%%%%%%%%%%%%%%%%%%%%%%%%%%%%%%%%%%
\section{Continuation and Regge Limit: The Hexagon}
\label{sec:continuation6}

In this section, we now turn to the central theme of this work, namely the
continuation of the collinear remainder function from the main sheet into
non-trivial Mandelstam regions. Our goal here is to illustrate the general
procedure at the simplest example, namely the hexagon.%
\footnote{The analytic continuation of the perturbative expansion of
the hexagon Wilson loop OPE has been explored before: Using a specific
choice of continuation path, Hatsuda~\cite{Hatsuda:2014oza} could
match the analytically continued collinear expansion against known
Regge-limit data up to the next-to-next-to-leading logarithmic
approximation (N$^2$LLA) at five loops, and produced predictions for
N$^3$LLA and N$^4$LLA. That work preceded the all-order continuation
of the hexagon carried out in~\cite{Basso:2014pla}.}
After a brief review
of the relevant coordinates and limits, we will discuss existing results for
the only Mandelstam region in which the Regge limit of the
hexagon remainder function is non-vanishing and can be computed to any
desired order. We will state the known results for two loops in the combined
multi-Regge collinear limit. Then we turn to the continuation in collinear kinematics.
Starting from the main sheet, we discuss how to reach the Mandelstam region
and compute the relevant cut contributions. We shall show how the continuation
allows to recover the two-loop result for the combined multi-Regge collinear limit
of the remainder function from the leading terms in the Wilson loop OPE.

%%%%%%%%%%%%%%%%%%%%%%%%%%%%%%%%%%%%%%%%%%%%%%%%%%%%%%%%%%%%
\subsection{Variables and Limits}

Let us now specialize the discussion of kinematics in \secref{sec:collinear} to the
case of $n=6$ external gluons, for which the finite remainder function depends on just
three cross ratios, which we denote as
$$ u_1=U_{25}\ ,\quad  u_2 = U_{36}\ , \quad  u_3 = U_{14}\ , $$
omitting the first index $j=5$ in eqs.~\eqref{eq:uj}. According to the general
prescription, the multi-Regge regime is reached when the ``large'' cross
ratio $u_1$ approaches $u_1= 1$, while the ``small'' cross ratios  go to zero,
\ie
\begin{equation}
u_{1}\to 1
\,,\qquad
u_{2}\to 0
\,,\qquad
u_{3}\to 0
\,.
\end{equation}
The ratios of vanishing terms remain finite, and are used to define
a single pair of anharmonic ratios $w= w_1 $ and $\wb= \wb_1$
as
\begin{equation}
\frac{u_{2}}{1-u_{1}}\to\frac{1}{\abs{1+w}^2}
\,,\qquad
\frac{u_{3}}{1-u_{1}}\to\frac{\abs{w}^2}{\abs{1+w}^2}
\,.
\end{equation}
For $n=6$ the approach to the Regge regime is controlled through a
single parameter,\footnote{Note that our normalization of the parameter
$\eps$ differs from the one used by Dixon \etal~\cite{Dixon:2011pw}, see
footnote 2 in~\cite{Bargheer:2015djt}, but it is consistent with
eqs.~\eqref{eq:reggef1} and~\eqref{eq:reggef0}.}
\begin{equation}
\eps=u_{2}u_{3}\,.
\end{equation}
that tends to zero in the limit.

Next we want to discuss the relation with the kinematic variables we have used
in our discussion of the Wilson loop OPE. The three cross ratios
$u_i$ are expressed through the kinematical variables $T\equiv T_1$, $S\equiv S_1$
and $F\equiv F_1=\exp(i\varphi)$ introduced in \secref{sec:collinear} as
\begin{equation}
u_1 = U_{2,5} = \frac{1}{1+S^2+T^2+2ST\cos\varphi} \,, \quad
u_2 = U_{3,6} = \frac{S^2\,u_1}{1+T^2} \,, \quad
u_3 = U_{1,4} = \frac{T^2}{1+T^2}
\,.
\label{eq:ujpar}
\end{equation}
In the collinear limit $T\to0$, the relations~\eqref{eq:ujpar} read
\begin{equation}
u_1 = \frac{1}{1+S^2} + \order{T}
\,, \quad
u_2 = \frac{S^2}{1+S^2} + \order{T}
\,, \quad
u_3 = T^2 + \order{T^4}
\,.
\label{eq:uihex}
\end{equation}
When parametrized in terms of $S$,
$T$, and $F=\exp(i\varphi)$, the Regge limit is taken by sending both $T$ and $S$ to
zero, while keeping $r=S/T$ finite, see eq.~\eqref{eq:Frw678}. In the Regge limit, the remainder function
depends on the finite variables $r$ and $F$ along with the quantity $T$ that
vanishes in the limit. These are related to $w, \wb$ and $\eps$ through
\begin{equation}
r^2=\frac{1}{w\wb}\,,
\qquad
F^2=\frac{w}{\wb}\,,
\qquad
{S^2}{T^2} = r^2 T^4 = \eps \,,
\end{equation}
as in~\eqref{eq:Frw678} and~\eqref{eq:reps678}.
This concludes our brief summary of the relevant variables and limits that are used
in the subsequent analysis of the hexagon remainder function.

%%%%%%%%%%%%%%%%%%%%%%%%%%%%%%%%%%%%%%%%%%%%%%%%%%%%%%%%%%%%
\subsection{The Remainder Function in Multi-Regge Kinematics}

For the hexagon $n=6$, the remainder function is well known to possess only one
Mandelstam region with a non-trivial Regge limit, namely the region $\varrho
=(\m\m)$. In this region, the two-loop contribution to the finite remainder
function in multi-Regge kinematics reads
\begin{equation}
R^{\m\m}_{6,(2)}(\eps,w)
= 2\pi i \, f(\eps;w)
= 2\pi i\bigbrk{f_1(w)\log\eps+f_0(w)}\ ,
\label{eq:f}
\end{equation}
where the leading logarithmic term contains the coefficient
\begin{equation}
f_1(w)=\frac{1}{2}\log\abs{1+w}^2\log\lrabs{\frac{1+w}{w}}^2
\,,
\label{eq:reggef1}
\end{equation}
while the next-to-leading logarithmic term is given by
\begin{multline}
f_0(w)=
-4\Li_3(-w)-4\Li_3(-\wb)
+2\log\abs{w}^2\bigbrk{\Li_2(-w)+\Li_2(-\wb)}\\[2mm]
+\frac{1}{3}\log^2\abs{1+w}^2\log\frac{\abs{w}^6}{\abs{1+w}^4}
-\frac{1}{2}\log{\abs{1+w}^2}\log\lrabs{\frac{1+w}{w}}^2\log\frac{\abs{w}^2}{\abs{1+w}^4}
\,.
\label{eq:reggef0}
\end{multline}
We use the variables that were introduced in the previous subsection. This formula
can be derived from the general expression for the six-gluon remainder
function due to~\cite{Bartels:2008sc} that in its original version
encodes at least the leading logarithmic (LL) terms to all loop orders. The result
of Bartels \etal parametrizes
the multi-Regge limit of the remainder function in terms of two functions of the
coupling $g$, the impact factor and the so-called BFKL eigenvalue. These possess
a power series expansion in $g$. In order to construct the LL contributions of the
remainder function at any loop order, it is sufficient to know the leading terms in
these expansions. It is not too difficult to reconstruct the LLA in eq.~\eqref{eq:f}
from the results in~\cite{Bartels:2008sc}, and in fact to carry out
these computations to higher loop orders, and even beyond the leading logarithmic
order, see~\cite{Dixon:2012yy} for an extensive discussion. Explicit expressions
for the impact factor and BFKL eigenvalue in NLLA were first given in~\cite{Fadin:2011we}.

If we apply the combined multi-Regge collinear limit, \ie send $r^2=1/(w\wb)\to\infty$
while keeping the ratio $w/\wb = F^2$ finite, these formulas reduce to
\begin{equation}
\lrsbrk{\sfrac{1}{2\pi i}R_{6,(2)}^{\m\m}}^\CL=
C\,2\log(r)\log(\eps)+C\bigbrk{8+8\log(r)+4\log(r)^2}\,,
\label{eq:Rmmcoll}
\end{equation}
with
\begin{equation}
C=\frac{\cos(\varphi)}{r}\,.
\label{eq:Ccos6}
\end{equation}
Let us stress that our formulas for the remainder function in multi-Regge kinematics and
the combined multi-Regge collinear limit contain terms from LLA and NLLA. At this order in
the weak-coupling expansion, the expression we state is complete. At higher orders, the
multi-Regge limit of the remainder function is also known exactly from the amplitude
bootstrap~\cite{Dixon:2014voa} and ultimately to all loops due to the work by Basso
\etal~\cite{Basso:2014pla}. We will not need such extensions here.

%%%%%%%%%%%%%%%%%%%%%%%%%%%%%%%%%%%%%%%%%%%%%%%%%%%%%%%%%%%%
\subsection{Analytic Continuation}
\label{ssec:hexContinuation}

Let us now see how we can recover the results we reviewed in the previous subsection,
and in particular eq.~\eqref{eq:Rmmcoll}, through analytic continuation from the
collinear $n=6$ remainder function on the main sheet.
The relevant Mandelstam region $\varrho=(--)$ is reached by some curve along which
only the large cross ratio $u_1 = U_{2, 5}$ has non-trivial winding number
\begin{equation}
n_{25}
=\frac{1}{4}\brk{\varrho_4-\varrho_3}\brk{\varrho_6-\varrho_5}
=\frac{1}{4}\brk{-1-1}\brk{1+1}
=-1
\label{eq:hexwind}
\end{equation}
around $u_1 = 0$, see
eq.~\eqref{eq:winding}. In order to initiate our analysis, let us display the
collinear limit of the two-loop remainder function on the main
sheet. From the Wilson loop OPE one finds
\begin{equation} \label{eq:Ropehex}
    R_{6, \brk{2}} \brk{S,T,F}
    = 2 T \cos{\brk{\varphi}} \f^{\brk{0}}_2\brk{S}
    + 2 T \cos{\brk{\varphi}} \log{\brk{T}} \f^{\brk{1}}_2\brk{S}
    - \bigbrk{\log \WBDS_g}^{\brk{2}}
    + \order{T^2}\ ,
\end{equation}
where
\begin{multline}
    \f^{(0)}_2\brk{S}
    = g^4 \brk{S+S^{-1}} \biggsbrk{
        \frac{\brk{12 + \pi^2}\log S^2}{3 \brk{1+S^2}}
        + \log\brk{1 + S^{-2}} \brk{4 - 2 \log\brk{S^2}}
        \\
        + \bigbrk{\log\brk{S^2}-2}\log^2\brk{1 + S^{-2}}
        + \frac23 \log^3\brk{1 + S^{-2}}
        - 2 \Li_3\brk{-S^{-2}}
    }
    \label{eq:f20}
\end{multline}
and
\begin{align}
    \f^{(1)}_{2}\brk{S}
    = 2 g^4 (S+S^{-1}) \biggsbrk{
        - \frac{2 \log\brk{S^2}}{1 + S^2}
        + \bigbrk{\log\brk{S^2}-2} \log\brk{1 + S^{-2}}
        + \log\brk{1 + S^{-2}}
    }.
    \label{eq:f21}
\end{align}
The BDS part is given by
\begin{align}
    \bigbrk{\log \WBDS_g}^{\brk{2}}
    = g^4 \frac{2 \pi^2}{3} T \brk{S + S^{-1}} \biggsbrk{
        \frac{\log\brk{S^2}}{1 + S^2}
        + \log\brk{1 + S^{-2}}
    }.
    \label{eq:W6BDS}
\end{align}
There are a number of comments we would like to make about these expressions. First of
all, a closer look at the arguments of the (poly-)logarithms reveals that the remainder
function possesses $r=2$ branch points in the complex $S^2$-plane. These are the
points $S^2 = 0$ and $S^2+1 = 0$. Of course we see this here only to the
given order of the expansion, but the statement remains true for higher loops, see \eg
the explicit three-loop expression in~\cite{Dixon:2013eka}. It is important to note that,
to leading order in the collinear limit, the cross ratios $u_i$ we listed in the
previous subsection can be built as products of the two functions $S^2$ and $1+S^2$
and their inverses. This makes it particularly easy to switch between curves in the
$S^2$-plane and in the space of cross ratios. Let us also note that the remainder
function vanishes in the multi-Regge limit $S \rightarrow 0$ before we
analytically continue from the main sheet into other Mandelstam regions.

Our task is to derive the eqs.~\eqref{eq:Rmmcoll} from the formula~\eqref{eq:Ropehex}
through analytic continuation along some specific curve in the space of kinematic
variables. Since we will focus on the collinear limit, our paths will remain in this
limit, \ie we will not vary $T$. This is justified by the fact that, by eq.~\eqref{eq:uihex},
$u_1$ and $u_2$ only depend on $S^2$ in the collinear limit, whereas
$u_3=U_{1,4}=T^2$ has winding number
$n_{1,4}\sim(\varrho_5-\varrho_4)=0$ by eq.~\eqref{eq:winding}.
As we can see from the explicit formulas above, the two-loop
contributions to the collinear limit are analytic in $\varphi$. This statement actually
remains true at any finite loop order. Hence, we can restrict
to paths along which $\varphi$ is kept constant. It remains to study paths in the
complex $S^2$-plane that start and end in the region where $S^2 > 0$.
As we pointed out before, the remainder function possesses branch points at $S^2=-1$
and $S^2=0$. Equivalence classes of paths are therefore parametrized by the fundamental
group $\pi_1(\mathbb{C}\setminus\{0,-1\})$. This group is generated by two elements
$p_1$ and $p_2$. The precise choice is a matter of convention. Let us agree that
the generator $p_1$ is associated with a curve $\gen_1$ that
starts at some point $S^2 > 0$, runs slightly above the real axis, surrounds $S^2 = 0$
in counterclockwise direction before running back to its starting point where $S^2>
0$. As for $p_2$, we make a similar choice, except that now the curve $\gen_2$ runs
from $S^2>0$ towards
$S^2=-1$ above the real axis, surrounds the point $S^2 = -1$ in counterclockwise
direction and runs back \textit{below} the real axis to its starting point. The
two curves are depicted in \tabref{table:pathgenerators}.
\begin{table}
\centering
    \begin{tabular}{cr}
        \toprule
        generator &
        $S^2$ \hspace{1.9cm}
        \\
        \midrule
        $\gen_1$ &
        \includegraphics[align=c]{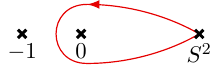}
        \\
        $\gen_2$ &
        \includegraphics[align=c]{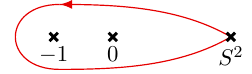}
        \\
        \bottomrule
    \end{tabular}
    \caption{Path generators for the hexagon.}
    \label{table:pathgenerators}
\end{table}

Since the cross ratios are rational functions of $S^2$, they acquire corresponding
phase shifts which are easy to work out,
\begin{alignat}{3}
\gen_1 u_1 & =  u_1
\,,& \quad
\gen_1 u_2 & = e^{2\pi i} u_2
\,,& \quad
\gen_1 u_3 & = u_3
\nn\\[2mm]
\gen_2 u_1 & = e^{-2\pi i} u_1
\,,& \quad
\gen_2 u_2 & = u_2
\,,& \quad
\gen_2 u_3 & = u_3
\,.
\label{eq:hexgenwind}
\end{alignat}
Here and in the following, the symbol $\gen_i f(u)$ means the value of a
function $f$ of the cross ratios at the endpoint of an analytic continuation along
the curve $\gen_i$. The value of the cross ratio $u_3$ is unaffected by
the continuation as it does not depend on $S^2$.

Now we can study the behavior of the remainder function upon continuation. For the
continuation along the curves $\gen_1$ and $\gen_2$ one finds
\begin{equation}
    % \label{eq:Cicont}
    \bigsbrk{\gen_i  \Rsixope}\supMRL  = (1+2\pi i\Delta_i) \Rsixope
\end{equation}
with
\begin{equation}
\label{eq:D1}
\Delta_2\Rsixope =C \bigbrk{8 + 8 \log(r) + 4 \log(r)^2} +
C \, 2 \log(r) \log(\eps)\,,
\end{equation}
and $\Delta_1 \Rsixope = 0$. Non-vanishing cut contributions for more general
elements $g \in \pi_1$ in our fundamental group can also be worked out by
combining the following building blocks
\begin{align}
\Delta_2\Delta_1\Rsixope
&=C\bigbrk{4 -4i\pi+2 \log(r)} - C\,\log(\eps)
\,,\nn\\[2mm]
\Delta_1\Delta_2\Rsixope
&=C \bigbrk{4 + 4i\pi + 6 \log(r)} + C\,\log(\eps)
\,,\nn\\[2mm]
\Delta_2\Delta_2\Delta_1\Rsixope&=
-\Delta_1\Delta_1\Delta_2\Rsixope=
-4C\,.
\label{eq:discmrlsix}
\end{align}
All other discontinuities, and in particular those beyond triple
discontinuities, vanish at this loop order. Let us stress once again
that our symbols $\Delta_i$ combine an analytic continuation with
taking the Regge limit. Before taking the Regge limit, a continuation
along $\gen_1$ produces a nontrivial cut contribution, which vanishes
only in the Regge limit. If, on the other hand, we continue this
cut contribution along $\gen_2$, new terms appear that possess a
non-trivial Regge limit.

Our first and most important observation is that continuation of the
collinear remainder function along the curve $\gen_2$ gives
a single non-vanishing cut contribution, namely $\Delta_2 \Rsixope$, which
agrees exactly with the expected formula~\eqref{eq:Rmmcoll} for the
collinear limit of the remainder function in the Mandelstam region
$\varrho=(--)$. This is fully consistent with the kinematics, as the
cross ratio $u_1$ has winding number $-1$ under the generator
$\gen_2$ of eq.~\eqref{eq:hexgenwind}, as required for the $\varrho=(\m\m)$
region by eq.~\eqref{eq:hexwind}.

\medskip

We can carry the analysis a bit further and ask to which extent
the expression~\eqref{eq:Rmmcoll} determines the curve. In  order to
investigate this issue, let us first write the collinear limit of the
hexagon remainder in the non-trivial Mandelstam region in terms of
cut contributions,
\begin{equation}
\lrsbrk{R_{6,(2)}^{\m\m}}^\CL = \biggbrk{
 2\pi i\sum_i\cc{i}\Delta_i
+(2\pi i)^2\sum_{i,j}\cc{i,j}\Delta_i\Delta_j
+(2\pi i)^3\sum_{i,j,k}\cc{i,j,k}\Delta_i\Delta_j\Delta_k}\Rsixope\ .
\label{eq:disclc}
\end{equation}
Using our list~\eqref{eq:discmrlsix} one may infer the following four
constraints on the coefficients
\begin{equation} \label{eq:fourconstr}
\cc{2}=1
\, ,\quad
\cc{1,2}=  0 = \cc{2,1}\ , \quad  \cc{1,1,2} = \cc{2,2,1}   \ .
\end{equation}
It is now easy to see that curves of the form
$\gen_1^{\alpha_1}\gen_2^{\alpha_2}\gen_1^{\alpha_3}$ are consistent with eq.~\eqref{eq:Rmmcoll}
if and only if $\alpha_1=\alpha_3=0$ and $\alpha_2=1$, \ie within this family of curves, $\gen_2$
is the only solution. On the other hand, there do exist other curves that give rise to the
same cut contributions, \eg
\begin{align}
    \gen_1^k \gen_2^{1 - k} \gen_1^{-1} \gen_2^k \gen_1^{1 - k}
    \label{eq:hexCurve}
\end{align}
for any value of $k \in \mathbb{Z}$. For $k \neq 0$, these curves are not the
same as $\gen_2$, even though they have the same winding numbers around
$u_1=0$ and $u_2=0$. It would be interesting to investigate whether these
curves are excluded by higher-order corrections in the loop or collinear
expansion.
The curve of eq.~\eqref{eq:hexCurve} is actually unique in the class
of curves that involve only two powers of $\gen_2$ generators that
are separated by some power of $\gen_1$, if we impose both the LLA and
NLLA constraints by matching formula~\eqref{eq:disclc}
against~\eqref{eq:Rmmcoll}. The conditions from LLA alone (\ie
$\log\varepsilon$ terms) dictate only $\cc{1,2} =
\cc{2, 1}$ and $\cc{1,1,2}=\cc{2,2,1}$. Eq.~\eqref{eq:fourconstr} is a consequence of the anticommutator
lemma~\eqref{eq:acommut} together with the total winding conditions
$\cc{1} = 1 - \cc{2} = 0$, which follow from eq.~\eqref{eq:hexgenwind}. If we allow elements of the fundamental group
that involve three powers of $\gen_2$, separated by some powers of
$\gen_1$, the LLA and NLLA constraints can be solved by
\begin{align}
    \gen_1^{\alpha_1}
    \gen_2^{\alpha_2}
    \gen_1^{-\alpha_1 - \alpha_2}
    \gen_2^{\alpha_1}
    \gen_1^{\alpha_2}
    \gen_2^{1 - \alpha_1 - \alpha_2}
\end{align}
for ${\alpha_1, \alpha_2} < -1$. We conclude that the two-loop
formulas in LLA and NLLA impose strong constraints on the curve of continuation, but in
order to fix the curve completely, one needs additional assumptions on
its form.

%%%%%%%%%%%%%%%%%%%%%%%%%%%%%%%%%%%%%%%%%%%%%%%%%%%%%%%%%%%%
%%%%%%%%%%%%%%%%%%%%%%%%%%%%%%%%%%%%%%%%%%%%%%%%%%%%%%%%%%%%
\section{Continuation and Multi-Regge Limit: The Heptagon}
\label{sec:continuation7}

We now approach the main goal of this work, namely to repeat the analysis
outlined in the previous section for the heptagon. In this case, the analysis
is richer because there exist four Mandelstam regions in which the remainder
function possesses a non-trivial multi-Regge limit. After a short discussion
of the relevant kinematical variables, we will review known results about the
multi-Regge limit of the heptagon remainder function in all four Mandelstam
regions. In the final subsections, their collinear limit will be reproduced
by analytic continuation from the Wilson loop OPE.

%%%%%%%%%%%%%%%%%%%%%%%%%%%%%%%%%%%%%%%%%%%%%%%%%%%%%%%%%%%%
\subsection{Variables and Limits}
\label{sec:heptagon-crs-and-mrl}

Before we start our analysis, it is again useful to review some of the formulas we
derived in \secref{sec:mrlfromope} for the heptagon. When $n=7$, there are
six independent cross ratios~\eqref{eq:uj}, two of which approach $u_{1,1} = u_{2,1} = 1$ while the
other four vanish in multi-Regge kinematics. Consequently, the approach to the
multi-Regge limit is now controlled by two small $\eps$ parameters
\begin{equation}
\eps_1 = u_{1,2} u_{1,3}
\,, \qquad
\eps_2 = u_{2,2} u_{2,3}
\,,
\end{equation}
while the limit itself is parametrized by four anharmonic ratios $w_1$, $\wb_1$
and $w_2$, $\wb_2$.

Once again, we can parametrize the heptagons in general kinematics in terms of
the variables
\begin{equation}
\brc{T_j,S_j,F_j}
=
\brc{e^{-\tau_j},e^{\sigma_j},e^{i\varphi_j}}
\,,\qquad
j=1,\dots,n-5
\label{eq:TSFhept}
\end{equation}
introduced in \secref{sec:collinear}.
The precise relation between these variables and our set of heptagon cross ratios~\eqref{eq:uj}
may be worked out from the formulas in \appref{app:bsvtessdetail},
\begin{align}
u_{1,3}=U_{1,4}&=\frac{T_1^2}{1+T_1^2}
\,,\nn\\[2mm]
u_{1,1}=U_{2,5}&=\frac{1+T_2^2}{1+S_1^2+2c_1S_1T_1+T_1^2+T_2^2}
\,,\nn\\[2mm]
u_{1,2}=U_{3,7}&=\frac{S_1^2}{(1+T_1^2)(1+T_2^2)}\cdot\frac{U_{2,5}}{U_{3,6}}
\,,\nn\\[2mm]
u_{2,1}=U_{3,6}&=\frac{
S_1^2(1+T_2^2)+S_2^2(1+T_1^2)+S_1^2S_2^2+2c_1S_1T_1S_2^2+2c_2S_1^2S_2T_2+2c_+S_1T_1S_2T_2
}{
\lrbrk{1+S_1^2+2c_1S_1T_1+T_1^2+T_2^2}
\lrbrk{1+S_2^2+2c_2S_2T_2+T_1^2+T_2^2}
}
\,,\nn\\[2mm]
        U_{2,6}&=\frac{S_2^2}{(1+T_1^2)(1+T_2^2)}\cdot\frac{U_{4,7}}{U_{3,6}}
\,,\nn\\[2mm]
u_{2,2}=U_{4,7}&=\frac{1+T_1^2}{1+S_2^2+2c_2S_2T_2+T_1^2+T_2^2}
\,,\nn\\[2mm]
u_{2,3}=U_{1,5}&=\frac{T_2^2}{1-T_1^2}
\,,
\label{eq:cr7opevar}
\end{align}
with the shorthand notation
\begin{equation}
c_1=\cos(\varphi_1)
\,,\qquad
c_2=\cos(\varphi_2)
\,,\quad
c_+=\cos(\varphi_1+\varphi_2)
\,.
\end{equation}
Besides the six cross ratios $u_{j,i}$, we have also listed $U_{2,6}$.
The latter is not independent, but related to the set $u_{j,i}$ through a
non-rational Gram determinant relation. Together, the seven
cross ratios~\eqref{eq:cr7opevar} constitute a multiplicative basis
for the set of all conformal cross ratios of the heptagon. For the purpose of
understanding analytic continuation paths in the space of cross
ratios, it is therefore sufficient to consider these seven.

Given the formulas~\eqref{eq:cr7opevar}, it is straightforward to obtain the following expressions for the
leading terms in the cross ratios as $T_1$ and $T_2$ are sent to zero,
\begin{gather}
u_{1,1}=U_{25}=\frac{1}{1+S_1^2}+\order{T_i} \nn \\[2mm]
u_{1,2}=U_{37}=\frac{S_1^2(1+S_2^2)}{S_1^2+S_2^2+S_1^2S_2^2}+\order{T_i} \,,\qquad
u_{1,3}=U_{14}=T_1^2+\order{T_1^4} \,,\nn \\[2mm]
u_{2,1}=U_{36}=\frac{S_1^2+S_2^2+S_1^2S_2^2}{(1+S_1^2)(1+S_2^2)} + \order{T_i} \label{eq:crs7coll}\\[2mm]
u_{2,2}=U_{47}=\frac{1}{1+S_2^2}+\order{T_i} \,,\qquad
u_{2,3}=U_{15}=T_2^2+ \order{T_2^4}\nn \\[2mm]
U_{26}=\frac{(1+S_1^2)S_2^2}{S_1^2+S_2^2+S_1^2S_2^2} + \order{T_i} \,. \nn
\end{gather}
Setting $S_1=r_1T_1$ and $S_2=1/(r_2T_2)$ in accordance
with~\eqref{eq:Frw678}, the multi-Regge limit is attained
for $T_j\to0$, keeping $r_j$ finite. For the heptagon, one finds
\begin{alignat}{3}
r_1^2&=\frac{S_1^2}{T_1^2}=\frac{1}{w_1\wb_1}\,,
\qquad &
F_1^2&=\frac{w_1}{\wb_1}\,,
\qquad &
{S_1^2}{T_1^2}&= r_1^2 T_1^4 = \eps_1\,,
\nn\\
r_2^2&=\frac{1}{S_2^2T_2^2}=\frac{1}{w_2\wb_2}\,,
\qquad &
F_2^2&=\frac{w_2}{\wb_2}\,,
\qquad &
\frac{T_2^2}{S_2^2} &= r_2^2 T_2^4 = \eps_2\,.
\label{eq:frtow7}
\end{alignat}
Note that the equations for $F_i$ require to expand the expressions~\eqref{eq:cr7opevar}
for the cross ratios to higher
orders in small $T_i$, beyond the terms stated in eq.~\eqref{eq:crs7coll}.
One way to derive eq.~\eqref{eq:frtow7} is to solve the system of quadratic
equations~\eqref{eq:uofw} for $\brc{w_1,\wb_1, w_2,\wb_2}$ in terms
of $\brc{T_i, S_i,F_i=\exp(i\varphi_i)}$ using eqs.~\eqref{eq:uj} and~\eqref{eq:cr7opevar}.
Once this is done, one can insert the result into the right
hand side of eq.~\eqref{eq:frtow7} \brk{after picking appropriate branches
of square roots} to obtain the left hand side as leading terms in the
collinear $T_i \to 0$ expansion.

From the multi-Regge limit, the combined multi-Regge collinear limit
is attained for $r_1,r_2\to\infty$. If we start in general kinematics
$\brc{T_j,S_j,F_j}$, we reach the collinear limit when we send $T_j
\to0$ while keeping $S_j$ and $F_j$ finite. We can then continue to
the combined multi-Regge collinear limit by setting $S_1=r_1T_1$ and $S_2=1/(r_2T_2)$, and taking
the limit $r_j\to\infty$, keeping $T_j\ll 1/r_j$.

%%%%%%%%%%%%%%%%%%%%%%%%%%%%%%%%%%%%%%%%%%%%%%%%%%%%%%%%%%%%
\subsection{The Remainder Function in Multi-Regge Kinematics}

The heptagon remainder function is known to possess a non-trivial multi-Regge
limit in four Mandelstam regions. These regions are associated with the four
different sign choices of the energies $p_i^0$, $i=4,5,6$, in which at least two energies
are flipped. At least for three of these regions, the remainder function in
multi-Regge kinematics at two loops is known.%
\footnote{For the last region $(\m\p\m)$, the remaining free
coefficients where fixed in~\cite{DelDuca:2018raq}, see below.}
As in the case of the hexagon, the two-loop
result receives contributions from both LLA and NLLA. The regions $\varrho=
(\m\m\p)$ and  $\varrho=(\p\m\m)$ are the easiest, because the answer involves
exactly the same information that appears for the multi-Regge limit of the
hexagon, \ie using the variables defined in eqs.~\eqref{eq:TSFhept} and~\eqref{eq:frtow7},
the multi-Regge limit of the remainder function reads~\cite{Bartels:2014jya}
\begin{equation}
  R_{7,(2)}^{\m\m\p}=f(\eps_1;w_1)
  \,,\qquad
  R_{7,(2)}^{\p\m\m}=f(\eps_2;w_2)
  \,,
  \label{eq:R7mmppmm}
\end{equation}
where $f$ is the function we defined in eq.~\eqref{eq:f}. For $\varrho=(\m\m\m)$, the
two-loop remainder function is also known in the Regge limit, but it involves a new
function $g$~\cite{Bartels:2014jya,Prygarin:2011gd,Bargheer:2015djt},
\begin{equation}
 R_{7,(2)}^{\m\m\m} =f(\eps_1;v_1)+f(\eps_2;v_2)+g(v_1,v_2)
\,,
\label{eq:R7mmm}
\end{equation}
where the variables
\begin{equation}
v_1=\frac{w_1 w_2}{1+w_2}
\equiv-1/y
\,,\qquad
v_2=\brk{1+w_1}w_2
\equiv-x
\label{eq:vofw}
\end{equation}
combine pairs of adjacent particles into clusters~\cite{Bartels:2011ge,Bargheer:2015djt}.
The symbol of the function $g$ was determined in~\cite{Bargheer:2015djt}, and it can be used
to constraint the function $g$. Based on symmetry arguments, it was fixed up to $25$
unfixed rational coefficients in~\cite{Bargheer:2015djt}. If additional constraints
from single-valuedness, symmetries and collinear limits are taken into account, the
function $g$ can be shown to take the following form, see \appref{app:gfunction} and~\cite{DelDuca:2018hrv},
\begin{align}
g(x,y)&=
-1/2\,\Gsshort{x}_{0}\Gsshort{\cy}_{0}\Gsshort{\cy}_{1}
+1/2\,\Gsshort{x}_{0}\Gsshort{x}_{1}\Gsshort{\cy}_{1}
+1/2\,\Gsshort{\cy}_{0}\Gsshort{x}_{1}\Gsshort{\cy}_{1}
-1/2\,\Gsshort{x}_{0}\Gsshort{x}_{1}\Gsshort{\cy}_{x}
\nn\\[1mm]&\quad
+1/2\,\Gsshort{\cy}_{0}\Gsshort{\cy}_{1}\Gsshort{\cy}_{x}
-\Gsshort{\cy}_{1}\Gsshort{x}_{0,1}
+\Gsshort{\cy}_{x}\Gsshort{x}_{0,1}
+\Gsshort{x}_{0}\Gsshort{\cy}_{0,1}
-\Gsshort{x}_{1}\Gsshort{\cy}_{0,1}
-\Gsshort{\cy}_{x}\Gsshort{\cy}_{0,1}
\nn\\[1mm]&\quad
+\Gsshort{x}_{1}\Gsshort{\cy}_{0,x}
-\Gsshort{\cy}_{0}\Gsshort{\cy}_{1,x}
-\Gsshort{x}_{1}\Gsshort{\cy}_{1,x}
+\Gsshort{\cy}_{1}\Gsshort{\cy}_{1,x}
+2\,\Gsshort{\cy}_{0,1,x}
-2\,\Gsshort{\cy}_{1,1,x}
\nn\\[1mm]&\quad
+\cg{0}\zeta_2\,\Gsshort{\cy}_{x}
+2\pi i\bigsbrk{
    \cg{1}\bigbrk{
       \brk{\Gsshort{x}_0-\Gsshort{x}_1}\Gsshort{x}_1
       +\brk{\Gsshort{\cy}_0-\Gsshort{\cy}_1}\Gsshort{\cy}_1
       }
       \nn\\[1mm] &\quad \mspace{170mu}
       +\cg{2}\brk{\Gsshort{x}_0-\Gsshort{x}_1}\Gsshort{\cy}_{1}
       +\cg{3}\brk{\Gsshort{x}_0-\Gsshort{\cy}_0}\Gsshort{\cy}_{x}
}
\,.
\label{eq:g2constrained}
\end{align}
Here and in the following, we use the condensed notation
$\Gsshort{z}_{a_1,\dots,a_n}\equiv\Gs(a_1,\dots,a_n;z)$, and
$\cy\equiv1/y$. The functions $\Gs$ are obtained by applying
the single-valued map~\cite{Brown:2013gia} to multiple or Goncharov
polylogarithms~\cite{Goncharov:2001iea}, which can be defined
recursively as iterated integrals
\begin{equation}
\label{eq:Gsdef}
G(a_1,\ldots,a_n;z)
\equiv
\begin{cases}
\displaystyle
\frac{1}{n!}\log^n z & \text{if }a_1=\ldots=a_n=0\,,\\[2ex]
\displaystyle
\int_0^z \frac{dt}{t-a_1}G(a_2,\ldots,a_n;t) & \text{otherwise,}
\end{cases}
\end{equation}
with $G(;z)=1$.
The expansion of this function in the collinear limit takes the form
\begin{align}
    \sbrk{g(v_1,v_2)}^\CL
    &= C_- \Bigbrk{2 \log(r_2)+4 \pi i (\cg1-\cg3)}
    \nn\\
    &+ C_+ \Bigbrk{
        (-6-8 \pi i (2 \cg1+\cg2)) \log(r_2)
        -8 \pi i (\cg1-\cg3) \log(r_1)
    \nn\\
    &+ 4 \pi i (\cg1-\cg3)
        +\frac{\pi^2 \cg{0}}{3}
        -4 \log^2(r_2)
        -4 \log(r_1) \log(r_2)-4
    }
    \nn\\
    &+ C_1 \Bigbrk{-\frac{\pi^2 \cg0}{3}-8 \pi i \cg3 \log(r_1)}
    -8 \pi i \cg1 C_2 \log(r_2)\ ,
\end{align}
where we have used the abbreviated notation
\begin{equation}
C_1=\frac{\cos(\varphi_1)}{r_1}\,,
\quad
C_2=\frac{\cos(\varphi_2)}{r_2}\,,
\quad
C_+=\frac{\cos(\varphi_1+\varphi_2)}{r_1r_2}\,,
\quad
C_-=\frac{\cos(\varphi_1-\varphi_2)}{r_1r_2}\,.
\label{eq:Ccos}
\end{equation}
We see that all remaining coefficients $\cg0,\dots,\cg3$ in the function $g$ survive the
collinear limit, and that they can be fixed by collinear data. In~\cite{DelDuca:2018hrv},
the values of the coefficients for the function $g$ were determined to be%
\footnote{The analysis in~\cite{DelDuca:2018hrv} assumes a certain
path of analytic continuation from the $(\p\p\p)$
to the $(\m\m\m)$ region.}
\begin{equation}
\cg0=\cg1=\cg2=\cg3=0
\,.
\label{eq:cvals}
\end{equation}
There is one more Mandelstam region we need to discuss, namely the
region $\varrho=(\m\p\m)$. In this case, the multi-Regge limit of the
remainder function is known to take the form~\cite{Bartels:2014jya,Bargheer:2015djt},
\begin{equation}
 R_{7,(2)}^{\m\p\m} =f(\eps_1,v_1) + f(\eps_2,v_2) - f(\eps_1,w_1) - f(\eps_2,w_2) +
 \tilde g(v_1,v_2) \ . \label{eq:R7mpmg}
\end{equation}
Here, $f$ is the same function~\eqref{eq:f} that appears for the hexagon. The function $\tilde g$
was not known until quite recently~\cite{DelDuca:2018raq}. Our analysis provides a
different route to determining it. We shall use that the symbols of the remainder
functions in the various Mandelstam regions satisfy the following linear
relation~\cite{Bargheer:2015djt}
\begin{equation}
S\sbrk{R_{7,(2)}^{\m\p\m}}=S\sbrk{R_{7,(2)}^{\m\m\m}}-S\sbrk{R_{7,(2)}^{\m\m\p}}-S\sbrk{R_{7,(2)}^{\p\m\m}}
\,,
\label{eq:Rsymbolid}
\end{equation}
and therefore the symbols of $g$ and $\tilde{g}$ are identical:
\begin{equation}
S\sbrk{g(v_1,v_2)}=S\sbrk{\tilde{g}(v_1,v_2)}
\,.
\end{equation}
Given that $g$ and $\tilde g$ possess the same symbol, and that the constraints we
imposed in order to obtain the expression~\eqref{eq:g2constrained} for
$g$ did not make any reference to
a specific Mandelstam region, we conclude that the general Ansatz~\eqref{eq:g2constrained}
is also valid for $\tilde g$. Of course, the values of the four free parameters within
this Ansatz do depend on the Mandelstam region, and hence are expected to differ from those we
stated for the Mandelstam region $\varrho = (\m\m\m)$
in eq.~\eqref{eq:cvals}. In other words,
\begin{equation}
\tilde{g}(v_4,v_1)
=g(v_4,v_1)\big|_{\cg{i}\to\cgtilde{i}}
\,.
\end{equation}
We will determine the values of the parameters
$\cgtilde0,\dots,\cgtilde3$ for the region $\varrho = (\m\p\m)$
below.
The collinear expansions of the relevant hexagon functions are
\begin{align}
    \sbrk{f\brk{\eps_1, w_1}}^\CL
    &= 2 C_1 \log(\eps_1) \log(r_1)
    + C_1 \bigbrk{4 \log^2(r_1)+8 \log(r_1)+8}
    \,,\\[2mm]
    \sbrk{f\brk{\eps_2, w_2}}^\CL
    &= 2 C_2 \log(\eps_2) \log(r_2)
    +C_2 \bigbrk{4 \log^2(r_2)+8 \log(r_2)+8}
    \,,\\[2mm]
    \sbrk{f\brk{\eps_1, v_1}}^\CL
    &= C_+ \log(\eps_1) \bigbrk{2 \log(r_1)+2 \log(r_2)}
    \\[2mm]
    &+ C_+ \bigbrk{4 \log^2(r_1)+4 \log^2(r_2)+8 \log(r_2) \log(r_1)+8 \log(r_1)+8 \log(r_2)+8}
    \,,\nn\\[2mm]
    \sbrk{f\brk{\eps_2, v_2}}^\CL
    &= \log(\eps_2) \bigbrk{-C_-+C_+ (2 \log(r_2)-1)+2 C_2 \log(r_2)}
    +C_- \bigbrk{-4 \log(r_2)-4}
    \nn\\[2mm]
    &+C_+ \bigbrk{4 \log^2(r_2)+4 \log(r_2)+4}
    +C_2 \bigbrk{4 \log^2(r_2)+8 \log(r_2)+8}
    \,.
\end{align}
Combining the above, we find that the
collinear limit of the multi-Regge remainder function in
the four different Mandelstam regions is
\begin{align}
\sbrk{R_{7,(2)}^{\m\m\p}}^\CL&=
    2 C_1 \log(\eps_1) \log(r_1)
    +C_1 \brk{4 \log^2(r_1)+8 \log(r_1)+8}
\,,
\label{eq:Rmmpcoll}\\[2mm]
\sbrk{R_{7,(2)}^{\p\m\m}}^\CL&=
    2 C_2 \log(\eps_2) \log(r_2)
    +C_2 \brk{4 \log^2(r_2)+8 \log(r_2)+8}
\,,
\label{eq:Rpmmcoll}\\[2mm]
\sbrk{R_{7,(2)}^{\m\m\m}}^\CL&=
    \log(\eps_1) C_+ \bigbrk{2 \log(r_1)+2 \log(r_2)}
    +\log(\eps_2) \bigbrk{-C_-+C_+ (2 \log(r_2)-1)+2 C_2 \log(r_2)}
\nn\\[2mm] &\quad    +C_1 \bigbrk{-2 \cg0 \zeta_2-8 \pi i \cg3 \log(r_1)}
+C_2 \bigbrk{(8-8 \pi i \cg1) \log(r_2)+4 \log^2(r_2)+8}
\nn\\[2mm] &\quad    +C_+ \bigbrk{
        (8-8 \pi i (\cg1-\cg3)) \log(r_1)
        +(6-8 \pi i (2 \cg1+\cg2)) \log(r_2)
\nn\\[2mm] &\quad \mspace{50mu}         +2 (\cg0 \zeta_2+4)
        +4 \pi i (\cg1-\cg3)
        +4 \log^2(r_1)
        +4 \log^2(r_2)
        +4 \log(r_2) \log(r_1)
    }
\nn\\[2mm] &\quad    +C_- \bigbrk{4 \pi i (\cg1-\cg3)-2 \log(r_2)-4}
\,,
\label{eq:Rmmmcoll}\\[2mm]
\sbrk{R_{7,(2)}^{\m\p\m}}^\CL &=
    \log(\eps_1) \brk{C_+ (2 \log(r_1)+2 \log(r_2))-2 C_1 \log(r_1)}
    +\log(\eps_2) \brk{C_+ (2 \log(r_2)-1)-C_-}
\nn\\[2mm]&\quad    +C_1 \bigbrk{(-8-8 \pi i \cgtilde3) \log(r_1)-2 (\cgtilde0 \zeta_2+4)-4 \log^2(r_1)}
-C_2 \, 8 \pi i \, \cgtilde1 \log(r_2)
\nn\\[2mm] &\quad    +C_+ \bigbrk{
        (8-8 \pi i (\cgtilde1-\cgtilde3)) \log(r_1)
        +(6-8 \pi i (2 \cgtilde1+\cgtilde2)) \log(r_2)
\nn\\[2mm]&\quad \mspace{50mu}       +2 (\cgtilde0 \zeta_2+4)
        +4 \pi i (\cgtilde1-\cgtilde3)
        +4 \log^2(r_1)
        +4 \log^2(r_2)
        +4 \log(r_2) \log(r_1)
    }
\nn\\[2mm] &\quad    +C_- \bigbrk{4 \pi i (\cgtilde1-\cgtilde3)-2 \log(r_2)-4}
\,.
\label{eq:Rmpmcoll}
\end{align}
This concludes our brief review of known results on the two-loop heptagon remainder  function in
multi-Regge kinematics. We are now prepared to compare with what we obtain when we continue the
collinear heptagon remainder function into the various Mandelstam regions.

%%%%%%%%%%%%%%%%%%%%%%%%%%%%%%%%%%%%%%%%%%%%%%%%%%%%%%%%%%%%
\subsection{Analytic Continuation}

As explained in \secref{sec:mrlfromope}, we can reach all Mandelstam
regions from the $(\p\p\p)$ region by analytic continuation of some of the
forward energy variables $p^0_i$, $i=4,5,6$. For the four non-trivial regions of the
heptagon, these continuations entail the windings of the seven cross ratios
around the origin displayed in \tabref{table:crsWinding}.
\begin{table}
\centering
\begin{tabular}{cccccccc}
\toprule
Region     & $u_{1,1}$ & $u_{1,2}$ & $u_{1,3}$ & $u_{2,1}$ & $u_{2,2}$ & $u_{2,3}$ & $U_{2,6}$ \\
\midrule
$(\m\m\p)$ & $-1$      & $0   $    & $0   $    & $0 $      & $-1/2$    & $1/2 $    & $0$\\
$(\p\m\m)$ & $0 $      & $1/2 $    & $-1/2$    & $-1$      & $0   $    & $0   $    & $0$\\
$(\m\m\m)$ & $0 $      & $0   $    & $0   $    & $0 $      & $0   $    & $0   $    & $-1$\\
$(\m\p\m)$ & $1 $      & $-1/2$    & $1/2 $    & $1 $      & $1/2 $    & $-1/2$    & $-1$\\
\bottomrule
\end{tabular}
\caption{%
    Winding numbers of basis cross ratios as one continues from
    the $(\p\p\p)$ to into the four different non-trivial Mandelstam regions.%
}
\label{table:crsWinding}
\end{table}
The entries of this table are produced with the help of our formula~\eqref{eq:winding}.
We note that in contrast to the hexagon, some of the cross ratios possess half-windings around the
origin. Let us also stress that in the $(\m\m\m)$ region, only the cross ratio $U_{2,6}$ possesses
a non-vanishing winding number around the origin.

%%%%%%%%%%%%%%%%%%%%%%%%%%%%%%
\paragraph{Generators.}

As in the previous section, we first need to determine a generating set of curves that
can expose all the branch cuts in the collinear limit. In order to do so, let us begin
by listing all the branch cuts of the two-loop collinear remainder function. The
component functions in eq.~\eqref{eq:R72ope} are linear combinations of products
of logarithms $\log(x_i)$ and polylogarithms $\Li_2(1-y_i)$, $\Li_3(1-y_i)$.
The arguments $x_i$ are ratios with factors%
\footnote{In their original form, the component functions also contain
$\log(S_i)$, which we can safely rewrite as $1/2\log(S_i^2)$. The
logarithm arguments also contain factors $(S_1^2+S_2^2)$, but these
are spurious: They cancel out upon expanding all logarithms.}
\begin{equation}
x_i\in\lrbrc{S_1^2,1+S_1^2,S_2^2,1+S_2^2,S_1^2+S_2^2+S_1^2S_2^2}
\,.
\label{eq:argfactors}
\end{equation}
In addition, there are five different polylogarithm arguments $y_i$:
\begin{equation}
\begin{tabular}{@{\,}r@{\quad}c@{\;\;\;\;}c@{\;\;\;\;}c@{\;\;\;\;}c@{\;\;\;\;}c@{\,}}
\toprule &
$y_1$ &
$y_2$ &
$y_3$ &
$y_4$ &
$y_5$
\\
\midrule
$\displaystyle y_i$\;: &
$\displaystyle \frac{1 + S_1^{2}}{S_1^{2}}$ &
$\displaystyle \frac{1 + S_2^{2}}{S_2^{2}}$ &
$\displaystyle \frac{S_1^2 + S_2^{2} + S_1^{2} S_2^{2}}{(1 + S_1^{2})S_2^2}$ &
$\displaystyle \frac{S_1^2 + S_2^{2} + S_1^{2} S_2^{2}}{S_1^2(1 + S_2^{2})}$ &
$\displaystyle \frac{(1 + S_1^2) (1 + S_2^2)}{S_1^2 + S_2^2 + S_1^2 S_2^2}$
\\[2.5ex]
$\displaystyle (1-y_i)$\;: &
$\displaystyle -\frac{1}{S_1^2}$ &
$\displaystyle -\frac{1}{S_2^2}$ &
$\displaystyle -\frac{S_1^2}{(1+S_1^2)S_2^2}$ &
$\displaystyle -\frac{S_2^2}{S_1^2(1+S_2^2)}$ &
$\displaystyle -\frac{1}{S_1^2+S_2^2+S_1^2S_2^2}$
\\
\bottomrule
\end{tabular}
\label{eq:1mplogArg}
\end{equation}
All expressions $y_i$ and $1-y_i$ are ratios of $x_i$. Since
$\Li_n(z)$ has branch points at $z=1$ and $z=0$, the space of complex $S_1^2$ and $S_2^2$ contains $r=5$ branch
points at $x_i = 0$. Consequently, the fundamental group can be generated
by five elements $p_\nu$, $\nu = 1, \dots, 5$. We will now describe the
precise curves $\gen_\nu$ we shall use in order to represent
the set of generators, see \tabref{table:hepGens}.
\begin{table}
    \centering
    \begin{tabular}{crr}
        \toprule
        generator &
        $S_1^2$ \hspace{2cm} &
        $S_2^2$ \hspace{2cm}
        \\
        \midrule
        $\gen_1$ & \includegraphics[align=c]{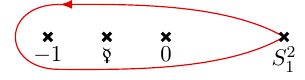} &
        fix \hspace{2cm}
        \\
        $\gen_2$ &
        \includegraphics[align=c]{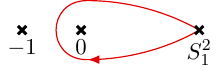} &
        \includegraphics[align=c]{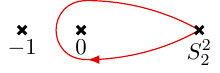}
        \\
        $\gen_3$ &
        fix \hspace{2cm} &
        \includegraphics[align=c]{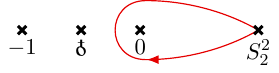}
        \\
        $\gen_4$ &
        \includegraphics[align=c]{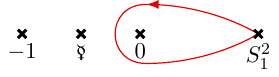} &
        fix \hspace{2cm}
        \\
        $\gen_5$ &
        fix \hspace{2cm} &
        \includegraphics[align=c]{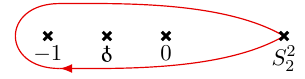}
        \\
        \bottomrule
        \\[-2ex]
        &
        $\text{\Mercury} = -{S_2^2}/\brk{1 + S_2^2}$ \hspace{1cm}
        &
        $\text{\Earth} = -{S_1^2}/\brk{1 + S_1^2}$ \hspace{0.8cm}
    \end{tabular}
    \caption{
        Tables of generators of the fundamental group.
        All generators except $\gen_2$, move only one of the $S_j^2$ variables.
        The $\gen_2$ generator winds both $S_1^2$ and $S_2^2$ at the same time,
        so that $S_1^2 + S_2^2 + S_1^2 S_2^2$ winds exactly once in the clockwise direction.}
    \label{table:hepGens}
\end{table}

All curves start and end at positive values of $S_1^2$ and $S_2^2$,
since all the branch points $x_i=0$ are located where either
$S_1^2\leq0$ or $S_2^2\leq0$.
The first curve
$\gen_1$ keeps $S_2^2$ constant. In the space of $S_1^2$,
it looks similar to the curve $\gen_2$ we introduced in the discussion
of the hexagon, see \tabref{table:pathgenerators}, except that now there
is one more branch point at
$S_1^2 = - S_2^2/(1+S_2^2) \in (-1,0)$, \ie in between the two branch points
at $S_1^2 =0$ and $S_1^2=-1$.
$\gen_2$ moves both variables $S_1^2$ and $S_2^2$ at the same time, in such a way that $S_1^2 + S_2^2 + S_1^2 S_2^2$
winds in the clockwise direction once.
For $\gen_3$, we keep $S_1^2$ fixed and rotate $S_2^2$ around $0$,
just like $\gen_1$ of the hexagon in \tabref{table:pathgenerators}.
$\gen_4$ is similar: it rotates $S_1^2$ around $0$ and fixes $S_2^2$.
$\gen_5$ is the same as $\gen_1$, but with the roles of $S_1^2$ and $S_2^2$ interchanged.

It turns out that we can model these generators $\gen_\nu$ with just circular movements
of $S_1^2$ and $S_2^2$ around $0$. More importantly, the generators $\gen_\nu$ are
chosen such that if an argument $1-y_i$ of a polylogarithm
winds around $y_i=0$ under the action of $\gen_\nu$, then it also winds
around $y_i=1$, along a path that is
homotopically equivalent to a circle in the $y_i$ plane.
This makes it easy to consistently pick the branches of the produced
logarithms. Namely, for all our generators $\gen_\nu$:
\begin{equation}
\gen_\nu \Li_n\brk{1-y_i}
= \Li_n\brk{1-y_i}
- s_\nu(y_i)\frac{2 \pi i}{\Gamma\brk{n}}
  \bigbrk{\log\brk{y_i-1} + \mathinner{s_\nu(y_i)}\pi i}^{n-1}
  \,,
\label{eq:contnLi}
\end{equation}
where $s_\nu(y_i)=\pm1$ is the counterclockwise winding number of
$y_i$ around $y_i=0$ under the action of the curve $\gen_\nu$. In
addition, we have the obvious continuation rule
\begin{equation}
\gen_\nu \log\brk{x}
= \log\brk{x} + s_\nu(x)2 \pi i
\,.
\label{eq:contnLog}
\end{equation}
From our description of the curves $\gen_\nu$, it is not difficult to infer the
winding numbers of the variables $x_i$ and of the cross ratios $U_{ij}$. The
winding numbers of the arguments $x_i$ and $y_i$ around zero are given in the following tables:
\begin{equation}
\begin{tabular}{cccccc}
\toprule
$x_i$ & $\gen_1$ & $\gen_2$ & $\gen_3$ & $\gen_4$ & $\gen_5$\\
\midrule
$S_1^2$                  & $1$ & $-1$ & $ 0$ & $1$ & $ 0$ \\
$1+S_1^2$                & $1$ & $ 0$ & $ 0$ & $0$ & $ 0$ \\
$S_2^2$                  & $0$ & $-1$ & $-1$ & $0$ & $-1$ \\
$1+S_2^2$                & $0$ & $ 0$ & $ 0$ & $0$ & $-1$ \\
$S_1^2+S_2^2+S_1^2S_2^2$ & $1$ & $-1$ & $ 0$ & $0$ & $-1$ \\
\bottomrule
\end{tabular}
\qquad
\begin{tabular}{cccccc}
\toprule
      & $\gen_1$ & $\gen_2$ & $\gen_3$ & $\gen_4$ & $\gen_5$\\
\midrule
$y_1$ & $0$      & $1$      & $0$      & $-1$     & $0$ \\
$y_2$ & $0$      & $1$      & $1$      & $0$      & $0$ \\
$y_3$ & $0$      & $0$      & $1$      & $0$      & $0$ \\
$y_4$ & $0$      & $0$      & $0$      & $-1$     & $0$ \\
$y_5$ & $0$      & $1$      & $0$      & $0$      & $0$ \\
\bottomrule
\end{tabular}
\end{equation}
Here, a ``1'' means that the corresponding factor winds once around
the origin, in the mathematically positive sense (counterclockwise).
The curves we have introduced have the virtue that each of them winds
exactly one of the cross ratios~\eqref{eq:crs7coll} around the origin,
as shown in \tabref{table:genCrs} on the left. Each cycle begins and ends
in the region $S_1^2,S_2^2>0$, and therefore $0<U_{ij}<1$ for all of
the cross ratios~\eqref{eq:crs7coll}. Besides winding one cross ratio
around the origin, each generator also winds two other cross ratios
around $U_{ij}=1$, as shown in \tabref{table:genCrs} on the right.
\begin{table}
\begin{center}
\begin{tabular}{cccccc}
\toprule
 & $\gen_1$ & $\gen_2$ & $\gen_3$ & $\gen_4$ & $\gen_5$\\
\midrule
$u_{1,1}$ & $-1$ & $0$ & $0$ & $0$ & $0$ \\
$u_{2,1}$ & $0$ & $-1$ & $0$ & $0$ & $0$ \\
$U_{26}$  & $0$ & $0$ & $-1$ & $0$ & $0$ \\
$u_{1,2}$ & $0$ & $0$ & $0$ & $1$ & $0$ \\
$u_{2,2}$ & $0$ & $0$ & $0$ & $0$ & $1$ \\
\bottomrule
\end{tabular}
\qquad
\begin{tabular}{cccccc}
\toprule
 & $\gen_1$ & $\gen_2$ & $\gen_3$ & $\gen_4$ & $\gen_5$\\
\midrule
$1-u_{1,1}$ & $0$  & $-1$ & $0$ & $-1$ & $0$ \\
$1-u_{2,1}$ & $-1$ & $0$  & $0$ & $0$  & $-1$ \\
$1-U_{26}$  & $0$  & $0$  & $0$ & $-1$ & $-1$ \\
$1-u_{1,2}$ & $1$  & $0$  & $1$ & $0$  & $0$ \\
$1-u_{2,2}$ & $0$  & $1$  & $1$ & $0$  & $0$ \\
\bottomrule
\end{tabular}
\end{center}
    \caption{Winding numbers of cross ratios under the action of generators $\gen_i$.}
    \label{table:genCrs}
\end{table}
The cross ratios $u_{1,3}$ and $u_{2,3}$ are not displayed, since their
leading collinear term does only depend on $T_1$ and $T_2$, respectively. In addition to
these full rotations of cross ratios $U_{ij}$ that start and end on the positive real line,
we also need to include possible curves that have half-integer winding numbers, as dictated
by the kinematics in \tabref{table:crsWinding}.
That means that we allow cross ratios to move to the negative real
line above (winding number $+1/2$)
or below (winding number $-1/2$) the
origin, and we choose to append such region-dependent continuations to the end of
our curves, so that we can just specify the correction terms to discontinuities
generated by $\gen_\nu$ for each region separately \brk{see \appref{app:mpmDisc}}.
For the continuation of the cross ratios $u_{1, 3}$ and $u_{2, 3}$,
that implies substitutions of the
collinear variables $T_j \to \pm i T_j$, where the sign depends on the sign of the half winding
in \tabref{table:crsWinding}.
That way, we need to consider continuations to negative arguments only of pure logarithms
produced by eq.~\eqref{eq:contnLi}.

%%%%%%%%%%%%%%%%%%%%%%%%%%%%%%
\paragraph{Discontinuities.}

Now let us focus on the discontinuities relevant for continuations
into the various Mandelstam regions.
Using the variables~\eqref{eq:TSFhept,eq:frtow7}, we find the following non-zero single
discontinuities of the two-loop, near-collinear remainder function in
the combined multi-Regge collinear limit:
\begin{align}
\Delta_1\Rope &=
    2 \log(\eps_1)\, C_1 \log(r_1)
    + C_1\bigbrk{4 \log^2(r_1)+8 \log(r_1)+8}\,,\nn\\[2mm]
\Delta_2\Rope &=
    2 \log(\eps_2)\, C_2 \log(r_2)
    + C_2\bigbrk{4 \log^2(r_2)+8 \log(r_2)+8}\,,\nn\\[2mm]
\Delta_3\Rope &=
    \log(\eps_1)\, C_+ \bigbrk{2 \log(r_1)+2 \log(r_2)}
    \nn\\[2mm] & \quad
    +\log(\eps_2) \bigbrk{2 C_2 \log(r_2)+ C_+(2 \log(r_2)-1)-C_-}
    \nn\\[2mm] & \quad
    + C_2\bigbrk{4 \log^2(r_2)+8 \log(r_2)+8}
    \nn\\[2mm] & \quad
    + C_+\bigbrk{4 \log^2(r_1)+4 \log(r_2) \log(r_1)+8 \log(r_1)+4 \log^2(r_2)+6 \log(r_2)+8}
    \nn\\[2mm] & \quad
    - C_-\bigbrk{2 \log(r_2)+4}
\,,
\label{eq:delta}
\end{align}
where we have used the abbreviated notation~\eqref{eq:Ccos}.
In particular, the discontinuities $\Delta_4\Rope$ and $\Delta_5\Rope$
vanish. That is, the generators $\gen_4$ and $\gen_5$ act trivially on
$\Rope$ in the multi-Regge limit, which immediately shows that the representation of the
fundamental group is not faithful.
Since $\gen_4$ and $\gen_5$ let only small cross ratios wind, this
shows that the remainder function on the main $(\p\p\p)$ sheet has
trivial monodromy in the combined multi-Regge collinear limit when these
small cross ratios wind around the origin. This confirms earlier
findings~\cite{Bargheer:2015djt}. For the double discontinuities $\Delta_{i,j} \equiv
\Delta_i \Delta_j\Rope$, we find:
\begin{align}
\Delta_{1,2} =
    &-\log(\eps_1)\, C_1
    -\log(\eps_2)\, C_+
    % \nn\\[2mm] &
    + C_1\bigbrk{-6 \log(r_1)+4 \pi i-4}
    + C_+\bigbrk{4 \log(r_1)-2 \log(r_2)-4 \pi i}
\,,\nn\\[2mm]
\Delta_{2,1} =
    &\log(\eps_1) \bigbrk{C_1-C_+}
    % \nn\\[2mm] &
    + C_1\bigbrk{-2 \log(r_1)+4 \pi i-4}
    + C_+\bigbrk{2 \log(r_1)-4 \pi i+2}
    -2 C_-
\,,\nn\\[2mm]
\Delta_{1,4} =
    &\log(\eps_1)\, C_1
    + C_1\bigbrk{6 \log(r_1)+4 \pi i+4}
\,,\nn\\[2mm]
\Delta_{4,1} =
    &-\log(\eps_1)\, C_1
    + C_1\bigbrk{2 \log(r_1)-4 \pi i+4}
\,,\nn\\[2mm]
\Delta_{2,5} =
    &\log(\eps_2)\, C_2
    + C_2\bigbrk{6 \log(r_2)+4 \pi i+4}
\,,\nn\\[2mm]
\Delta_{5,2} =
    &-\log(\eps_2)\, C_2
    + C_2\bigbrk{2 \log(r_2)-4 \pi i+4}
\,,\nn\\[2mm]
\Delta_{3,4} =
    &\log(\eps_1)\, C_+
    + C_+\bigbrk{6 \log(r_1)+4 \log(r_2)+4 \pi i+4}
\,,\nn\\[2mm]
\Delta_{4,3} =
    &-\log(\eps_1)\, C_+
    -\log(\eps_2)\, C_+
    + C_+\bigbrk{2 \log(r_1)+2 \log(r_2)-4 \pi i+4}
\,,\nn\\[2mm]
\Delta_{3,5} =
    &\log(\eps_1)\, C_+
    +\log(\eps_2) \bigbrk{C_2+C_+}
    \nn\\[2mm] &
    + C_2\bigbrk{6 \log(r_2)+4 \pi i+4}
    + C_+\bigbrk{2 \log(r_1)+6 \log(r_2)+4 \pi i+2}
    -2 C_-
\,,\nn\\[2mm]
\Delta_{5,3} =
    &-\log(\eps_2) \bigbrk{C_2+C_+}
    + C_+\bigbrk{2 \log(r_2)-4 \pi i}
    + C_2\bigbrk{2 \log(r_2)-4 \pi i+4}
\,.
\label{eq:ddelta}
\end{align}
Notably, the generators $\gen_4$ and $\gen_5$ act non-trivially on
some of the single discontinuities. The triple discontinuities
$\Delta_{i,j,k} \equiv \Delta_i \Delta_j \Delta_k\Rope$ are:
\begin{alignat}{2}
\Delta_{1,2,5} &
= \Delta_{1,5,2}
= \Delta_{5,1,2}
= \Delta_{5,2,1}
= \Delta_{4,1,3}
= \Delta_{4,3,1} &&
\nn\\[2mm]\nn &
= -\sfrac12 \Delta_{3,4,4}
= \sfrac12 \Delta_{4,3,3}
= -\Delta_{3,4,5}
= -\Delta_{3,5,4}
= && -2 C_+
\,,\nn\\[2mm]
\Delta_{1,2,4} &
= \Delta_{1,4,2}
= 2 C_+ - 4 C_1
\,,
&& \mspace{-70mu}
\Delta_{1,2,2}
= \Delta_{2,1,1}
= 4 C_1-4 C_+
\,,\nn\\[2mm]
\Delta_{3,5,5} &
= -\Delta_{5,3,3}
= 4 C_2+4 C_+
\,,
&& \mspace{-70mu}
\Delta_{5,2,3}
= \Delta_{5,3,2}
= -2 C_+-4 C_2
\,,\nn\\[2mm]
\Delta_{1,4,4} &
= -\Delta_{4,1,1}
= 4 C_1
\,,
&& \mspace{-70mu}
\Delta_{2,5,5}
= -\Delta_{5,2,2}
= 4 C_2
\,.
\label{eq:dddelta}
\end{alignat}
%

%%%%%%%%%%%%%%%%%%%%%%%%%%%%%%
\paragraph{Continuation Results.}

Having listed all the relevant discontinuities that remain non-trivial in
multi-Regge kinematics, we can now compute the collinear remainder function
in the various Mandelstam regions. For the regions $\varrho = (\m\m\p)$,
$(\p\m\m)$, and $(\m\m\m)$, immediate candidates for admissible continuations
are those along the curves $\gen_1$, $\gen_2$, and $\gen_3$,
respectively. Indeed, one finds that
\begin{gather} \label{eq:pmmpcurve}
\bigsbrk{R_{7,(2)}^{\m\m\p}}^\CL=\bigsbrk{\gen_1\Rope}\supMRL
\,,\qquad
\bigsbrk{R_{7,(2)}^{\p\m\m}}^\CL=\bigsbrk{\gen_2\Rope}\supMRL
\,,\\[2mm]
\bigsbrk{\Rmmm}^\CL=\bigsbrk{\gen_3\Rope}\supMRL \,,
\label{eq:mmmCurve}
\end{gather}
where we use the simple shorthand $\bar\gen_i\equiv\gen_i^{-1}$ for the
inverse of the curve $\gen_i$, \ie the curve with opposite orientation.
The right hand side of these three equations is computed from the discontinuities we listed
above (and the corrections for appended half-windings that are listed in \appref{app:mpmDisc}). What one
obtains reproduces exactly the expressions in eqs.~\eqref{eq:Rmmpcoll}--\eqref{eq:Rmmmcoll},
including the values~\eqref{eq:cvals} of the parameters $\cg0, \dots, \cg3$ that cannot be
determined by general constraints such as single-valuedness, symmetries, and collinear
limits.

For the region $\varrho=(\m\p\m)$, finally, we infer from \tabref{table:crsWinding}
and \tabref{table:genCrs} that the corresponding curves must contain $\bar \gen_1$, $\bar
\gen_2$, and $\gen_3$. Of course the order in which we put these generators matters for
the continuation, but it cannot be determined from the winding numbers alone. It turns
out that we can obtain the correct result in several ways, including
\begin{equation} \label{eq:result1heptagon}
\bigsbrk{\Rmpm}^\CL
=\bigsbrk{\bar\gen_2\gen_3\bar\gen_1\Rope}\supMRL
=\bigsbrk{\bar\gen_2\bar\gen_1\gen_3 \Rope}\supMRL
=\bigsbrk{\gen_3\bar\gen_2\bar\gen_1\Rope}\supMRL
\,.
\end{equation}
Once again, the continuation along the three paths on the right hand side is computed
with the help of the cut contributions we listed above, along with the
corrections for appended
half-windings from \appref{app:mpmDisc}. In all three cases, the result is the
same, and it agrees with formula~\eqref{eq:Rmpmcoll}, with the parameters $\cgtilde{i}$
given by
\begin{equation}  \label{eq:ctildevalues}
    \cgtilde{0} = \cgtilde{1} = \cgtilde{2} = \cgtilde{3} - \frac12 = 0\ .
\end{equation}
These values are in agreement with the recent results in~\cite{DelDuca:2018hrv}. Let
us note that in all three paths in~\eqref{eq:result1heptagon}, the generators $\bar \gen_1$
and $\bar \gen_2$ appear in  the same order. But there are three more permutations in which the order
of these two generators is reversed. For these remaining three paths, it is necessary to
involve additional generators $\gen_4$ and $\gen_5$ in order to get the correct result,
\eg
\begin{equation} \label{eq:result2heptagon}
\bigsbrk{\Rmpm}^\CL
=\bigsbrk{\bar\gen_5\gen_4\bar\gen_1\bar\gen_2\gen_3\bar\gen_4\gen_5\Rope}\supMRL
\,.
\end{equation}
It turns out that the precise dressing of the path with the generators $\gen_4$ and
$\gen_5$ can be determined by matching to the LLA on the left hand side. Once the
dressing is known, one can use it to compute the NLLA, and one finds again full
agreement with eqs.~\eqref{eq:Rmpmcoll} and~\eqref{eq:ctildevalues}. The same is
true for the remaining two orders in which we can place our three generators
$\bar \gen_1, \bar \gen_2$ and $\gen_5$. Some more detailed comments on the
derivation of these results are collected in the final subsection.

%%%%%%%%%%%%%%%%%%%%%%%%%%%%%%%%%%%%%%%%%%%%%%%%%%%%%%%%%%%%
\subsection{Analysis of Continuation Paths}
\label{ssec:curves}

\begin{figure}
    \centering
    \includegraphics[align=c]{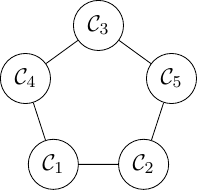}
    \caption{%
        Commutation relations between generators.
        An edge between two generators $\gen_i$ and $\gen_j$ corresponds to
        a non-trivial commutator $\comm{\gen_i}{\gen_j}$. Notably, we
        find that whenever a commutator vanishes,
        $\comm{\gen_i}{\gen_j}$=0, the double discontinuities
        $\Delta_{i,j}$ and $\Delta_{j,i}$ also vanish.%
    }
    \label{fig:pentagon}
\end{figure}

This final subsection contains a number of detailed comments on the results we
summarized in eqs.~\eqref{eq:result1heptagon}--\eqref{eq:result2heptagon} at
the end of the previous subsection. As we explained above, each of the
Mandelstam regions is reached by a curve that must satisfy the total winding
conditions of \tabref{table:crsWinding}. The latter comes purely from kinematical
considerations. We view the curves as monomials in noncommutative generators
$\gen_\nu$, so that the total winding fixes the total power of each generator
in these monomials. In case some cross ratios undergo half-windings, these are
added at the end, after all the full windings have been carried out. The
noncommutative structure of the fundamental group starts to reveal itself at
the level of the double discontinuities. Let us note that some generators
$\gen_i$ commute in our two-loop analysis. In fact, two generators $\gen_i$,
$\gen_j$ commute iff $\Delta_{A,i,j,B}=\Delta_{A,j,i,B}$ for all (possibly
empty) $A$ and $B$. From the tables of discontinuities~\eqref{eq:ddelta}
and~\eqref{eq:dddelta}, we see that for example $\gen_3$ commutes with both
$\gen_1$ and $\gen_2$. We collect the commutation relations among all generators
in \figref{fig:pentagon}. After these introductory comments, let us now briefly
consider the four different Mandelstam regions.
%

%%%%%%%%%%%%%%%%%%%%%%%%%%%%%%
\subsubsection{The Mandelstam Regions \texorpdfstring{$(\m\m\p)$}{(--+)} and
\texorpdfstring{$(\p\m\m)$}{(+--)}}

Even though the combined multi-Regge collinear limit~\eqref{eq:Rmmpcoll}
and~\eqref{eq:Rpmmcoll} of the remainder function in the regions
$(\m\m\p)$ and $(\p\m\m)$ reproduces formula~\eqref{eq:Rmmcoll}
for the hexagon, the derivation through analytic continuation
from the Wilson loop OPE is not quite the same. On the one hand, the heptagon
case involves some half-windings that need to be taken into account, see
\tabref{table:crsWinding}. On the other hand, the two-loop functions $h_2$, $\bar{h}_2$ in
the expression~\eqref{eq:R72ope} for the collinear heptagon remainder
function constitute a new ingredient that does not appear for the hexagon. In
addition, there are also more generators to consider for the heptagon.

It turns out that there is a natural choice of the continuation curve, dictated by
the structure of the OPE prediction~\eqref{eq:R72ope} that agrees with the previous analysis.
In the $\brk{\m\m\p}$ and $\brk{\p\m\m}$ regions,
as one can see from the lists of discontinuities in eq.~\eqref{eq:delta} together with the
half-winding corrections shown in~\eqref{eq:deltapmm} and~\eqref{eq:deltammp},
the contributions from $h_2$ and $\bar{h}_2$ to the analytic
continuations along the curves $\gen_1$ and $\gen_2$ respectively
are trivial in the combined multi-Regge collinear limit:
\begin{align}
    \Delta_1^{\brk{\m\m\p}} h_2
    = \Delta_1^{\brk{\m\m\p}} \bar{h}_2
    = \Delta_2^{\brk{\p\m\m}} h_2
    = \Delta_2^{\brk{\p\m\m}} \bar{h}_2
    = 0.
\end{align}
This leaves only the hexagonal functions $\f_1$ and $\f_2$ to contribute to these
simplest continuation curves, which then leads to exactly the same form shown in
eq.~\eqref{eq:Rmmpcoll} and~\eqref{eq:Rpmmcoll}
of the heptagonal remainder function in these two regions.
Hence, in the end, the results of the continuation along the simplest curves for the $\brk{\m\m\p}$
and $\brk{\p\m\m}$ regions do boil down to the hexagon case, albeit in
a somewhat nontrivial manner.

%%%%%%%%%%%%%%%%%%%%%%%%%%%%%%
\subsubsection{The Mandelstam Region \texorpdfstring{$(\m\m\m)$}{(---)}}

The collinear limit of the remainder
function~\eqref{eq:Rmmmcoll} contains four free parameters $\cg0, \dots,
\cg3$ that are not fixed by general considerations. Under some
assumptions on the complexity of the path of continuation,
we can fix these unknowns. In order to do so, we first constrain the paths
using only information about the LLA in eq.~\eqref{eq:Rmmmcoll}. Subsequently,
we can use any such LLA-admissible path to determine the free parameters in
NLLA. As in the hexagon case~\eqref{eq:disclc}, we accomplish this by constructing
a general $\Z$-linear combination of discontinuities,
\begin{equation}
\lrsbrk{R_{7,(2)}^{\m\m\m}}^\CL = \biggbrk{
 2 \pi i\sum_i\cc{i}\Delta_i
+(2 \pi i)^2\sum_{i,j}\cc{i,j}\Delta_i\Delta_j
+(2 \pi i)^3\sum_{i,j,k}\cc{i,j,k}\Delta_i\Delta_j\Delta_k}\Rope
\,,
\label{eq:disclcHep}
\end{equation}
and match it against~\eqref{eq:Rmmmcoll}.
Here, we include only non-zero discontinuities, see~\eqref{eq:delta,
eq:ddelta, eq:dddelta}, which means that we need to find 5 linear coefficients
$\cc{i}$, 10 quadratic coefficients $\cc{i, j}$ and 22 cubic
coefficients $\cc{i, j, k}$. The total
winding conditions from \tabref{table:crsWinding}, together with the
dictionary in \tabref{table:genCrs}, fix the linear
coefficients $\cc{i}$ to
\begin{align}
    \brc{\cc{1}, \cc{2}, \cc{3}, \cc{4}, \cc{5}}
    =
    \brc{0, 0, 1, 0, 0} \ .
    \label{eq:mmmTotal}
\end{align}
Matching at LLA, we obtain four constraints on the quadratic coefficients $\cc{i, j}$.
Next, we exploit that only special linear combinations in eq.~\eqref{eq:disclcHep}
can actually come from some continuation curve, therefore there must be additional
constraints on the coefficients $\cc{i, j}$ and $\cc{i, j, k}$. One such condition
is discussed in eq.~\eqref{eq:acommut}: One can express $\cc{i, j} = \cc{i} \cc{j}
- \cc{j, i}$, which reduces the number of unknowns by 5.

We fix the remaining uncertainty in the coefficients by considering an Ansatz for the
curve of continuation. From the structure of the commutation relations shown in \figref{fig:pentagon},
we see that the most general continuation curve that has only one instance of~$\gen_3$ and
does not involve~$\gen_1$ or~$\gen_2$ is
\begin{align}
    \gamma =
    \gen_4^{\alpha_1}
    \gen_5^{\alpha_2}
    \gen_3
    \gen_4^{\alpha_3}
    \gen_5^{\alpha_4}\ .
\end{align}
Assuming such a curve,
all coefficients $\brc{\cc{i}, \cc{i, j}, \cc{i, j, k}}$
can be expressed in terms of the exponents $\brc{\alpha_1, \alpha_2,
\alpha_3, \alpha_4}$ via~\eqref{eq:discRnlope}.
Using the conditions~\eqref{eq:disclcHep} in LLA with the values~\eqref{eq:mmmTotal}
put in, the solution is unique:
\begin{align}
    \alpha_1 = \alpha_2 = \alpha_3 = \alpha_4 = 0
    \,,
\end{align}
which motivates our choice of the simplest continuation curve for the $\brk{\m\m\m}$
region in~\eqref{eq:mmmCurve}. With this curve, we can then compute the NLLA part
of the remainder function $\Rmmm$, and fix the four coefficients $\cg{i}$
in the Ansatz~\eqref{eq:g2constrained} to assume the
values~\eqref{eq:cvals} we stated before. Thereby, we have determined all the
unknowns in our Ansatz for $\Rmmm$. The numerical values for the $\cg{i}$ we obtain
are in full agreement with~\cite{DelDuca:2018hrv}.

%%%%%%%%%%%%%%%%%%%%%%%%%%%%%%
\subsubsection{The Mandelstam Region \texorpdfstring{$(\m\p\m)$}{(-+-)}}

After our detailed discussion of the region $\brk{\m\m\m}$, we can be rather brief for the
remaining case $\brk{\m\p\m}$. For the latter, the winding number conditions force us to
continue along a composite curve that must involve $\bar\gen_1$, $\bar \gen_2$ and
$\gen_3$. In addition, it can certainly also contain $\gen_4$ and $\gen_5$, as in the
previous subsection. \Tabref{table:crsWinding} shows that the Mandelstam
region $\varrho=\brk{\m\p\m}$ also involves some half-windings, so that the continuation
requires results from \appref{app:mpmDisc} in addition to the discontinuities
we listed in the previous subsection.

Assuming that the generators $\gen_3$, $\bar\gen_1$, and $\bar\gen_2$ each appear
only once in the path-monomial, and taking into account the commutation relations
of \figref{fig:pentagon}, we find that there exist six different classes of possible
curves
\begin{align}
    & \gamma_{123} =
    \gen_4^{\alpha_1}
    \gen_5^{\alpha_2}
    \bar\gen_1
    \bar\gen_2
    \gen_4^{\alpha_3}
    \gen_5^{\alpha_4}
    \gen_3
    \gen_4^{\alpha_5}
    \gen_5^{\alpha_6}\ ,
    & \gamma_{213} =
    \gen_4^{\alpha_1}
    \gen_5^{\alpha_2}
    \bar\gen_2
    \bar\gen_1
    \gen_4^{\alpha_3}
    \gen_5^{\alpha_4}
    \gen_3
    \gen_4^{\alpha_5}
    \gen_5^{\alpha_6}\ ,
    \\
    & \gamma_{132} =
    \gen_4^{\alpha_1}
    \gen_5^{\alpha_2}
    \bar\gen_1
    \gen_4^{\alpha_3}
    \gen_3
    \gen_5^{\alpha_4}
    \bar\gen_2
    \gen_4^{\alpha_5}
    \gen_5^{\alpha_6}\ ,
    & \gamma_{231} =
    \gen_4^{\alpha_1}
    \gen_5^{\alpha_2}
    \bar\gen_2
    \gen_5^{\alpha_3}
    \gen_3
    \gen_4^{\alpha_4}
    \bar\gen_1
    \gen_4^{\alpha_5}
    \gen_5^{\alpha_6}\ ,
    \\
    & \gamma_{312} =
    \gen_4^{\alpha_1}
    \gen_5^{\alpha_2}
    \gen_3
    \gen_4^{\alpha_3}
    \gen_5^{\alpha_4}
    \bar\gen_1
    \bar\gen_2
    \gen_4^{\alpha_5}
    \gen_5^{\alpha_6}\ ,
    & \gamma_{321} =
    \gen_4^{\alpha_1}
    \gen_5^{\alpha_2}
    \gen_3
    \gen_4^{\alpha_3}
    \gen_5^{\alpha_4}
    \bar\gen_2
    \bar\gen_1
    \gen_4^{\alpha_5}
    \gen_5^{\alpha_6}\ .
\end{align}
By matching at LLA, for $\gamma_{123}$, $\gamma_{132}$ and $\gamma_{312}$ we obtain the
following constraints on the exponents of the generators $\gen_4$ and $\gen_5$:
\begin{align}
    \alpha_1
    = \alpha_2
    = \alpha_3
    = \alpha_4
    = \alpha_5
    = \alpha_6
    = 0\ .
\end{align}%
Similarly, one can also evaluate the constraints from the LLA for the paths $\gamma_{213}$,
$\gamma_{231}$ and $ \gamma_{321}$ to find that
\begin{align}
    -\alpha_1
    = \alpha_2
    = \alpha_5
    = - \alpha_6
    &= 1
    \\[2mm]
    \alpha_3 = \alpha_4 &= 0\ .
\end{align}
With these constraints implemented, one can now proceed and compute the NLLA through
analytic continuation of eq.~\eqref{eq:R72ope}. For all six curves, one obtains the
same result~\eqref{eq:Rmpmcoll} with the NLLA coefficients~\eqref{eq:ctildevalues}.
Our analysis has indeed confirmed the paths we anticipated in
eqs.~\eqref{eq:result1heptagon, eq:result2heptagon}.

%%%%%%%%%%%%%%%%%%%%%%%%%%%%%%%%%%%%%%%%%%%%%%%%%%%%%%%%%%%%
%%%%%%%%%%%%%%%%%%%%%%%%%%%%%%%%%%%%%%%%%%%%%%%%%%%%%%%%%%%%
\section{Conclusions and Outlook}
\label{sec:concl}

In this work, we have proposed a new tool to determine the finite remainder
function for all Mandelstam regions in multi-Regge kinematics. Our constraints
were obtained through analytic continuation of known expressions for the
remainder function on the main sheet in collinear kinematics. This input
into our analysis is provided by the Wilson loop OPE, and is in principle available for
any number of external gluons and any loop order. We illustrated the
general procedure in two examples, namely the hexagon and the heptagon at
two loops. While the hexagon case admits only a single non-trivial Mandelstam
region, there are four such regions for the heptagon. For one of these
regions, the multi-Regge limit of the two-loop finite remainder was only
determined quite recently in~\cite{DelDuca:2018raq}, though the leading
logarithmic terms were known for some time~\cite{Bartels:2011ge}.

Pushing this analysis to higher loops is not that difficult in principle. As
we stressed above, the relevant input from the Wilson loop OPE is available.
Here, we made some effort to express the two-loop collinear remainder function
in terms of ordinary polylogarithms. That helped with the analytic continuation,
but is not crucial. We could have been content with expressions involving
Goncharov's multiple polylogarithms. The relevant expressions are considerably more bulky,
but they can still be continued with computer algebra techniques~\cite{Chestnov:2019thesis}.
For the hexagon, there is nothing new to learn, since the multi-Regge
limit of the remainder function is known to all orders~\cite{Basso:2014pla}.
On the other hand, higher-loop results for the multi-Regge limit of the
remainder function with more than six external gluons are scarce. For the
heptagon remainder function, for example, only the LLA is known
to all orders  in the coupling and for all Mandelstam
regions~\cite{Bartels:2013jna,Bartels:2014jya}. It would therefore be very
interesting to extend our analysis of the heptagon to higher loops. Of course,
the ultimate hope would be to obtain all-loop expressions for all Mandelstam
regions by analytically continuing the Wilson loop OPE at finite
coupling, \ie without expanding in the 't~Hooft coupling, as done
in~\cite{Basso:2014pla} for the hexagon.
Right now, this seems difficult, especially
for the $\varrho = (\m\p\m)$ region. The perturbative analysis we carried out
above and its extension to higher orders is a rather pedestrian approach that
could help to obtain insights into the appropriate continuation paths, and
thereby valuable input for the more ambitious finite-coupling analysis.

The potential issues with finding the correct path of analytic continuation
were seen at strong coupling already. At infinite 't~Hooft coupling, the
most non-trivial contribution to the remainder function may be interpreted
as the free energy of an integrable one-dimensional quantum
system~\cite{Alday:2009dv}. For generic kinematics, this is still difficult to
compute, because the elementary particles of the one-dimensional system are
screened by clouds of excitations. As was shown
in~\cite{Bartels:2010ej,Bartels:2012gq}, however, the multi-Regge
limit in the gauge theory
amounts to sending all masses in the one-dimensional auxiliary system to
infinity, and hence it suppresses the difficult quantum fluctuations. As a
result, in computing the free energy, one only has to solve for a finite
set of Bethe roots rather than for a set of particle densities that are
determined as solutions of a coupled set of non-linear integral
equations~\cite{Bartels:2014mka}. Which roots contribute for a given Mandelstam
region, however, depends on the continuation path. For the hexagon, and for
three of the four Mandelstam regions in the heptagon, the relevant solutions
were found in~\cite{Bartels:2010ej,Bartels:2014ppa}. But in the case of
the region $\varrho = (\m\p\m)$, the suggested continuation path
was shown to be associated with the trivial solution of the Bethe Ansatz
equations, even though its winding numbers correctly satisfy the kinematic
constraints~\eqref{eq:winding}. In the light of our analysis in
\secref{sec:continuation7}, and in particular the insight we gained into
the choice of paths close to the collinear limit, it would be interesting
to revisit this issue at strong coupling.

Our perturbative analysis above was restricted to the leading terms in the
collinear limit. It would certainly be of interest to include higher-order
corrections in the variables $T_i$. This requires to consider contributions from  gluon
bound-state excitations of the GKP string, extending the resummations
of~\cite{Drummond:2015jea}. Continuing additional terms in the expansion
around the collinear limit could provide additional information on the
remainder function in the various Mandelstam regions.

There is actually another way in which our analysis may be extended to obtain
further information on the multi-Regge remainder function. The combined
multi-Regge collinear limit with the coordinates introduced
in \secref{sec:collinear} probes the neighborhood of a particular codimension-two
subregion of the heptagon multi-Regge limit. We can probe different regions by
cyclically shifting the kinematics $x_i\to x_{i-k}$ on the Wilson loop OPE side.
The collinear limit described by the Wilson loop OPE will not have an overlap
with the multi-Regge limit for all shifts $k$, but it does have a non-trivial
overlap for $k=1$ and for $k=4$ (for a non-trivial overlap, the ``small''
cross ratios $u_{j,k}$, $j=5,6$, $k=2,3$ must become small when
$T_1,T_2\to0$.). For the heptagon, the kinematics with a shift $k=1$ are related to $k=0$
by a combination of target-projectile symmetry ($p_i\to p_{3-i}$) and
reversing the OPE variables $\brc{F,S,T}_1\leftrightarrow\brc{F,S,T}_{2}$ and
hence does not yield independent information on the Regge
limit. The shift $k=4$ does contain
new information. In addition, it is a very symmetric choice,
see \figref{fig:shiftedkinematics}.
For the analysis in this paper, the constraints from $k=0$ were
sufficient to fix all unknowns, and we merely used the independent
constraints from $k=4$ to cross-check our results. But at higher loop orders it might be
useful to combine the constraints from $k=0$ and $k=4$ to determine
the remainder function.
\begin{figure}
\centering
\begin{tabular}{c@{\qquad}c@{\qquad}c}
\includegraphics[align=b]{FigShifted0}
&
\includegraphics[align=b]{FigShifted1}
&
\includegraphics[align=b]{FigShifted4}
\\[1ex]
$k=0$ & $k=1$ & $k=4$
\end{tabular}
\caption{Shifted heptagon kinematics: The collinear limit of the Wilson-loop OPE (tessellation
shown in red) has a non-trivial overlap with the
multi-Regge limit (kinematics shown in black) for three different
cyclic shifts $x_i\to x_{n-k}$, where $k=0,1,4$. The shifts $k=0,1$
are related by symmetries, but the shift $k=4$ is an independent
combined multi-Regge collinear limit.}
\label{fig:shiftedkinematics}
\end{figure}

Let us finally also mention the extension to higher numbers of external gluons. The
multi-Regge limit of the remainder function $R_8$ for $n=8$ external gluons probes a
new cut that can be associated with the eigenvalue of a non-compact Heisenberg spin
chain of length three~\cite{Bartels:8points}.
While in leading logarithmic order this eigenvalue is simply
the sum of eigenvalues for a spin chain of length two, higher orders give rise to
terms which represent a new three-body interaction between reggeized
gluons~\cite{Bartels:2012sw}. Again, the detailed composition of $R_8$ depends upon the
kinematic region. The next extensions of the spin chain are expected to be seen in
the $n=10$ point scattering process, $2 \to 8$, the $n=12$ point process, $2 \to 10$,
\etc. One of the challenges will be to find, beyond the LLA,  the eigenvalues of this
spin chain. The three-body interaction found in~\cite{Bartels:2012sw} raises some
doubts whether they are simply obtained from the sum of two-body interactions. This
issue certainly deserves further investigation.

%%%%%%%%%%%%%%%%%%%%%%%%%%%%%%%%%%%%%%%%%%%%%%%%%%%%%%%%%%%%
\subsection*{Acknowledgments}

We wish to thank
Joachim Bartels,
Benjamin Basso,
Simon Caron-Huot,
Lance Dixon,
Claude Duhr,
Sven-Olaf Moch,
Georgios Papathanasiou,
and Amit Sever
for useful discussions.
This work
was supported in part by the Deutsche Forschungsgemeinschaft
under Germany‘s Excellence Strategy -- EXC 2121 „Quantum Universe“
-- 390833306.

%%%%%%%%%%%%%%%%%%%%%%%%%%%%%%%%%%%%%%%%%%%%%%%%%%%%%%%%%%%%
%%%%%%%%%%%%%%%%%%%%%%%%%%%%%%%%%%%%%%%%%%%%%%%%%%%%%%%%%%%%
\appendix

% Show only section headings in toc from here on:
\makeatletter
\protect\addtocontents{toc}{\global \c@tocdepth 1\relax}
\makeatother

%%%%%%%%%%%%%%%%%%%%%%%%%%%%%%%%%%%%%%%%%%%%%%%%%%%%%%%%%%%%
%%%%%%%%%%%%%%%%%%%%%%%%%%%%%%%%%%%%%%%%%%%%%%%%%%%%%%%%%%%%
\section{Explicit BSV-Like Tessellation Variables}
\label{app:bsvtessdetail}

In \secref{sec:collinear}, we recalled the parametrization of general
null polygons in terms of conformal transformations that preserve
internal null tetragons of the tessellated polygon, as well as the
relation between the associated tetragon variables
\begin{equation}
T_j = e^{-\tau_j}
\,,\quad
S_j = e^{\sigma_j}
\,, \quad
F_j = e^{i\varphi_j}
\label{eq:FSTapp}
\end{equation}
and the multi-Regge limit. In the following, we will define these
variables explicitly. Both the external and the internal null
lines of the tessellation are conveniently parametrized by
four-component momentum twistors $Z_j$, $j=1,\dots,n$, such
that $x_i\simeq(Z_{n,i},Z_{n,i+1})$~\cite{Hodges:2009hk}. The momentum twistors $Z_j$ are
obtained by acting with the conformal transformations parametrized by
$\brc{F_i,S_i,T_i}$ on a fixed reference polygon given by reference
momentum twistors $\mathbf{Z}_j$.
Following and generalizing Appendix~A of~\cite{Basso:2013aha}, we
choose for our reference polygons for $n=6,\dots,9$:
\begin{align}
                                   \mathbf{Z}_{7,1}                  =\mathbf{Z}_{8,1}                  =\mathbf{Z}_{9,5}                  &=( 0,2,-1, 1)\,,\nn\\
\mathbf{Z}_{6,1}                  =\mathbf{Z}\suprm{int}_{7,\text{c}}=\mathbf{Z}\suprm{int}_{8,\text{c}}=\mathbf{Z}\suprm{int}_{9,\text{c}}&=( 0,1,-1, 1)\,,\nn\\
\mathbf{Z}_{6,2}                  =\mathbf{Z}_{7,2}                  =\mathbf{Z}_{8,2}                  =\mathbf{Z}_{9,6}                  &=( 0,1, 0, 0)\,,\nn\\
\mathbf{Z}_{6,3}                  =\mathbf{Z}_{7,3}                  =\mathbf{Z}_{8,3}                  =\mathbf{Z}_{9,7}                  &=( 0,1, 1, 0)\,,\nn\\
\mathbf{Z}_{6,4}                  =\mathbf{Z}_{7,4}                  =\mathbf{Z}_{8,4}                  =\mathbf{Z}\suprm{int}_{9,\text{e}}&=( 1,0, 1, 1)\,,\nn\\
\mathbf{Z}_{6,5}                  =\mathbf{Z}_{7,5}                  =\mathbf{Z}_{8,5}                  =\mathbf{Z}_{9,1}                  &=( 1,0, 0, 0)\,,\nn\\
\mathbf{Z}_{6,6}                  =\mathbf{Z}_{7,6}                  =\mathbf{Z}_{8,6}                  =\mathbf{Z}_{9,2}                  &=(-1,0, 0, 1)\,,\nn\\
\mathbf{Z}\suprm{int}_{6,\text{a}}=\mathbf{Z}\suprm{int}_{7,\text{a}}=\mathbf{Z}\suprm{int}_{8,\text{a}}=\mathbf{Z}\suprm{int}_{9,\text{a}}&=( 0,0, 1, 0)\,,\nn\\
\mathbf{Z}\suprm{int}_{6,\text{b}}=\mathbf{Z}\suprm{int}_{7,\text{b}}=\mathbf{Z}\suprm{int}_{8,\text{b}}=\mathbf{Z}\suprm{int}_{9,\text{b}}&=( 0,0, 0, 1)\,,\nn\\
                                   \mathbf{Z}_{7,7}                  =\mathbf{Z}\suprm{int}_{8,\text{d}}=\mathbf{Z}\suprm{int}_{9,\text{d}}&=(-1,1,-1, 3)\,,\nn\\
                                                                      \mathbf{Z}_{8,7}                  =\mathbf{Z}_{9,3}                  &=( 0,1,-1, 2)\,,\nn\\
                                                                      \mathbf{Z}_{8,8}                  =\mathbf{Z}_{9,4}                  &=( 1,0, 1,-3)\,,\nn\\
                                                                                                         \mathbf{Z}_{9,8}                  &=( 1,1, 3, 1)\,,\nn\\
                                                                                                         \mathbf{Z}_{9,9}                  &=( 2,0, 1, 1)\,.
\label{eq:ZBSVref}
\end{align}
Here, $\mathbf{Z}_{n,i}$,
$i=1,\dots,n$, parametrizes the reference $n$-gon, and the momentum twistors
$\mathbf{Z}\suprm{int}_{n,x}$ are associated to the internal lines
$x=\text{a},\dots,\text{e}$, see \figref{fig:BSVtessellations}. The full $n$-gon parametrization is obtained by
acting with the stabilizing matrices of the internal tetragons as
follows:
\begin{align}
Z_{6,3}&=\mathbf{Z}_{6,3}M_1      \,, & Z_{8,5}&=\mathbf{Z}_{8,5}M_2               \,, & Z_{9,8}&=\mathbf{Z}_{9,8}M_4M_1            \,,\nn\\
Z_{6,4}&=\mathbf{Z}_{6,4}M_1      \,, & Z_{8,4}&=\mathbf{Z}_{8,4}M_1M'_2           \,, & Z_{9,9}&=\mathbf{Z}_{9,9}M_4M_1            \,,\nn\\
&                                     & Z_{8,3}&=\mathbf{Z}_{8,3}M_1M'_2           \,, & Z_{9,7}&=\mathbf{Z}_{9,7}M_1               \,,\nn\\
Z_{7,3}&=\mathbf{Z}_{7,3}M_1      \,, & Z_{8,\text{a}}&=\mathbf{Z}_{8,\text{a}}M_2 \,, & Z_{9,\text{e}}&=\mathbf{Z}_{9,\text{e}}M_1 \,,\nn\\
Z_{7,4}&=\mathbf{Z}_{7,4}M_1      \,, & Z_{8,7}&=\mathbf{Z}_{8,7}M_3^{-1}          \,, & Z_{9,3}&=\mathbf{Z}_{9,3}M_3^{-1}M_2^{-1}  \,,\nn\\
Z_{7,7}&=\mathbf{Z}_{7,7}M_2^{-1} \,, & Z_{8,8}&=\mathbf{Z}_{8,8}M_3^{-1}          \,, & Z_{9,4}&=\mathbf{Z}_{9,4}M_3^{-1}M_2^{-1}  \,,\nn\\
Z_{7,1}&=\mathbf{Z}_{7,1}M_2^{-1} \,, &&                                               & Z_{9,5}&=\mathbf{Z}_{9,5}M_2^{-1}          \,,\nn\\
&                                     &&                                               & Z_{9,\text{d}}&=\mathbf{Z}_{9,\text{d}}M_2^{-1} \,,
\label{eq:ZBSV}
\end{align}
with all other $Z_{n,i}=\mathbf{Z}_{n,i}$.
In terms of the variables~\eqref{eq:FSTapp}, the
stabilizing matrices $M_j$ are defined by the relation
\begin{equation}
\begin{pmatrix*}[l]
\mathbf{Z}\subrm{right}\\\mathbf{Z}\subrm{left}\\\mathbf{Z}\subrm{bottom}\\\mathbf{Z}\subrm{top}
\end{pmatrix*}
M_j=\sqrt{F_j}
\begin{pmatrix}
1/(F_jS_j) &0 &0 &0 \\
0 &S_j/F_j &0 &0 \\
0 &0 &T_j &0 \\
0 &0 &0 &1/T_j
\end{pmatrix}
\begin{pmatrix*}[l]
\mathbf{Z}\subrm{right}\\\mathbf{Z}\subrm{left}\\\mathbf{Z}\subrm{bottom}\\\mathbf{Z}\subrm{top}
\end{pmatrix*}
\,.
\label{eq:MjeqBSV}
\end{equation}
Here, the momentum twistors $\mathbf{Z}$ parametrize the respective
internal square, see \figref{fig:BSVtessellations}. Note that compared to Appendix~A
of~\cite{Basso:2013aha}, top and bottom as well as left and right are
reversed. Also, due to the alternating nature of the tessellation, the
notions ``left'' and ``right'' interchange from one internal tetragon
to the next. Explicitly, the stabilizing matrices are given by%
\footnote{The stabilizing matrix $M_4$ has a relatively simple form
compared to $M_3$, because we chose to extend the octagon at the top
rather than at the bottom. This choice simplifies the expressions for
the cross ratios, in particular in general kinematics, away from the
collinear limit.}
\begin{align}
M_1&=\sqrt{F_1}\begin{pmatrix}
\frac{S_1}{F_1} & 0                & 0             & 0 \\
0               & \frac{1}{F_1S_1} & 0             & 0 \\
0               & 0                & \frac{1}{T_1} & 0 \\
0               & 0                & 0             & T_1
\end{pmatrix}
\,,\qquad
M_2=\sqrt{F_2}\begin{pmatrix}
\frac{1}{F_2S_2} & 0                   & 0   & -\frac{1}{F_2S_2}+\frac{1}{T_2} \\
0                & \frac{S_2}{F_2}     & 0   & 0 \\
0                & \frac{S_2}{F_2}-T_2 & T_2 & \frac{1}{T_2}-T_2 \\
0                & 0                   & 0   & \frac{1}{T_2}
\end{pmatrix}
\,, \nn\\[1ex]
M_3&=\sqrt{F_3}\begin{pmatrix}
\frac{2 S_3}{F_3}-T_3 & T_3-\frac{1}{T_3}                   & \frac{1}{T_3}-T_3                    & -\frac{2 S_3}{F_3}+3 T_3-\frac{1}{T_3} \\[.8ex]
0                     & \frac{2}{F_3 S_3}-\frac{1}{T_3}     & \frac{1}{T_3}-\frac{1}{F_3 S_3}      & \frac{1}{F_3 S_3}-\frac{1}{T_3} \\[.8ex]
\frac{S_3}{F_3}-T_3   & T_3+\frac{2}{F_3 S_3}-\frac{3}{T_3} & -T_3-\frac{1}{F_3 S_3}+\frac{3}{T_3} & \frac{1-S_3^2}{F_3S_3}+3 T_3-\frac{3}{T_3} \\[.8ex]
\frac{S_3}{F_3}-T_3   & T_3-\frac{1}{T_3}                   & \frac{1}{T_3}-T_3                    & -\frac{S_3}{F_3}+3 T_3-\frac{1}{T_3}
\end{pmatrix}
\,, \nn\\[1ex]
M_4&=\sqrt{F_4}\begin{pmatrix}
\frac{1}{F_4 S_4}               & 0               & 0                   & 0 \\
0                               & \frac{S_4}{F_4} & \frac{S_4}{F_4}-T_4 & 0 \\
0                               & 0               & T_4                 & 0 \\
\frac{1}{T_4}-\frac{1}{F_4 S_4} & 0               & \frac{1}{T_4}-T_4   & \frac{1}{T_4}
\end{pmatrix}
\,.
\label{eq:M1234}
\end{align}
Using the four-bracket combinations $\vev{i, j, k, l} \defas \det\brk{Z_i Z_j Z_k Z_l}$ of these variables,
one can express the cross ratios shown in eq.~\eqref{eq:Uij} as:
\begin{align}
    U_{ij} = \frac{
        \vev{i, i + 1, j + 1, j + 2}
        \vev{i + 1, i + 2, j, j + 1}
    }{
        \vev{i, i + 1, j, j + 1}
        \vev{i + 1, i + 2, j + 1, j + 2}
    }\,.
\label{eq:UofZdet}
\end{align}
Plugging the momentum twistors $Z_i$ into this formula, one recovers
the hexagon and heptagon cross ratios stated in eqs.~\eqref{eq:ujpar}
and~\eqref{eq:cr7opevar}. For convenience, we reproduce their
collinear limits~\eqref{eq:uihex} and~\eqref{eq:crs7coll} here: For
$T_1\to0$, the hexagon cross ratios become:
\begin{equation}
u_{1,1} = U_{25} \to \frac{1}{1+S^2}
\,, \quad
u_{1,2} = U_{36} \to \frac{S^2}{1+S^2}
\,, \quad
u_{1,3} = U_{14} \to T^2
\label{eq:hexagoncrsapp}
\end{equation}
The heptagon cross ratios in the collinear limit $T_i\to0$ read:
\begin{align}
u_{1,1}=U_{25}&\to \frac{1}{1+S_1^2} \,, &
u_{1,2}=U_{37}&\to \frac{S_1^2(1+S_2^2)}{S_1^2+S_2^2+S_1^2S_2^2} \,, &
u_{1,3}=U_{14}&\to T_1^2 \,, \nn \\
u_{2,1}=U_{36}&\to \frac{S_1^2+S_2^2+S_1^2S_2^2}{(1+S_1^2)(1+S_2^2)} \,, &
u_{2,2}=U_{47}&\to \frac{1}{1+S_2^2} \,, &
u_{2,3}=U_{15}&\to T_2^2 \,, \nn \\
        U_{26}&\to \frac{(1+S_1^2)S_2^2}{S_1^2+S_2^2+S_1^2S_2^2} \,. &&&&
\label{eq:heptagoncrsapp}
\end{align}
For the octagon and nonagon cross ratios, we will only state their
collinear limits, as their full expressions can easily be reproduced
from the above formul\ae. We will use the following shorthands:
\begin{align}
S_{12  }^2 &= S_1^2+S_2^2+S_1^2S_2^2\,, \nn \\
S_{23  }^2 &= S_2^2+S_3^2+S_2^2S_3^2\,, \nn \\
S_{14  }^2 &= S_1^2+S_4^2+S_1^2S_4^2\,, \nn \\
S_{123 }^2 &= S_2^2+S_1^2S_2^2+S_1^2S_3^2+S_2^2S_3^2+S_1^2S_2^2S_3^2\,, \nn \\
S_{124 }^2 &= S_1^2+S_1^2S_2^2+S_1^2S_4^2+S_2^2S_4^2+S_1^2S_2^2S_4^2\,, \nn \\
S_{1234}^2 &= S_1^2S_2^2+S_1^2S_3^2+S_2^2S_4^2+S_1^2S_2^2S_3^2+S_1^2S_2^2S_4^2+S_1^2S_3^2S_4^2+S_2^2S_3^2S_4^2+S_1^2S_2^2S_3^2S_4^2
\,.
\end{align}
With these abbreviations, the octagon cross ratios in the collinear
limit $T_i\to0$ become:
\begin{align}
% u_{1,1}=
% u_{1,2}=
% u_{1,3}=
% u_{2,1}=
% u_{2,2}=
% u_{2,3}=
% u_{7,1}=
% u_{7,2}=
% u_{7,3}=
U_{25}&\to \frac{1}{1+S_1^2} \,, &
U_{38}&\to \frac{S_1^2(1+S_2^2)}{S_{12}^2} \,, &
U_{14}&\to T_1^2 \,, \nn \\
U_{36}&\to \frac{S_{123}^2}{(1+S_1^2)S_{23}^2} \,, &
U_{48}&\to \frac{1}{1+S_2^2} \,, &
U_{15}&\to T_2^2 \,, \nn \\
U_{47}&\to \frac{(1+S_2^2)S_3^2}{S_{23}^2} \,, &
U_{58}&\to T_3^2 \,, &
U_{16}&\to \frac{1}{1+S_3^2} \,, \nn \\
U_{26}&\to \frac{(1+S_1^2)S_2^2(1+S_3^2)}{S_{123}^2} \,, &
U_{37}&\to \frac{S_{12}^2S_{23}^2}{(1+S_2^2)S_{123}^2} \,, &
U_{27}&\to \frac{S_{123}^2}{S_{12}^2(1+S_3^2)} \,.
\label{eq:octagoncrs}
\end{align}
The nonagon cross ratios in the collinear limit read:
\begin{align}
% u_{1,1}=
% u_{1,2}=
% u_{1,3}=
% u_{2,1}=
% u_{2,2}=
% u_{2,3}=
% u_{7,1}=
% u_{7,2}=
% u_{7,3}=
% u_{8,1}=
% u_{8,2}=
% u_{8,3}=
U_{25}&\to \frac{1}{1+S_3^2} \,, &
U_{39}&\to \frac{(1+S_2^2)S_3^2}{S_{23}^2} \,, &
U_{14}&\to T_3^2 \,, \nn \\
U_{36}&\to \frac{S_{123}^2}{S_{12}^2(1+S_3^2)} \,, &
U_{49}&\to \frac{1}{1+S_2^2} \,, &
U_{15}&\to T_2^2 \,, \nn \\
U_{47}&\to \frac{S_{124}^2}{S_{12}^2(1+S_4^2)} \,, &
U_{59}&\to T_1^2 \,, &
U_{16}&\to \frac{1}{1+S_1^2} \,, \nn \\
U_{58}&\to \frac{1}{1+S_4^2} \,, &
U_{69}&\to T_4^2 \,, &
U_{17}&\to \frac{(1+S_1^2)S_4^2}{S_{14}^2} \,, \nn \\
U_{26}&\to \frac{(1+S_1^2)S_2^2(1+S_3^2)}{S_{123}^2} \,, &
U_{37}&\to \frac{S_{12}^2S_{1234}^2}{S_{123}^2S_{124}^2} \,, &
U_{48}&\to \frac{S_1^2(1+S_2^2)(1+S_4^2)}{S_{124}^2} \,, \nn \\
U_{27}&\to \frac{S_{123}^2S_{14}^2}{(1+S_1^2)S_{1234}^2} \,, &
U_{38}&\to \frac{S_{23}^2S_{124}^2}{(1+S_2^2)S_{1234}^2} \,, &
U_{28}&\to \frac{S_{1234}^2}{S_{23}^2S_{14}^2} \,.
\label{eq:nonagoncrs}
\end{align}
All expressions in~\eqref{eq:hexagoncrsapp}-\eqref{eq:nonagoncrs} hold
up to linear terms in the $T_i$ variables, except for the cross ratios
of the form $T_j^2$, which further expand as
$T_j^2+\order{T_j^3}+\sum_{i\neq j}\order{T_i}$.

Starting with the cross ratio expressions in general kinematics, we can express
all $S_i$ in terms of $r_iT_i$, and further all $r_i$ and $F_i$ in
terms of $w_i$ and $\wb_i$ via~\eqref{eq:Frw678} and~\eqref{eq:Frw9}.
Subsequently taking the limit $T_i\to0$ (\ie a double-scaling limit in
$T_i$ and $S_i$), we recover the multi-Regge-limit
parametrizations~\eqref{eq:ujaofwwb} of the reduced cross ratios
$u_{j,2}/(1-u_{j,1})$ and $u_{j,3}/(1-u_{j,1})$.

%%%%%%%%%%%%%%%%%%%%%%%%%%%%%%%%%%%%%%%%%%%%%%%%%%%%%%%%%%%%
%%%%%%%%%%%%%%%%%%%%%%%%%%%%%%%%%%%%%%%%%%%%%%%%%%%%%%%%%%%%
\section{A Multi-Regge-Friendly Tessellation for any \texorpdfstring{$n$}{n}}
\label{app:mrl-tessellation}

The Wilson loop OPE of~\cite{Basso:2013vsa,Basso:2013aha} relies on a
tessellation of the null polygon into internal null squares. As shown
in \secref{sec:collinear} and \appref{app:bsvtessdetail}, the
multi-collinear limit, where the Wilson loop OPE applies, has a
non-trivial overlap with the multi-Regge limit for up to nine points.
Beyond nine points, the collinear $T_i\to0$ limit of the
``alternating'' tessellation employed by the OPE has no overlap with
the multi-Regge limit.

In the following, we will describe a slightly modified tessellation
whose multi-collinear $T_i\to0$ limit does feature
an overlap with the $2\rightarrow n-2$ multi-Regge limit for any number $n$ of edges,
with canonical relations between all relevant parameters and cross ratios.
Even though an OPE formula for this tessellation is not known to date,
the tessellation variables do provide a multi-Regge friendly
parametrization of the general $n$-gon and thus might be of use in the
future.
It would be very interesting to generalize the Wilson loop OPE to this
type of tessellation.

\begin{figure}
\centering
\includegraphics[align=c]{FigInnerHexagon.mps}
\qquad \quad
\includegraphics[align=c]{FigInnerHexagonZ.mps}
\quad
\parbox{5cm}{%
\begin{align*}
Z_j\suprm{int}
=\mbox{}&\vev{1,2,3,j}Z_{j+1}\\
&-\vev{1,2,3,j+1}Z_j
\end{align*}}
\caption{A multi-Regge friendly tessellation.
\emph{Left:} Inner tetragon (red) stabilized by $\tau_j$, $\sigma_j$, and
$\varphi_j$, together with the surrounding hexagon (green) that defines
the cross ratios~\protect\eqref{eq:ujprime}. The primed coordinates
all lie on the line $x_{12}$. For an $n$-sided polygon, there are in
total $n-5$ internal tetragons, indexed by $j=1,\dots,n-5$. \emph{Middle:} The corresponding
momentum-twistor configuration.
\emph{Right:} Formula for the
``internal'' momentum twistors $Z\suprm{int}_j$ that define the primed
coordinates $x'_j$.}
\label{fig:innerhexagon}
\end{figure}
%

%%%%%%%%%%%%%%%%%%%%%%%%%%%%%%
\paragraph{The Tessellation.}

As in the tessellation employed in~\cite{Basso:2013vsa,Basso:2013aha}
and in the main text of this paper, our modified tessellation
decomposes the $n$-gon into $(n-5)$
internal tetragons and two boundary tetragons. The tessellation is
defined by the (unique) null lines
from cusps $x_4,\dots,x_{n-1}$ to line $x_{12}$, intersecting the line
$x_{12}$ at points $x_4',\dots,x_{n-1}'$, see \figref{fig:configuration}
and \figref{fig:innerhexagon} (left). As before, each internal null
tetragon $(x_{j+3},x_{j+4},x_{j+4}',x_{j+3}')$ is preserved by three
conformal transformations that we parametrize by variables
$\tau_j$, $\sigma_j$, $\varphi_j$, $j=1,\dots,n-5$,
following~\cite{Alday:2010ku,Sever:2011pc,Basso:2013vsa,Basso:2013aha}.
Acting with these conformal transformations on all cusps $x_i$, $i\in\brc{j+5,\dots,n,1}$
that lie below that internal
tetragon generates all conformally inequivalent configurations.%
\footnote{Unlike in \appref{app:bsvtessdetail}, we choose to let
the three conformal transformations of each tetragon act on the bottom
part of the polygon. Acting instead with the inverse
conformal transformations on the top part of the polygon yields
a conformally equivalent configuration. Consistently, in the
definition of the conformal transformations via~\eqref{eq:Mjeq} below,
the orientation of the internal tetragon (``top'' and ``bottom''
momentum twistors) is flipped relative to~\eqref{eq:MjeqBSV}.}
The tetragon variables $\tau_j$, $\sigma_j$, $\varphi_j$ can be expressed in terms of the
cross ratios $u_{j,1}'$, $u_{j,2}'$, and $u_{j,3}'$ of the surrounding
internal hexagon (see \figref{fig:innerhexagon}),
\begin{equation}
\begin{aligned}
u_{j,1}'&=U_{(j+2)',j+4,j+2,j+5}
\,,\\
u_{j,2}'&=U_{j+2,j+5,j+3,(j+5)'}
\,,\\
u_{j,3}'&=U_{(j+5)',j+3,(j+2)',j+4}
\,,
\end{aligned}
\qquad
U_{i,j,k,l}=\frac{x_{il}^2x_{jk}^2}{x_{ij}^2x_{kl}^2}
\,,
\qquad
x_{ij}\equiv \abs{x_i-x_j}
\,,
\label{eq:ujprime}
\end{equation}
via
\begin{equation} \label{eq:ujprimepar}
u_{j,1}'=\frac{1-e^{-2\tau_j}}{1+e^{2\sigma_j}+2\mathinner{e^{\sigma_j-\tau_j}}\cos\varphi_j}
\,,\qquad
\frac{u_{j,2}'}{u_{j,1}'}=e^{2\sigma_j}
\,,\qquad
u_{j,3}'=e^{-2\tau_j}
\, .
\end{equation}
As before, we shall also use the exponentiated variables~\eqref{eq:TSF}
\begin{equation}
T_j = e^{-\tau_j}
\,,\quad
S_j = e^{\sigma_j}
\,,
\quad
F_j = e^{i\varphi_j}
\,.
\end{equation}
The limit in which all $\tau_j$ are large, that is $T_j\ll1$, again is a multi-collinear
limit. The multi-Regge
regime, on the other hand, corresponds to a double scaling limit where $T_j \ll 1$
and $S_j \ll 1$ while the ratios $S_j/T_j \equiv r_j$ are kept finite.

\begin{figure}
\centering
\includegraphics[align=c]{FigPolyCR}
\quad
{\Large $\xrightarrow{\;T_j\,\to\,0\;}$}
\quad
\includegraphics[align=c]{FigPolyCRLimit}
\caption{One inner null tetragon (gray dashed lines) with the
associated ``large'' cross ratio $u_{j,1}$ (orange, solid) and ``small'' cross ratios
$u_{j,2}$ (blue, dashed) and $u_{j,3}$ (green, dotted). Upon sending $T_j\to0$, the lower part
of the polygon flattens to the bottom of the tetragon, implying
$u_{j,3}\sim T_j^2\to0$. Further sending $S_j\to\infty$ shifts all points
on the lower null line to the right. Conversely, sending
$S_j\to0$ shifts all points to the left. One can easily see that
$u_{j,1}\to1$, $u_{j,2}\to0$ when $S_j\to0$, as required by the
multi-Regge limit.}
\label{fig:polycr}
\end{figure}

% The arguments of large  logarithms in the multi-Regge limit are given
% by~\eqref{eq:epsjdef}
% %
% \begin{equation}
% \eps_j=u_{j,2}u_{j,3}\,,
% \qquad
% j=5,\dots,n-1\,.
% \end{equation}
% %
Our choice of tessellation is such that each triple of cross ratios
$\brc{u_{j,1},u_{j,2},u_{j,3}}$ is associated to one of the inner null
tetragons, namely the tetragon that is invariant under $\tau_j$,
$\sigma_j$, and $\varphi_j$. This is illustrated in \figref{fig:polycr}:
in the limit $T_j\to0$, the lower part of the polygon is flattened to
the bottom of the tetragon, which sends $u_{j,3}\to0$. In this limit, the
variable $S_j$ controls the positions of all cusps on the upper line
of the tetragon. Upon sending $S_j\to0$, all cusps on that line
approach the point $x_{j+4}$ on the left, which implies that $u_{j,1}\to1$
and $u_{j,2}\to0$, as required by the multi-Regge limit.%
\footnote{Conversely, all points on the lower line approach $x_{1}$
on the right when sending $S_j\to\infty$.}
More precisely, for fixed $j$ we have
\begin{align}
u_{j,1} &\to
1+\order{T_j^2}
\,, &
u_{j,2}&\to
\order{T_j^2}
\,, &
u_{j,3} & \to
\order{T_j^2}
&
\text{for }\  T_j&\to0 \text{ with } S_j/T_j \text{ fixed.}
\end{align}
Even though not apparent in \figref{fig:polycr}, all other cross
ratios $u_{i,k}$, $i\neq j$, remain finite in this limit. The exact
expression of $u_{j,k}$ in terms of $r_j=S_j/T_j$ and $F_j$ depend on
the normalizations of $\tau_j$, $\sigma_j$, and $\varphi_j$, as well as
on the choice of ``reference polygon'' obtained when
$\tau_j,\sigma_j,\varphi_j=0$. We found it possible to choose
normalizations and reference polygons such that with
\begin{equation}
r_j^2=\frac{S_j^2}{T_j^2}=\frac{1}{{w_j\wb_j}}
\,,\qquad
F_j^2={\frac{w_j}{\wb_j}}
\,,\qquad
S_j^2T_j^2=r_j^2T_j^4=\varepsilon_j
\,,
\label{eq:FSTtoweps}
\end{equation}
the cross ratios satisfy the wanted Regge limit
relations~\eqref{eq:uofw} and~\eqref{eq:epsjdef}, namely
\begin{gather}
\frac{u_{j,2}}{1-u_{j,1}}
=\frac{1}{\abs{1+w_j}^2}
+\sum_i\order{T_i^2}
\,,\qquad
\frac{u_{j,3}}{1-u_{j,1}}
=\frac{\abs{w_j}^2}{\abs{1+w_j}^2}
+\sum_i\order{T_i^2}
\,,\nn\\[2mm]
u_{j,2}\,u_{j,3}
=\varepsilon_j
\biggbrk{1+\sum_{i\neq j}\order{T_i^2}}
+\order{T_j^5}
\,,\qquad
j=1,\dotsc,n-5
\label{eq:crsmrl}
\end{gather}
when $T_j\to0$ with $r_j=S_j/T_j$ fixed for all $j$. On the other
hand, in the multi-collinear limit $T_j\to0$ with $S_j$ finite for
all $j$, the relevant cross ratios become
\begin{gather}
u_{j,1}=\frac{1+S_{j-1}^2(1+S_j^2)}{(1+S_{j-1}^2)(1+S_j^2)}
+\sum_i\order{T_i}
\,,\nn\\[2mm]
u_{j,2}=\frac{A_j}{1+A_j}
+\sum_i\order{T_i}
\,,\qquad
u_{j,3}=T_j^2
\biggbrk{1+\sum_{i\neq j}\order{T_i^2}}
\,,
\label{eq:crscoll}
\end{gather}
where
\begin{equation}
S_j^2=\frac{A_j}{1+A_{j+1}}
\,,\qquad
j=1,\dots,n-5
\,.
\label{eq:Ajdef}
\end{equation}
In these formulas, the index $j$ runs through $j=1, \dots,n-5$ and we need to
supply the boundary conditions
\begin{equation}
A_{n-4}=S_0=0
\end{equation}
to make our formulas well defined. Subleading terms in the
expansions~\eqref{eq:crsmrl} and~\eqref{eq:crscoll} are provided
in~\eqref{eq:crscollmore} and~\eqref{eq:crsmrlmore} below.

The parametrization of the multi-Regge cross ratios~\eqref{eq:uj} in
general kinematics in terms of the tetragon
variables $T_j,S_j,F_j$ is well
adapted to the combined multi-Regge collinear limit
for any number $n$ of external gluons.  We do not state general formulas for
the $u_{j,a}$ in terms of the tetragon variables here, but will do so for $n=6,7$
below.%
\footnote{In fact we can write an exact expression at least for
$u_{j,3}$, see~\eqref{eq:uj3exact} below.}
 All this requires is to work out relations between the $u'_{j,a}$ defined
in eq.~\eqref{eq:ujprime} and the cross ratios~\eqref{eq:uj} we use in the
multi-Regge limit.

%%%%%%%%%%%%%%%%%%%%%%%%%%%%%%
\paragraph{Momentum Twistors.}

We will now state our parametrization explicitly.
As before, we employ four-component momentum twistors $Z_j$ (see
\figref{fig:innerhexagon}), which yield all conformally invariant
cross ratios via~\eqref{eq:UofZdet}.
For the $j$'th internal null tetragon, the relevant momentum twistors are:
\begin{equation}
\includegraphics[align=c]{FigSquare}
\end{equation}
Conformal transformations act linearly on momentum twistors. Again, the three
transformations $\tau_j$, $\sigma_j$, and $\varphi_j$ are defined in
terms of the matrix $M_j=M_j(\tau_j,\sigma_j,\varphi_j)$ via (see~\eqref{eq:MjeqBSV})
\begin{equation}
\begin{pmatrix}
Z_{j+4}\\Z_2\\Z_{j+3}\suprm{int}\\Z_{j+4}\suprm{int}
\end{pmatrix}
M_j=\sqrt{F_j}
\begin{pmatrix}
1/(F_jS_j) &0 &0 &0 \\
0 &S_j/F_j &0 &0 \\
0 &0 &T_j &0 \\
0 &0 &0 &1/T_j
\end{pmatrix}
\begin{pmatrix}
Z_{j+4}\\Z_2\\Z_{j+3}\suprm{int}\\Z_{j+4}\suprm{int}
\end{pmatrix}
\,,
\label{eq:Mjeq}
\end{equation}
where the column of $Z$'s is understood as a $4\times4$ matrix.
Let us illustrate how all conformally inequivalent null polygons are
constructed from the parameters $\brc{\tau_j,\sigma_j,\varphi_j}$ by
considering the hexagon. In this case, there is only one internal
tetragon:
\begin{equation}
\includegraphics[align=c]{FigHexagon}
\quad
\longleftrightarrow
\quad
\includegraphics[align=c]{FigHexagonZ}
\end{equation}
Because conformally inequivalent null hexagons have three degrees of
freedom, and because all null tetragons are conformally equivalent,
all conformally inequivalent hexagons are generated by starting with
an arbitrary fixed reference hexagon, and acting with the conformal
transformations $M_1$ that preserve the inner tetragon on the upper
part of the hexagon, \ie on the momentum twistors $Z_3$ and $Z_4$.
Concretely, we use:
\begin{equation}
\begin{aligned}
           Z_2 &=(1,0,0,0)  \,, \\
           Z_3 &=(1,0,1,1) \,, \\
           Z_4 &=(0,1,0,-1) \,, \\
           Z_5 &=(0,1,0,0)  \,, \\
           Z_6 &=(0,1,-1,0){\colred M_1}  \,, \\
           Z_1 &=(-1,0,1,1){\colred M_1}  \,, \\
Z_4\suprm{int} &=(0,0,0,1) \,, \\
Z_5\suprm{int} &=(0,0,1,0) \,,
\end{aligned}
\qquad
\begin{aligned}
{\colred M_1}&=\sqrt{F_1}
\begin{pmatrix}
S_1/F_1 & 0          & 0     & 0 \\
0       & 1/(F_1S_1) & 0     & 0 \\
0       & 0          & 1/T_1 & 0 \\
0       & 0          & 0     & T_1 \\
\end{pmatrix}
\,,
\end{aligned}
\label{eq:Zhex}
\end{equation}
where $M_1$ is the solution to eq.~\eqref{eq:Mjeq}.
We construct the heptagon by extending the hexagon at the bottom, such
that $Z_1$ of the hexagon becomes $Z_6\suprm{int}$ of the heptagon:
\begin{equation}
\includegraphics[align=c]{FigHeptagon}
\quad
\longleftrightarrow
\quad
\includegraphics[align=c]{FigHeptagonZ}
\end{equation}
Concretely, we choose:
\begin{equation}
\begin{aligned}
           Z_2 &=(1,0,0,0)  \,, \\
           Z_3 &=(1,0,1,1){\colred M_1^{-1}} \,, \\
           Z_4 &=(0,1,0,-1){\colred M_1^{-1}} \,, \\
           Z_5 &=(0,1,0,0)  \,, \\
           Z_6 &=(0,1,-1,0)  \,, \\
           Z_7 &=(1,1,-2,-1){\colgreen M_2}  \,, \\
           Z_1 &=(2,0,0,-1){\colgreen M_2}  \,, \\
Z_4\suprm{int} &=(0,0,0,1) \,, \\
Z_5\suprm{int} &=(0,0,1,0) \,, \\
Z_6\suprm{int} &=(-1,0,1,1) \,,
\end{aligned}
\qquad
\begin{aligned}
{\colred M_1}&=\sqrt{F_1}
\begin{pmatrix}
\frac{S_1}{F_1} & 0                & 0             & 0 \\
0               & \frac{1}{F_1S_1} & 0             & 0 \\
0               & 0                & \frac{1}{T_1} & 0 \\
0               & 0                & 0             & T_1 \\
\end{pmatrix}
\,, \\
{\colgreen M_2}&=\sqrt{F_2}
\begin{pmatrix}
\frac{S_2}{F_2}               & 0                & 0                    & 0 \\
0                             & \frac{1}{F_2S_2} & T_2-\frac{1}{F_2S_2} & 0 \\
0                             & 0                & T_2                  & 0 \\
\frac{S_2}{F_2}-\frac{1}{T_2} & 0                & \frac{1-T_2^2}{T_2}  & \frac{1}{T_2} \\
\end{pmatrix}
\,,
\end{aligned}
\label{eq:Zhept}
\end{equation}
where again $M_1$ and $M_2$ are the solutions to eq.~\eqref{eq:Mjeq}, and
we have chosen to act with the transformation $M_1^{-1}$ on $Z_3$ and $Z_4$
instead of acting with $M_1$ on $Z_6$, $Z_7$, $Z_1$, and
$Z_6\suprm{int}$, the difference being just an overall conformal
transformation.

Note that the choice of numerical reference polygon
in eqs.~\eqref{eq:Zhex} and~\eqref{eq:Zhept} is
arbitrary. Our choice is constructed in such a way that the ``large''
and ``small'' cross ratios~\eqref{eq:uj}
\begin{equation}
u_{j,i}
\,,\qquad
j=1,\dotsc,n-5
\,,\qquad
i=1,2,3
\,,
\label{eq:uijapp}
\end{equation}
when expressed in terms of $T_j$, $S_j$, and $F_j$,
in the multi-Regge limit take the canonical form~\eqref{eq:crsmrl},
with the simple identifications~\eqref{eq:FSTtoweps}
\begin{equation}
S_j^2=\frac{T_j^2}{w_j\wb_j}
\,,\qquad
F_j^2=\frac{w_j}{\wb_j}
\,.
\label{eq:SFtowwb}
\end{equation}
For other choices of reference polygons, the relation between $w_j$,
$\wb_j$ and $T_j$, $S_j$, $F_j$ might get more complicated. More
concretely, extending the $(n{-}1)$-gon to the $n$-gon requires adding
two momentum twistors (that will become $Z_n$ and the new $Z_1$),
which have six degrees of freedom. Three of those degrees of freedom
are fixed by consistency with the $(n{-}1)$-gon, namely by requiring that
the equation (see \figref{fig:innerhexagon})
\begin{equation}
Z_j\suprm{int}
=\vev{1,2,3,j}Z_{j+1}
-\vev{1,2,3,j+1}Z_j
\label{eq:Zintapp}
\end{equation}
for the internal twistor $Z\suprm{int}_{n-1}$ (which was $Z_1$ in the
$(n{-}1)$-gon) is satisfied. The three remaining degrees of freedom
are fixed by imposing the multi-Regge limit
relations~\eqref{eq:crsmrl} with~\eqref{eq:SFtowwb}.

This construction can be iterated to any number of points. We have
solved the constraints explicitly for up to $n=18$, and then
recognized the pattern shown in \tabref{tab:ngonZ}.
\begin{table}
\centering
\fbox{%
\begin{minipage}{-0.5cm+\textwidth}
\centering
\parbox{5cm}{%
\begin{align*}
&\mspace{-48mu}
\text{For all }j\geq0:\\
\mathbf{Z}_{6j+4}&=\brk{6j,1,0,-1}\,,\\
\mathbf{Z}_{6j+5}&=\brk{6j,1,0,0}\,,\\
\mathbf{Z}_{6j+6}&=\brk{6j,1,-1,0}\,,\\
\mathbf{Z}_{6j+7}&=\brk{6j+1,1,-2,-1}\,,\\
\mathbf{Z}_{6j+8}&=\brk{6j+3,1,-2,-2}\,,\\
\mathbf{Z}_{6j+9}&=\brk{6j+5,1,-1,-2}\,,
\end{align*}
}
\qquad
\parbox{4.7cm}{%
\begin{align*}
\begin{pmatrix}
\mathbf{Z}_3\\
\mathbf{Z}\suprm{int}_4\\
\mathbf{Z}\suprm{int}_5\\
\mathbf{Z}\suprm{int}_6\\[-0.4ex]
\vdots\\
\mathbf{Z}\suprm{int}_{n-1}\\
\mathbf{Z}_1
\end{pmatrix}
=
\begin{pmatrix}
\brk{1,0,1,1}   \\
\brk{0,0,0,1}   \\
\brk{0,0,1,0}   \\
\brk{-1,0,1,1}  \\
\brk{2,0,0,-1}  \\
\brk{2,0,1,0}   \\[-0.5ex]
\vdots\\
\text{cyclic rep.}\\[-0.7ex]
\vdots
\end{pmatrix},
\end{align*}
}
\qquad
\parbox{2.9cm}{%
\begin{equation*}
\mathbf{Z}_2=\brk{1,0,0,0}
\,,
\end{equation*}
}
\\[-1.5ex]
\begin{alignat*}{2}
Z_j&=
\begin{cases}
\mathbf{Z}_j                                   & 2 \leq j\leq 5\,,\\
\mathbf{Z}_j M_{j-5}\dots M_{2}M_{1}           & 6 \leq j\leq n\,,
\end{cases}
\quad &
Z\suprm{int}_j&=
\begin{cases}
\mathbf{Z}\suprm{int}_j                        & 4 \leq j\leq 5\,,\\
\mathbf{Z}\suprm{int}_jM_{j-5}\dots M_{2}M_{1} & 6 \leq j\leq n-1\,,
\end{cases}\\
Z_1&=\mathbf{Z}_1 M_{n-5}\dots M_{2}M_{1}\,. & &
\end{alignat*}
\vspace{-4mm}
\end{minipage}}
\caption{Parametrization of the $n$-point polygon for any $n\geq6$.
The boldface momentum twistors $\mathbf{Z}_j$ define a fixed reference
polygon, where ``cyclic rep.'' stands for cyclic repetitions of the
six given momentum twistors. All conformally inequivalent polygons are
parametrized by the $Z$ in normal face. They are obtained from the
reference polygon by acting with each matrix $M_j$ stabilizing the
$j$'th internal tetragon on all momentum twistors below that tetragon
(see eqs.~\protect\eqref{eq:Zhex} and~\protect\eqref{eq:Zhept} for
examples). For reference, the matrices $M_j$ are given
in \protect\tabref{tab:ngonMt}. This specific choice of reference
polygon is engineered to satisfy the relations~\protect\eqref{eq:crsmrl}
with eq.~\protect\eqref{eq:SFtowwb} in the multi-Regge limit.}
\label{tab:ngonZ}
\end{table}
Here, the transformation matrices $M_j$, $j=1,\dots,n{-}5$, are defined
via eq.~\eqref{eq:Mjeq} with all other $M_i$, $i\neq j$ set to~$1$.
For completeness, we explicitly list these matrices in \tabref{tab:ngonMt}.
We have checked the consistency of this parametrization
with eqs.~\eqref{eq:crsmrl}, \eqref{eq:SFtowwb}, and~\eqref{eq:Zintapp} up
to $n=30$.
In order to avoid overly complicated expressions, one can
alternatively choose to act with the global conformal transformation $M_{1}^{-1}\dots M_{\floor{n/2}-2}^{-1}$
on the full polygon, which
evenly distributes the action of the matrices $M_j$ on the top and bottom of
the polygon and thus renders expressions a bit simpler.
\begin{table}
\centering
\fbox{%
\begin{minipage}{-0.3cm+\textwidth}
\centering
\begin{align*}
M_{j}&=\sqrt{F_j}
\begin{pmatrix}
 \frac{S_j}{F_j} & 0 & 0 & 0 \\
 \frac{(j-1) \brk{1-S_j^2}}{F_j S_j} & \frac{1}{F_j S_j} & 0 & 0 \\
 0 & 0 & \frac{1}{T_j} & 0 \\
 0 & 0 & 0 & T_j
\end{pmatrix}
\,,& j&\in6\Integers+1
\,,\\
M_{j}&=\sqrt{F_j}
\begin{pmatrix}
 \frac{S_j}{F_j} & 0 & 0 & 0 \\
 \frac{(j-2) \brk{1-S_j^2}}{F_j S_j} & \frac{1}{F_j S_j} & T_j-\frac{1}{F_j S_j} & 0 \\
 0 & 0 & T_j & 0 \\
 \frac{S_j}{F_j}-\frac{1}{T_j} & 0 & \frac{1}{T_j}-T_j & \frac{1}{T_j}
\end{pmatrix}
\,,& j&\in6\Integers+2
\,,\\
M_{j}&=\sqrt{F_j}
\begin{pmatrix}
 \frac{S_j}{F_j} & 0 & 0 & 0 \\
 -\frac{(j-2) \brk{S_j^2-1}}{F_j S_j}-2 T_j+\frac{2}{T_j} & \frac{1}{F_j S_j} & 2 T_j-\frac{2}{F_j S_j} & 2 T_j-\frac{1}{F_j S_j}-\frac{1}{T_j} \\
 -\frac{S_j}{F_j}-T_j+\frac{2}{T_j} & 0 & T_j & T_j-\frac{1}{T_j} \\
 \frac{2 S_j}{F_j}-\frac{2}{T_j} & 0 & 0 & \frac{1}{T_j}
\end{pmatrix}
\,,& j&\in6\Integers+3
\,,\\
M_{j}&=\sqrt{F_j}
\begin{pmatrix}
 \frac{S_j}{F_j} & 0 & 0 & 0 \\
 -\frac{(j-1) \brk{S_j^2-1}}{F_j S_j}-4 T_j+\frac{4}{T_j} & \frac{1}{F_j S_j} & \frac{2}{T_j}-\frac{2}{F_j S_j} & 2 T_j-\frac{2}{F_j S_j} \\
 \frac{2}{T_j}-\frac{2 S_j}{F_j} & 0 & \frac{1}{T_j} & 0 \\
 \frac{2 S_j}{F_j}-2 T_j & 0 & 0 & T_j
\end{pmatrix}
\,,& j&\in6\Integers+4
\,,\\
M_{j}&=\sqrt{F_j}
\begin{pmatrix}
 \frac{S_j}{F_j} & 0 & 0 & 0 \\
 -\frac{j \brk{S_j^2-1}}{F_j S_j}-2 T_j+\frac{2}{T_j} & \frac{1}{F_j S_j} & -T_j-\frac{1}{F_j S_j}+\frac{2}{T_j} & \frac{2}{T_j}-\frac{2}{F_j S_j} \\
 2 T_j-\frac{2 S_j}{F_j} & 0 & T_j & 0 \\
 \frac{S_j}{F_j}-2 T_j+\frac{1}{T_j} & 0 & \frac{1}{T_j}-T_j & \frac{1}{T_j}
\end{pmatrix}
\,,& j&\in6\Integers+5
\,,\\
M_{j}&=\sqrt{F_j}
\begin{pmatrix}
 \frac{S_j}{F_j} & 0 & 0 & 0 \\
 -\frac{j \brk{S_j^2-1}}{F_j S_j} & \frac{1}{F_j S_j} & 0 & \frac{1}{T_j}-\frac{1}{F_j S_j} \\
 T_j-\frac{S_j}{F_j} & 0 & T_j & T_j-\frac{1}{T_j} \\
 0 & 0 & 0 & \frac{1}{T_j}
\end{pmatrix}
\,,& j&\in6\Integers
\,.
\end{align*}
\vspace{-3mm}
\end{minipage}}
\caption{Matrices $M_j$, $1\,{\leq}\,j\,{\leq}\,(n-5)$ appearing
in \protect\tabref{tab:ngonZ} that stabilize the $j$'th internal
tetragon of the reference polygon. Each of these matrices is a
solution to the defining equation~\protect\eqref{eq:Mjeq} with all
other matrices set to $M_i=1$.}
\label{tab:ngonMt}
\end{table}
%

%%%%%%%%%%%%%%%%%%%%%%%%%%%%%%
\paragraph{Cross Ratios in Collinear and Regge Limits.}

The parametrization in \tabref{tab:ngonZ} entails uniform
expressions for the cross ratios~\eqref{eq:uijapp}. We quote their expansions in the
multi-collinear limit $T_j\to0$, $j=1,\dots,n-5$:
\begin{align}
u_{j,1} &
=\frac{1}{(1+S_{j-1}^2)(1+S_j^2)}\Biggsbrk{
1+S_{j-1}^2(1+S_j^2)
-\frac{2\cos(\varphi_{j-1})S_{j-1}^3S_j^2T_{j-1}}{1+S_{j-1}^2}
-\frac{2\cos(\varphi_j)S_jT_j}{1+S_j^2}
\nn\\ & \mspace{30mu}
-\frac{S_{j-1}^2\bigsbrk{1-2\cos(2\varphi_{j-1})S_{j-1}^2+S_{j-1}^4}S_j^2T_{j-1}^2}{\brk{1+S_{j-1}^2}^2}
-\frac{\bigsbrk{1-2\cos(2\varphi_j)S_j^2+S_j^4}T_j^2}{\brk{1+S_j^2}^2}
\nn\\ & \mspace{30mu}
+\frac{2S_{j-1}S_j\bigsbrk{\cos(\varphi_{j-1}+\varphi_j)\brk{1+S_j^2+S_{j-1}^2S_j^2}-\cos(\varphi_{j-1}-\varphi_j)S_{j-1}^2}T_{j-1}T_j}{\brk{1+S_{j-1}^2}\brk{1+S_j^2}}
}
+\order{T_i^3}
\,,\nn\\
u_{j,2} &
=\frac{A_j}{1+A_j}
+\frac{2}{\brk{1+A_j}^2}\Biggsbrk{
-\cos\varphi_j A_j S_j T_j
+\sum_{i=j+1}^{n-1}\cos\varphi_i\prod_{k=j}^{i-1}S_k^2\, S_i T_i
}+\order{T_i^2}
\,,
\label{eq:crscollmore}
\\
u_{j,3} &
=T_j^2\Biggsbrk{
1+\bigbrk{T_{j-1}^2+T_{j+1}^2}
+\bigbrk{
 T_{j-2}^2T_{j-1}^2
+T_{j-1}^4
+T_{j-1}^2T_{j+1}^2
+T_{j+1}^4
+T_{j+1}^2T_{j+2}^2
}+\order{T_{i\neq j}^6}
}
\nn
\end{align}
The leading terms in these expressions can be recognized in eq.~\eqref{eq:crscoll}.
Here, $\order{T_i^\ell}$ stands for any products of all $T_i$ of
total order at least $\ell$. The parameters $A_j$ are defined via~\eqref{eq:Ajdef}.
In all expressions above and below, we set
\begin{equation}
A_{n-4}=0
\,,\qquad
S_j=T_j=0
\quad\text{for}\quad
j\notin\brc{1,\dots,n-5}
\,.
\end{equation}
The last expression for $u_{j,3}$ in eq.~\eqref{eq:crscollmore} is exact
in $T_j$. In fact, we can write the full expression for $u_{j,3}$ in
general kinematics. It is a ratio of polynomials in the $T_i^2$ with
unit coefficients:
\begin{gather}
u_{j,3}=T_j^2
\frac{P_1^{j-2}\,P_{j+2}^{n-5}}{P_{1}^{j-1}\,P_{j+1}^{n-5}}
\,,\qquad
P_a^b=\sum_{I\in\,\mathcal{I}_a^b}
\prod_{\ell=1}^{\abs{I}}\brk{-T_{i_\ell}^2}
=1-\sum_{i=a}^bT_i^2+\dots
\,,\nn\\
\mathcal{I}_a^b=\bigbrc{\brk{i_1,\dots,i_k}\,|\,k\geq0\wedge a\leq i_\ell\leq b\wedge i_\ell+2\leq i_{\ell+1}}
\,.
\label{eq:uj3exact}
\end{gather}
In the multi-Regge limit, where all $T_i\to0$ with $r_i=S_i/T_i$
fixed, we find the expansions
\begin{align}
\label{eq:crsmrlmore}
u_{j,2}\,u_{j,3} &
=\Bigsbrk{
1 + T_{j-1}^2
+\lrbrk{2+2\cos(\varphi_{j+1}) r_{j+1} + r_{j+1}^2}T_{j+1}^2
+ \order{T_{i\neq j}^4}
}r_{j}^2 T_{j}^4 + \order{T_j}^6
\,,\\
\frac{u_{j,2}}{1-u_{j,1}} &
=\frac{r_j^2}{1+2\cos(\varphi_j)r_j+r_j^2}\biggsbrk{
1+\biggbrk{
2\frac{\cos(\varphi_{j-1}) + \cos(\varphi_{j-1} + \varphi_{j})r_{j}}{1 + 2\cos(\varphi_{j})r_{j} + r_{j}^2}r_{j-1}
+r_{j-1}^2
}T_{j-1}^2
\nn\\ & \mspace{240mu}
- T_{j}^2
+\Bigbrk{1 + 2\cos(\varphi_{j+1})r_{j+1} + r_{j+1}^2}T_{j+1}^2
} + \order{T_i^4}
\,,\nn\\
\frac{u_{j,3}}{u_{j,2}} &
=\frac{1}{r_j^2}\biggsbrk{
1+T_{j-1}^2+\lrsbrk{1+2\cos(\varphi_j)r_j+r_j^2}T_j^2
-\lrsbrk{2\cos(\varphi_{j+1})r_{j+1}+r_{j+1}^2}T_{j+1}^2
}+\order{T_i^4}
.\nn
\end{align}
Besides providing an unambiguous parametrization of conformally
inequivalent null polygons, the variables $\tau_j$, $\sigma_j$, and
$\varphi_j$ explained above are of course designed for the application of
the OPE for null polygon Wilson
loops~\cite{Alday:2010ku,Sever:2011pc,Basso:2013vsa,Basso:2013aha,Basso:2014koa,Basso:2014nra}.
The applicability of the latter crucially relies on the specific
``alternating'' tessellation of~\cite{Sever:2011pc} that we used in
the main body of this paper. It would be nice to find a generalization
of the Wilson loop OPE that applies to our multi-Regge friendly tessellation.

Note that the reference $n$-point polygon
of \tabref{tab:ngonZ} is somewhat singular: For $T_j=S_j=F_j=1$,
several internal distances $x_i-x_j$ among cusps are null, and some others are timelike.
The reference polygon therefore lies outside the Euclidean region
where the Wilson loop OPE is applicable. However, all internal
distances become spacelike when approaching the multi-collinear limit,
\ie when all parameters $T_j$ are small, and hence the Wilson loop OPE
can be applied in this region. In fact, we find experimentally that
all internal distances $x_i-x_j$ are spacelike as long as all
$T_j\leq1/2$, $j=1,\dots,n-5$.

%%%%%%%%%%%%%%%%%%%%%%%%%%%%%%
\paragraph{The Hexagon.}

In terms of the tessellation variables $T\equiv T_1$, $S\equiv S_1$,
and $\varphi\equiv\varphi_1$, the three cross ratios of the hexagon read
\begin{equation}
u_1 = U_{2,5} = \frac{1-T^2}{1+S^2+2ST\cos\varphi} \,, \quad
u_2 = U_{3,6} = \frac{S^2(1-T^2)}{1+S^2+2ST\cos\varphi} \,, \quad
u_3 = U_{1,4} = T^2
\,.
\label{eq:ujparhexapp}
\end{equation}
When comparing with our expressions in eq.~\eqref{eq:ujprimepar}, we use the
identifications $x'_3 = x_2$ and $x'_6 = x_1$ for $n=6$ to show that
\begin{gather*}
u'_1 = U_{3',5,3,6} = U_{2,5,3,6} = U_{2,5} = u_1
\,,\qquad
u'_2 = U_{3,6,4,6'} = U_{3,6,4,1}= U_{3,6} = u_2
\,,\\[2mm]
u'_3 = U_{6',4,3',5} = U_{1,4,2,5} = U_{1,4}= u_3
\,.
\end{gather*}
In the collinear limit, the relations~\eqref{eq:ujpar} read
\begin{equation}
u_1 = \frac{1}{1+S^2} + \order{T}
\,, \quad
u_2 = \frac{S^2}{1+S^2} + \order{T^3}
\,, \quad
u_3 = T^2
\,.
\label{eq:uihexapp}
\end{equation}
This coincides with our claim in eqs.~\eqref{eq:crscoll} if we take into account
that $S_0 = 0$ and $S^2 = S_1^2 = A_1$ for $n=6$. When parametrized in terms of $S$,
$T$, and $F=\exp(i\varphi)$, the Regge limit is taken by sending both $T$ and $S$ to
zero, while keeping $r=S/T$ finite, \ie in the Regge limit, the remainder function
depends on the finite variables $r$ and $F$ along with the quantity $T$ that
vanishes in the limit. These are related to $w, \wb$ and $\eps$ through~\eqref{eq:FSTtoweps}:
\begin{equation}
r^2=\frac{1}{w\wb}\,,
\qquad
F^2=\frac{w}{\wb}\,,
\qquad
{S^2}{T^2} = r^2 T^4 = \eps \,.
\end{equation}
%

%%%%%%%%%%%%%%%%%%%%%%%%%%%%%%
\paragraph{The Heptagon.}

Via the general formula~\eqref{eq:UofZdet}, our tessellation induces the following
formulas for the heptagon cross ratios:
\begin{align}
u_{1,3}=U_{1,4}&=\frac{T_1^2}{1-T_2^2}
\,,\nn\\[2mm]
u_{1,1}=U_{2,5}&=\frac{1-T_1^2}{1+S_1^2+2c_1S_1T_1}
\,,\nn\\[2mm]
u_{1,2}=U_{3,7}&=\frac{
S_1^2
\lrbrk{1-T_1^2-T_2^2}
}{
(1-T_2^2)
\lrbrk{1+S_1^2+2c_1S_1T_1}
}
\cdot\frac{1}{U_{3,6}}
\,,\nn\\[2mm]
u_{2,1}=U_{3,6}&=\frac{
S_1^2S_2^2(1-T_1^2)+(1-T_2^2)+S_1^2+2c_1S_1T_1+2c_2S_1^2S_2T_2+2c_+S_1T_1S_2T_2
}{
\lrbrk{1+S_1^2+2c_1S_1T_1}
\lrbrk{1+S_2^2+2c_2S_2T_2}
}
\,,\nn\\[2mm]
        U_{2,6}&=\frac{
\lrbrk{1-T_1^2-T_2^2}
}{
(1-T_1^2)
\lrbrk{1+S_2^2+2c_2S_2T_2}
}
\cdot\frac{1}{U_{3,6}}
\,,\nn\\[2mm]
u_{2,2}=U_{4,7}&=\frac{S_2^2(1-T_2^2)}{1+S_2^2+2c_2S_2T_2}
\,,\nn\\[2mm]
u_{2,3}=U_{1,5}&=\frac{T_2^2}{1-T_1^2}
\,,
\label{eq:cr7opevarapp}
\end{align}
where we have used the shorthand notation
\begin{equation}
c_1=\cos(\varphi_1)
\,,\qquad
c_2=\cos(\varphi_2)
\,,\quad
c_+=\cos(\varphi_1+\varphi_2)
\,.
\end{equation}
Given the formulas~\eqref{eq:cr7opevarapp}, it is straightforward to
obtain the following expressions for the
leading terms in the cross ratios as $T_1$ and $T_2$ are sent to zero,
\begin{align}
u_{1,1}=U_{25}&=\frac{1}{1+S_1^2} \,, &
u_{1,2}=U_{37}&=\frac{S_1^2(1+S_2^2)}{1+S_1^2+S_1^2S_2^2} \,, &
u_{1,3}=U_{14}&=T_1^2 \,, \nn \\[2mm]
u_{2,1}=U_{36}&=\frac{1+S_1^2+S_1^2S_2^2}{(1+S_1^2)(1+S_2^2)} \,, &
u_{2,2}=U_{47}&=\frac{S_2^2}{1+S_2^2} \,, &
u_{2,3}=U_{15}&=T_2^2 \,, \nn \\[2mm]
        U_{26}&=\frac{1+S_1^2}{1+S_1^2+S_1^2S_2^2} \,.
\label{eq:crs7collapp}
\end{align}
From the general kinematics~\eqref{eq:cr7opevarapp}, the multi-Regge limit is attained
by setting $S_j=r_jT_j$ and letting $T_j\to0$, keeping $r_j$ finite, with the
identifications~\eqref{eq:FSTtoweps}.
From the multi-Regge limit, the combined multi-Regge collinear limit
is attained for $r_1,r_2\to\infty$. Conversely, if we start in general
kinematics, we reach the collinear limit when we send $T_j
\to0$ while keeping $S_j$ and $F_j$ finite. We can then continue to
the combined multi-Regge collinear limit by letting $S_j\to0$,
keeping $S_j/T_j\gg1$.

%%%%%%%%%%%%%%%%%%%%%%%%%%%%%%%%%%%%%%%%%%%%%%%%%%%%%%%%%%%%
%%%%%%%%%%%%%%%%%%%%%%%%%%%%%%%%%%%%%%%%%%%%%%%%%%%%%%%%%%%%
\section{Performing the One-Loop Sums}
\label{app:sums}

In this appendix, we fill in some details about our evaluation of the residue sums
that appear in the one-loop contribution of one-gluon excitation shown in eq.~\eqref{eq:h1sum},
which we split into a double sum $\Sigma_2$ and a single sum $\Sigma_1$:
\begin{align}
    h_1^{\brk{0, 0}} = g^2 \frac{S_2}{S_1} \Bigbrk{\Sigma_2 + \Sigma_1}\,.
    \label{eq:h1sumSigma}
\end{align}
We begin with the double sum,
\begin{equation}
    \Sigma_2
    \defas
    \sum\limits_{k_{1,2} \in \Z_{\ge 1}}
    \frac{
        \brk{-S_1^{-2}}^{k_1} \brk{-S_2^{-2}}^{k_2}
    }{
        \brk{k_1 + 1} \brk{k_2 + 1} k_1 k_2
    }
    \times
    \frac{
        \Gamma\brk{1 + k_1 +k_2}
    }{
        \Gamma\brk{k_1} \Gamma\brk{k_2}
    }
    =
    \sum\limits_{j = 2}^{\infty}\sum\limits_{k_1 = 1}^{j - 1}
    \frac{
        \brk{S_1^{-2} S_2^2}^{-k_1} \brk{-S_2^{-2}}^j
    }{
        \brk{k_1 + 1} \brk{j - k_1 + 1}
    }
    \binom{j}{k_1}\,,
    \label{eq:resSum}
\end{equation}
where to reach the right hand side, we have introduced a new summation index
$j = k_1 + k_2$. Now we can split denominators in the following way:
\begin{equation}
    \frac{1}{\brk{k_1 + 1} \brk{j - k_1 + 1}}
    =
    \frac{1}{j + 2} \Bigbrk{
        \frac{1}{k_1 + 1}
        + \frac{1}{j - k_1 + 1}
    }
    \,.
\end{equation}
Note that if we let $k_1' = j - k_1$ and make use of $\binom{j}{j - k_1'} =
\binom{j}{k_1'}$, we can decompose eq.~\eqref{eq:resSum} into two sums that
only differ by the replacement $S_{1} \leftrightarrow S_{2}$,
\begin{equation}
    \Sigma_2
    =
    \sum\limits_{j = 2}^{\infty}
    \frac{\brk{-S_2^{-2}}^j}{j + 2}
    \Biggbrk{
        \sum\limits_{k_1 = 1}^j
        \brk{S_1^{-2} S_2^2}^{k_1}
        \binom{j}{k_1}
        \frac{1}{k_1 + 1}
        -
        \brk{S_1^{-2} S_2^2}^{j}
        \frac{1}{j + 1}
    }
    +
    \brk{S_1 \leftrightarrow S_2}
    \,.
\end{equation}
Next, we eliminate the $\brk{+1}$ offset in the denominator of the inner sum.
This can be done with the help of the binomial identity $\binom{j}{k - 1} =
\frac{k}{j + 1} \binom{j + 1}{k}$ (for a more general treatment, see eq. (44) in~\cite{Moch:2001zr}),
\begin{equation}
    \sum\limits_{k_1 = 1}^j
    \brk{S_1^{-2} S_2^2}^{k_1}
    \binom{j}{k_1}
    \frac{1}{k_1 + 1}
    =
    \sum\limits_{k = 2}^{j + 1}
    \brk{S_1^{-2} S_2^2}^{k - 1}
    \binom{j}{k - 1}
    \frac1k
    %=
    %\sum\limits_{k = 2}^{j + 1}
    %\Bigbrk{\frac{S_2^2}{S_1^2}}^{k - 1}
    %\binom{j + 1}{k}
    %\frac{1}{j + 1}
    =
    \lrbrk{1 + \frac{S_2^2}{S_1^{2}}}^{j + 1} \frac{S_1^2}{S_2^{2}} \frac{1}{j + 1} - 1\,.
\end{equation}
After these steps, the outer summation over $j$ is straightforward:
\begin{equation} \label{eq:sum1}
    \Sigma_2
    =
    \brk{S_1^2 + S_2^2 + S_1^2 S_2^2} \log{
        \frac{S_1^2 + S_2^2 + S_1^2 S_2^2}{\brk{1 + S_1^2} \brk{1 + S_2^2}}
    }
    + 1\,.
\end{equation}
Having evaluated the sum in eq.~\eqref{eq:resSum}, it remains to perform the single sum,
\begin{equation}
    \Sigma_1 \defas \sum\limits_{k_1 \in \Z_{\ge 1}}
    \frac{\brk{-S_1^{-2}}^{k_1} + \brk{-S_2^{-2}}^{k_1}}{k_1 + 1}
    + 1
    =
    S_1^2 \log{\brk{1 + S_1^{-2}}}
    + S_2^{2} \log{\brk{1 + S_2^{-2}}}
    - 1\,.
\end{equation}
Combining these $\Sigma_1$ and $\Sigma_2$ sums back into eq.~\eqref{eq:h1sumSigma},
we obtain \brk{up to an overall sign} eq.\ (125) from~\cite{Basso:2013aha} that was stated in eq.~\eqref{eq:h1intev}.

%%%%%%%%%%%%%%%%%%%%%%%%%%%%%%%%%%%%%%%%%%%%%%%%%%%%%%%%%%%%
%%%%%%%%%%%%%%%%%%%%%%%%%%%%%%%%%%%%%%%%%%%%%%%%%%%%%%%%%%%%
\section{The Function \texorpdfstring{$g$}{g}}
\label{app:gfunction}

At LLA, the MRL two-loop remainder function for all multiplicities and in all
kinematic regions can be expressed in terms of the six-point
function~\eqref{eq:f}. At NLLA, further functions $g$ and $\tilde{g}$
are required to express the seven-point remainder function in the kinematic
regions $(\m\m\m)$ and $(\m\p\m)$. Writing out the
expressions~\eqref{eq:R7mmm} and~\eqref{eq:R7mpmg}, one finds
\begin{align}
  R_{7,(2)}^{\m\m\m}&=\log(\eps_1)f_1(v_1)+\log(\eps_2)f_1(v_2)+f_0(v_1)+f_0(v_2)+g(v_1,v_2)
                      \,,\label{eq:R7mmmg}\\
  R_{7,(2)}^{\m\p\m}&=\log(\eps_1)\bigbrk{f_1(v_1)-f_1(w_1)}+\log(\eps_2)\bigbrk{f_1(v_2)-f_1(w_2)}
                      \nn\\
                    &\mspace{100mu}+f_0(v_1)-f_0(w_1)+f_0(v_2)-f_0(w_2)+\tilde{g}(v_1,v_2)
                      \,.\label{eq:R7mpmgapp}
\end{align}
At symbol level, the remainder functions in the various regions
satisfy the identity~\cite{Bargheer:2015djt}
\begin{equation}
S\sbrk{R_{7,(2)}^{\m\p\m}}=S\sbrk{R_{7,(2)}^{\m\m\m}}-S\sbrk{R_{7,(2)}^{\m\m\p}}-S\sbrk{R_{7,(2)}^{\p\m\m}}
\,,
\end{equation}
and therefore the symbols of $g$ and $\tilde{g}$ are identical:
\begin{equation}
S\sbrk{g(v_1,v_2)}=S\sbrk{\tilde{g}(v_1,v_2)}
\,.
\end{equation}
Based on its symbol as well as symmetry arguments, the function $g$
was determined up to $25$ unfixed rational
coefficients~\cite{Bargheer:2015djt}. Subsequently, the function $g$
was fully determined in~\cite{DelDuca:2018hrv}. The latter result
assumes a specific path of continuation from the Euclidean to the
$(\m\m\m)$ region, and is justified by Regge factorization. In the
following, we independently refine the determination of the functions $g$ and
$\tilde{g}$. Without making assumptions on the path of continuation,
we are able to determine both $g$ and $\tilde{g}$ up to a few coefficients.

%%%%%%%%%%%%%%%%%%%%%%%%%%%%%%
\paragraph{Function Space.}

Maximally helicity-violating amplitudes in $\superN=4$ super
Yang--Mills theory are rational polynomials in multiple
polylogarithms, $i\pi$, and (multiple) zeta values, where all occurring
monomials have the same (uniform) transcendental weight~\cite{Kotikov:2004er}.
Multiple polylogarithms, also called Goncharov
polylogarithms~\cite{Goncharov:2001iea}, can be defined
recursively as iterated integrals
\begin{equation}
\label{eq:Gdef}
G(a_1,\ldots,a_n;z)
\equiv
\begin{cases}
\displaystyle
\frac{1}{n!}\log^n z & \text{if }a_1=\ldots=a_n=0\,,\\[2ex]
\displaystyle
\int_0^z \frac{dt}{t-a_1}G(a_2,\ldots,a_n;t) & \text{otherwise,}
\end{cases}
\end{equation}
with $G(;z)=1$. The sequence of parameters $(a_1,\dots,a_n)$ is called
the weight vector, and the length of the weight vector equals the
transcendental weight (or transcendentality) of the function
$G(a_1,\dots,a_n;z)$. The parameters $a_i$ are also called letters,
and the set of all multiplicatively independent letters that occur in
a given function is called the alphabet of that function.
Multiple zeta values are defined in terms of
multiple polylogarithms evaluated at unity, and inherit their
transcendental weight: $\zeta_k$ has weight $k$,
$\zeta_{j,k}$ has weight $j+k$, and so forth. $\pi$ has weight $1$.

As noted in~\cite{Bargheer:2015djt}, using the variables~\eqref{eq:vofw},
the alphabet of the seven-point remainder
function in all Mandelstam regions becomes
\begin{equation}
\aleph_{xy}
=\brc{x,1-x,y,1-y,1-xy}\cup\brc{\text{c.c.}}
\,.
\end{equation}
Multiple polylogarithms whose symbols draw their entries from this
alphabet belong to the class of two-dimensional harmonic
polylogarithms (2dHPLs)~\cite{Gehrmann:2000zt}. An independent basis for
these is given by~\cite{RADFORD1979432}%
\footnote{The choice of basis is not unique. We used a
different basis in~\cite{Bargheer:2015djt}, but found the
choice~\eqref{eq:Gbasis} more suitable for the present analysis.}
\begin{equation}
\bigbrc{G(\vec a,x)\,|\,\vec a_i\in\lyndon\brc{0,1}}
\cup
\bigbrc{G(\vec a,1/y)\,|\,\vec a_i\in\lyndon\brc{0,1,x}}
\cup\brc{\text{c.c.}}
\,,
\label{eq:Gbasis}
\end{equation}
where $\brc{\text{c.c.}}$ stands for the complex conjugates of the
previous sets, and $\lyndon\brc{0,1}$ and $\lyndon\brc{0,1,x}$
denote the sets of Lyndon words formed from the ordered sets of
letters $\brc{0,1}$ and $\brc{0,1,x}$, respectively.

%%%%%%%%%%%%%%%%%%%%%%%%%%%%%%
\paragraph{Single-Valuedness.}

Besides the consistency with the known symbol, the functions $g$ and
$\tilde{g}$ have
to satisfy various constraints. One of them is single-valuedness: Due
to unitarity, a physical amplitude can only have branch points where
one of the cross ratios vanishes (or becomes infinite). Since the
cross ratios are expressed in terms of absolute squares of the complex
variables $w_1$ and $w_2$, see eq.~\eqref{eq:uofw}, a rotation
$(w_1-z,\wb_1-\bar z)\to(e^{+2\pi i}(w_1-z),e^{-2\pi i}(\wb_1-\bar z))$
around any point $z$ in the complex plane can never let
a cross ratio wind around zero (or infinity). The same is true for
rotations of $w_2$, and therefore also for rotations of $x$ and $y$.
The conclusion is that the remainder function in the multi-Regge limit
must be a single-valued function of the complex variables $x$ and $y$,
and thus the same must be true for the functions $g$ and $\tilde{g}$.
This property has been essential for the determination of the
six-point multi-Regge limit to high loop
orders~\cite{Dixon:2012yy,Pennington:2012zj,Dixon:2014voa,Dixon:2014xca}.

It turns out that
the single-valuedness constraint can be satisfied directly at the level
of the basis: Single-valued multiple polylogarithms were constructed
by Brown~\cite{Brown:2013gia}. A suitable basis of such functions for
multi-Regge amplitudes of any multiplicity was proposed in~\cite{DelDuca:2016lad}.
The single-valued basis can be constructed purely algebraically from
the basis of ordinary multiple polylogarithms~\eqref{eq:Gbasis} using
the Hopf algebra structure that underlies the multiple polylogarithm
algebra~\cite{Goncharov:2001iea}: Each holomorphic element $G$ of the
ordinary basis~\eqref{eq:Gbasis} gets promoted to a single-valued
function $\Gs$ by the single-valued map~\cite{Brown:2013gia}
\begin{equation}
\mathbf{s}:G(\vec a,z)
\mapsto
\Gs(\vec a,z)\equiv
(-1)^{\abs{\vec a}}\mu(\bar S\otimes\id)\Delta G(\vec a,z)
\,,
\label{eq:svmap}
\end{equation}
where $\Delta$ is the coproduct, $\id$ is the identity, $\bar S$
is the complex conjugate of the
antipode map of the Hopf algebra, and $\mu$ denotes the multiplication
operator $\mu(a\otimes b)=a\cdot b$. All details of the single-value
map are spelled out in
Section 3.4.3 of~\cite{DelDuca:2016lad}, and we will not reproduce
them here. For example, one finds%
\footnote{After the completion of this computation, the \mathematica\
package \soft{PolyLogTools} appeared~\cite{Duhr:2019tlz}, which can
perform many of the required manipulations of single-valued and
ordinary multiple polylogarithms. For example, single-valued MPLs can
be expanded in ordinary MPLs via the function \code{cGToG} of that
package. The expansion typically contains non-basis functions. Most of
these can be related back to basis functions via shuffle identities,
all others can be mapped to combinations of
basis functions by
numerical comparison, \eg via
\soft{GiNaC}~\cite{Bauer:2000cp} (as done by the function
\code{ToFibrationBasis} of \soft{PolyLogTools}).}
\begin{multline}
\Gs(0,x,1/y)=
-G(0,x)G(\bar x,1/\bar y)
-G(0,\bar x)G(\bar x,1/\bar y)
+G(0,1/y)G(\bar x,1/\bar y)
\\
+G(0,1/\bar y)G(\bar x,1/\bar y)
+G(0,x,1/y)
-G(0,\bar x,1/\bar y)
\,.
\end{multline}
The anti-holomorphic elements
of~\eqref{eq:Gbasis} can equally be promoted to single-valued functions,
which however are not independent from the single-valued functions
generated from the holomorphic elements. A full basis of single-valued
2dHPLs is therefore provided by the single-valued completions of the
holomorphic elements of the ordinary basis~\eqref{eq:Gbasis}. Since
this halves the size of the algebra basis, it significantly reduces
the number of linearly independent elements in a general Ansatz at any
fixed weight.

To summarize, the single-valued algebra basis that we will employ is
\begin{equation}
\bigbrc{\Gs(\vec a,x)|\vec a\in\lyndon\brc{0,1}}
\cup
\bigbrc{\Gs(\vec a,1/y)|\vec a\in\lyndon\brc{0,1,x}}
\,,
\label{eq:Gsbasis}
\end{equation}
where every single-valued function $\Gs(\vec a,z)$ is
constructed from the ordinary multiple polylogarithm $G(\vec a,z)$ according
to the algebraic prescription~\eqref{eq:svmap}.

%%%%%%%%%%%%%%%%%%%%%%%%%%%%%%
\paragraph{The Ansatz and Symbol Constraints.}

Here and in the following, we use the condensed notation
$\Gsshort{z}_{a_1,\dots,a_n}\equiv\Gs(a_1,\dots,a_n;z)$, and
$\cy\equiv1/y$.
For up to weight three, the basis~\eqref{eq:Gsbasis} has $19$ elements
and reads
\begin{align}
\big\{&
\Gsshort{x}_{0}\,,
\Gsshort{x}_{1}\,,
\Gsshort{\cy}_{0}\,,
\Gsshort{\cy}_{1}\,,
\Gsshort{\cy}_{x}\,,
\Gsshort{x}_{0,1}\,,
\Gsshort{\cy}_{0,1}\,,
\Gsshort{\cy}_{0,x}\,,
\Gsshort{\cy}_{1,x}\,,
\nn\\&
\Gsshort{x}_{0,0,1}\,,
\Gsshort{x}_{0,1,1}\,,
\Gsshort{\cy}_{0,0,1}\,,
\Gsshort{\cy}_{0,0,x}\,,
\Gsshort{\cy}_{0,1,1}\,,
\Gsshort{\cy}_{0,1,x}\,,
\Gsshort{\cy}_{0,x,1}\,,
\Gsshort{\cy}_{0,x,x}\,,
\Gsshort{\cy}_{1,1,x}\,,
\Gsshort{\cy}_{1,x,x}
\big\}\,
\end{align}
There are $65$ different weight-three products of the above basis functions.
At leading weight, the function $g(v_1,v_2)$ therefore has to be a linear combination of
these $65$ terms, with rational coefficients. Taking the symbol of this general
linear combination, and equating it to the known symbol of $g$ fixes
all $65$ coefficients, the result being
\begin{align}
g(x,y)&=
-1/2\,\Gsshort{x}_{0}\Gsshort{\cy}_{0}\Gsshort{\cy}_{1}
+1/2\,\Gsshort{x}_{0}\Gsshort{x}_{1}\Gsshort{\cy}_{1}
+1/2\,\Gsshort{\cy}_{0}\Gsshort{x}_{1}\Gsshort{\cy}_{1}
-1/2\,\Gsshort{x}_{0}\Gsshort{x}_{1}\Gsshort{\cy}_{x}
\nn\\&\quad
+1/2\,\Gsshort{\cy}_{0}\Gsshort{\cy}_{1}\Gsshort{\cy}_{x}
-\Gsshort{\cy}_{1}\Gsshort{x}_{0,1}
+\Gsshort{\cy}_{x}\Gsshort{x}_{0,1}
+\Gsshort{x}_{0}\Gsshort{\cy}_{0,1}
-\Gsshort{x}_{1}\Gsshort{\cy}_{0,1}
-\Gsshort{\cy}_{x}\Gsshort{\cy}_{0,1}
\nn\\&\quad
+\Gsshort{x}_{1}\Gsshort{\cy}_{0,x}
-\Gsshort{\cy}_{0}\Gsshort{\cy}_{1,x}
-\Gsshort{x}_{1}\Gsshort{\cy}_{1,x}
+\Gsshort{\cy}_{1}\Gsshort{\cy}_{1,x}
+2\,\Gsshort{\cy}_{0,1,x}
-2\,\Gsshort{\cy}_{1,1,x}
\nn\\&\quad
+\zeta_2\,g^{(1)}(x,y)
+d_0\zeta_3
+i\pi\,g^{(2)}(x,y)
+i\pi d_1\zeta_2
\,,
\label{eq:g2ansatz}
\end{align}
where $g^{(2)}$ and $g^{(1)}$, at this point, are general linear combinations of the
$19$ and $5$ possible weight-two and weight-one products of basis
functions, and $d_0$ and $d_1$ are rational constants.
These terms are not constrained by the symbol of $g$.

%%%%%%%%%%%%%%%%%%%%%%%%%%%%%%
\paragraph{Symmetries.}

While the terms with subleading functional weight are not seen by the
symbol, they can be constrained by symmetry requirements. Firstly, MHV
amplitudes are invariant under parity (spatial reflection), which is
realized by $w_i\leftrightarrow\wb_i$ in the multi-Regge
limit~\cite{Prygarin:2011gd}, that is
$x\leftrightarrow\bar x$ and $y\leftrightarrow\bar y$. Secondly, the
multi-Regge limit amplitude should be invariant under
target-projectile symmetry (exchange of the two incoming momenta), which
amounts to symmetry under $w_1\leftrightarrow
1/w_2$~\cite{Bartels:2014ppa}, that is $x\leftrightarrow y$ and $\bar
x\leftrightarrow\bar y$. The sums
of six-point terms in the expressions~\eqref{eq:R7mmmg}--\eqref{eq:R7mpmgapp}
are separately invariant under these transformations, and hence we can
require parity as well as target-projectile symmetry for the
function~$g$ by itself, and equally for~$\tilde{g}$. These symmetries significantly reduce the
number of free parameters, as summarized in \tabref{tab:g2params}.
\begin{table}
\centering
\begin{tabular}{lcccc}
\toprule
& $g^{(1)}$ & $g^{(2)}$ & $i\pi\zeta_2$ & $\zeta_3$ \\
\midrule
Ansatz~\eqref{eq:g2ansatz} & 19 & 5 & 1 & 1\\
Parity invariance & 15 & 5 & 1 & 1\\
TP-symmetry & 9 & 3 & 1 & 1\\
Collinear limit 1 & 6 & 2 & 0 & 0\\
Collinear limit 2 & 4 & 1 & 0 & 0\\
Second entry & 3 & 1 & 0 & 0\\
\bottomrule
\end{tabular}
\caption{The table shows the numbers of undetermined coefficients in
the different parts of the initial Ansatz~\protect\eqref{eq:g2ansatz} (first
line), as well as their reduction upon imposing various consistency
constraints.}
\label{tab:g2params}
\end{table}

Both parity and target-projectile symmetry are not trivially
implemented: The parity map replaces all holomorphic weight vectors
and arguments of our single-valued basis functions $\Gs$ with their
complex conjugates. Again using the Hopf algebra antipode, these conjugate
single-valued functions can be re-expressed in terms of single-valued
functions with holomorphic arguments~\cite{DelDuca:2016lad}, but those
will not necessarily be elements of the basis~\eqref{eq:Gsbasis}.
Similarly, the target-projectile inversion map $x\leftrightarrow y$
produces non-basis functions. In order to derive constraints for our
Ansatz coefficients, all non-basis functions need to be re-expressed
in terms of basis functions, which is possible due to the many
relations among multiple polylogarithms.
The single-valued map~\eqref{eq:svmap} is an algebra
homomorphism, hence every identity among ordinary multiple
polylogarithms lifts to a corresponding identity among single-valued
multiple polylogarithms. In this way, single-valued multiple
polylogarithms inherit the shuffle and stuffle algebra relations from
their ordinary counterparts, as well as the simpler rescaling property
\begin{equation}
\Gs(a_1,\dots,a_n;z)=\Gs(ca_1,\dots,ca_n;cz)
\qquad\text{for }
a_n\neq0\text{ and }c\neq0\,.
\label{eq:rescaleG}
\end{equation}
We list all relations among single-valued polylogarithms with
holomorphic arguments that
are needed to evaluate parity and target-projectile symmetry:
\begin{alignat}{3}
    \Gsshort{\check{x}}_{0} &= -\Gsshort{x}_{0}
    , &
    \Gsshort{y}_{0} &= -\Gsshort{\check{y}}_{0}
    , &
    \Gsshort{y}_{1} &= \Gsshort{\check{y}}_{1}-\Gsshort{\check{y}}_{0}
    , \nn\\
    \Gsshort{1}_{x} &= \Gsshort{x}_{1}-\Gsshort{x}_{0}
    , &
    \Gsshort{x}_{\tilde{x}} &= \Gsshort{\check{y}}_{1}-\Gsshort{\check{y}}_{0}
    , &
    \Gsshort{\check{y}}_{\tilde{x}} &= \Gsshort{x}_{1}-\Gsshort{x}_{0}
    , \nn\\
    \Gsshort{1}_{0,x} &= \tfrac{1}{2} \brk{\brk{\Gsshort{x}_{0}}}^2-\Gsshort{x}_{0,1}
    , &
    \Gsshort{x}_{0,\tilde{x}} &= \tfrac{1}{2} \brk{\Gsshort{\check{y}}_{0}}^2-\Gsshort{\check{y}}_{0,1}
    , &
    \Gsshort{\check{y}}_{0,\tilde{x}} &= \tfrac{1}{2} \brk{\Gsshort{x}_{0}}^2-\Gsshort{x}_{0,1}
    , \mspace{30mu} \nn\\
    \Gsshort{1}_{x,0} &= \Gsshort{x}_{0,1}-\tfrac{1}{2} \brk{\Gsshort{x}_{0}}^2
    , &
    \Gsshort{1}_{x,1} &= \Gsshort{x}_{0,1}+\tfrac{1}{2} \brk{\Gsshort{x}_{1}}^2-\Gsshort{x}_{0} \Gsshort{x}_{1}
    , \mspace{-30mu} &&\nn\\
    \Gsshort{\check{y}}_{\tilde{x},x} &= \Gsshort{\check{y}}_{0,x}-\Gsshort{\check{y}}_{1,x}-\Gsshort{x}_{0} \Gsshort{\check{y}}_{1}+\Gsshort{x}_{1} \Gsshort{\check{y}}_{1}
    , \mspace{-100mu} &&&& \nn\\
    \Gsshort{x}_{0,0,\tilde{x}} &= \Gsshort{\check{y}}_{0,0,1}-\tfrac{1}{6} \brk{\Gsshort{\check{y}}_{0}}^3
    , &
    \Gsshort{\check{y}}_{0,0,\tilde{x}} &= \Gsshort{x}_{0,0,1}-\tfrac{1}{6} \brk{\Gsshort{x}_{0}}^3
    , \nn && \\
    \Gsshort{\check{y}}_{0,\tilde{x},x} &= -\Gsshort{\check{y}}_{1} \Gsshort{x}_{0,1}-\Gsshort{\check{y}}_{1} \Gsshort{\check{y}}_{0,x}+\Gsshort{\check{y}}_{0,0,x}+\Gsshort{\check{y}}_{0,1,x}+\Gsshort{\check{y}}_{0,x,1}+\tfrac{1}{2} \brk{\Gsshort{x}_{0}}^2 \Gsshort{\check{y}}_{1}
    ,\nn \mspace{-400mu} &&&& \\
    \Gsshort{\check{y}}_{\tilde{x},\tilde{x},x} &= -\Gsshort{x}_{0} \Gsshort{\check{y}}_{0,1}+\Gsshort{x}_{1} \Gsshort{\check{y}}_{0,1}-\Gsshort{\check{y}}_{1} \Gsshort{\check{y}}_{0,x}+\Gsshort{\check{y}}_{0,0,x} +\Gsshort{\check{y}}_{0,x,1}+\Gsshort{\check{y}}_{1,1,x}+\tfrac{1}{2} \brk{\Gsshort{x}_{0}}^2 \Gsshort{\check{y}}_{1}
    \nn \mspace{-500mu} &&&& \\ & \qquad \qquad
    +\tfrac{1}{2} \Gsshort{x}_{0} \brk{\Gsshort{\check{y}}_{1}}^2-\Gsshort{x}_{1} \Gsshort{x}_{0} \Gsshort{\check{y}}_{1}-\tfrac{1}{2} \Gsshort{x}_{1} \brk{\Gsshort{\check{y}}_{1}}^2+\tfrac{1}{2} \brk{\Gsshort{x}_{1}}^2 \Gsshort{\check{y}}_{1}
    \mspace{-600mu} &&&&
    .
    \label{eq:GGtobasis}
\end{alignat}
where $\check{x} \defas 1/x$ and $\tilde{x} \defas x/y$.
The weight-one single-valued basis functions are individually parity
invariant, hence parity symmetry only affects the Ansatz for the
weight-two part $g^{(2)}$. It reduces the number of undetermined
coefficients from $19$ to $15$. Target-projectile symmetry further
reduces the uncertainty in $g^{(2)}$ and $g^{(1)}$ to $9$ and $3$
coefficients, respectively.

%%%%%%%%%%%%%%%%%%%%%%%%%%%%%%
\paragraph{Collinear Limit.}

Next, we want to expand our Ansatz function in the collinear limit. One way
to do so is: First write the single-valued 2dHPLs in terms of ordinary
Goncharov polylogarithms according to~\eqref{eq:svmap}, then
rewrite the Goncharov polylogarithms in terms of classical
polylogarithms, for example using~\cite{Frellesvig:2016ske}, and
finally expand in large $r_1$ and $r_2$, using~\eqref{eq:vofw},
and~\eqref{eq:frtow7}, which combine to
\begin{alignat}{2}
x &= -\frac{F_2 (F_1 + r_1)}{r_1 r_2}
\,,&\qquad
\bar x &= -\frac{1+F_1r_1}{F_1 F_2 r_1 r_2}
\,,\nn\\
y &= -\frac{r_1 (F_2+r_2)}{F_1 F_2}
\,,&
\bar y &= -F_1 r_1 (1+F_2r_2)
\,.
\label{eq:coll1}
\end{alignat}
While the Regge limit sits at $T_1,T_2,S_1\to0$, $S_2\to\infty$ with
$r_1\equiv S_1/T_1$ and $r_2\equiv1/(S_2T_2)$ fixed, the collinear limit is defined by $T_j\to0$
with $S_j$ finite. From the Regge limit, the combined Regge collinear
limit is therefore attained by letting $r_j\to\infty$, that is
\begin{equation}
x\approx-\frac{F_2}{r_2}\to0
\,,\quad
\bar x\approx-\frac{1}{F_2r_2}\to0
\,,\quad
y\approx-\frac{r_1r_2}{F_1F_2}\to\infty
\,,\quad
\bar y\approx-F_1F_2r_1r_2\to\infty
\,.
\label{eq:xycoll3}
\end{equation}
The expansion of the basis functions~\eqref{eq:Gbasis} in this case
is simple, since all arguments $x$, $\bar x$, $1/y$, and
$1/\bar y$ tend to zero.
After writing all single-valued
functions $\Gs$ in terms of ordinary multiple
polylogarithms and performing the expansion, we obtain the
expansions of the Ansatz~\eqref{eq:g2ansatz} for $g$
near the collinear limit~\eqref{eq:xycoll3}. In doing so, one has to be
careful in picking consistent branches for all occurring logarithms. Every single-valued multiple polylogarithm
of the basis~\eqref{eq:Gbasis}
expands to a power series in $\log(r_i)$ and $1/r_i$, where the series
coefficients are rational functions of $F_1$ and $F_2$ as presented below:
\begin{alignat}{2}
    \Gsshort{x}_{0} &=
        -2 \log\brk{r_2}
        % +\frac{F_1}{r_1}
        % +\frac{1}{F_1 r_1}
        % +\frac{2\cos\varphi_1}{r_1}
        +2C_1
        +\order{r^{-2}}
        , &
    \Gsshort{\check{y}}_{0} &=
        -2 \log\brk{r_1}
        -2 \log\brk{r_2}
        % -\frac{F_2}{r_2}
        % -\frac{1}{F_2 r_2}
        % -\frac{2\cos\varphi_2}{r_2}
        -2C_2
        +\order{r^{-2}}
        , \nn\\
    \Gsshort{x}_{1} &=
        % \frac{F_2}{r_2}
        % +\frac{1}{r_2 F_2}
        % \frac{2\cos\varphi_2}{r_2}
        2C_2
        % +\frac{F_2 F_1}{r_1 r_2}
        % +\frac{1}{F_1 F_2 r_1 r_2}
        % +\frac{2\cos(\varphi_1+\varphi_2)}{r_1r_2}
        + 2C_+
        +\order{r^{-3}}
        , &
    \Gsshort{\check{y}}_{1} &=
        % \frac{F_2 F_1}{r_1 r_2}
        % +\frac{1}{F_1 F_2 r_1 r_2}
        % \frac{2\cos(\varphi_1+\varphi_2)}{r_1r_2}
        2C_+
        +\order{r^{-3}}
        , \nn \\
        &&
    \Gsshort{\check{y}}_{x} &=
        % -\frac{F_1}{r_1}
        % -\frac{1}{r_1 F_1}
        % -\frac{2\cos\varphi_1}{r_1}
        -2C_1
        % +\frac{F_2 F_1}{r_1 r_2}
        % +\frac{1}{F_2 r_1 r_2 F_1}
        % +\frac{2\cos(\varphi_1+\varphi_2)}{r_1r_2}
        +2C_+
        +\order{r^{-3}}
        , \nn\\
    \Gsshort{x}_{0,1} &=
        -\frac{2 \log\brk{r_2}}{F_2r_2} \lrbrk{1+\frac{1}{F_1 r_1}}
        % +\frac{F_2 F_1}{r_1 r_2}
        % +\frac{F_1}{F_2 r_1 r_2}
        % +\frac{2\cos(\varphi_1+\varphi_2)}{r_1r_2}
        +2C_+
        +\frac{F_2}{r_2}
        -\frac{1}{F_2 r_2}
        +\order{r^{-3}}
        , \mspace{-270mu} \nn && \\
    \Gsshort{\check{y}}_{0,1} &=
        -\frac{2 \log\brk{r_1}}{F_1 F_2 r_1 r_2}
        -\frac{2 \log\brk{r_2}}{F_1 F_2 r_1 r_2}
        +\frac{F_1 F_2}{r_1 r_2}
        -\frac{1}{F_1 F_2 r_1 r_2}
        +\order{r^{-3}}
        , \mspace{-270mu} \nn && \\
    \Gsshort{\check{y}}_{0,x} &=
        \frac{2 \log\brk{r_1}}{F_1r_1} \lrbrk{1-\frac{1}{F_2 r_2}}
        -\frac{F_1}{r_1}
        +\frac{1}{r_1 F_1}
        +\frac{2F_2C_1}{r_2}
        +\order{r^{-3}}
        , \mspace{-270mu} \nn && \\
    \Gsshort{\check{y}}_{1,x} &=
        \frac{2 \log\brk{r_2}}{F_1 F_2 r_1 r_2}
        +\frac{2C_2}{F_1r_1}
        +\order{r^{-3}}
        , \mspace{-270mu} \nn && \\
    % \Gsshort{x}_{0,0,1} &=
    %     \bigbrk{\log\brk{r_2}}^2 \Bigsbrk{\frac{2}{F_2 r_2}+\frac{2}{F_1 F_2 r_1 r_2}}
    %     +\log\brk{r_2} \Bigsbrk{\frac{2}{F_2 r_2}-\frac{2 F_1}{F_2 r_1 r_2}}
    %     \nn\\
    %     &\qquad+\frac{F_2}{r_2}
    %     +\frac{1}{F_2 r_2}
    %     +\frac{F_1 F_2}{r_1 r_2}
    %     -\frac{F_1}{F_2 r_1 r_2}
    %     +\order{r^{-3}},
    %     \nn\\
    % \Gsshort{\check{y}}_{0,0,1} &=
    %     \frac{2 \bigbrk{\log\brk{r_1}}^2}{F_1 F_2 r_1 r_2}
    %     +\log\brk{r_1} \Bigsbrk{\frac{4 \log\brk{r_2}}{F_1 F_2 r_1 r_2}+\frac{2}{F_1 F_2 r_1 r_2}}
    %     +\frac{2 \bigbrk{\log\brk{r_2}}^2}{F_1 F_2 r_1 r_2}
    %     +\frac{2 \log\brk{r_2}}{F_1 F_2 r_1 r_2}
    %     \nn\\
    %     &\qquad+\frac{F_1 F_2}{r_1 r_2}
    %     +\frac{1}{F_1 F_2 r_1 r_2}
    %     +\order{r^{-3}},
    %     \nn\\
    % \Gsshort{\check{y}}_{0,1,x} &=
    %     \log\brk{r_1} \Bigsbrk{-\frac{4 \log\brk{r_2}}{F_1 F_2 r_1 r_2}-\frac{2 F_2}{F_1 r_1 r_2}-\frac{2}{F_1 F_2 r_1 r_2}}
    %     -\frac{4 \bigbrk{\log\brk{r_2}}^2}{F_1 F_2 r_1 r_2}
    %     -\frac{4 \log\brk{r_2}}{F_1 F_2 r_1 r_2}
    %     -\frac{2}{F_1 F_2 r_1 r_2}
    %     +\order{r^{-3}},
    %     \nn\\
    \Gsshort{\check{y}}_{0,1,x} &=
        \frac{
        -4 \log^2\brk{r_2}
        -4 \log\brk{r_1}\log\brk{r_2}
        -2\log\brk{r_1} \bigbrk{F_2^2+1}
        -4 \log\brk{r_2}
        -2
        }{F_1F_2r_1r_2}
        +\order{r^{-3}}
        , \mspace{-310mu} \nn && \\
    \Gsshort{\check{y}}_{1,1,x} &=
        -\frac{2 \log\brk{r_2}}{F_1 r_1 r_2^2}
        -\frac{F_2^2}{2 F_1 r_1 r_2^2}
        -\frac{2}{F_1 r_1 r_2^2}
        +\frac{1}{2 F_1 F_2^2 r_1 r_2^2}
        +\order{r^{-4}}
        \,, \mspace{-270mu} &&
    \label{eq:GGcoll3}
\end{alignat}
where we have used the abbreviations~\eqref{eq:Ccos}, and where
$\order{r^{-n}}$ stands for terms with $n$ or more inverse powers of $r_1$ or $r_2$.

In the Mandelstam regions that
we consider, the BDS amplitude correctly captures the
leading behavior in the collinear limit. Hence, the remainder function has to vanish in this limit.
That is, there should be no terms that are free of $1/r_i$ factors in
the collinear expansion of the remainder function.
Moreover, we can require
consistency with the general form of the Wilson loop OPE that governs
the remainder function in the collinear
limit~\cite{Basso:2013vsa,Basso:2013aha}. The general systematics of the
Wilson loop OPE predicts that the remainder function in the combined
Regge collinear limit (at two loops and in any kinematic region)
takes the form shown in eq.~\eqref{eq:R72ope}, whose multi-Regge expansion
via~\eqref{eq:frtow7} has the form
\begin{align}
R_7^{\mathrm{MRL-coll}}
&=\frac{\cos(\varphi_1)}{r_1}\f_1\bigbrk{\log(\eps_1),\log(r_1)}
+\frac{\cos(\varphi_2)}{r_2}\f_2\bigbrk{\log(\eps_2),\log(r_2)}
\nn\\
&+\frac{\cos(\varphi_1+\varphi_2)}{r_1r_2}
h\bigbrk{\log(\eps_1),\log(\eps_2),\log(r_1),\log(r_2)}
\nn\\
&+\frac{\cos(\varphi_1-\varphi_2)}{r_1r_2}
\bar h\bigbrk{\log(\eps_1),\log(\eps_2),\log(r_1),\log(r_2)}
+\order{r^{-2}}
\,,
\label{eq:opeform}
\end{align}
where $F_i=e^{i\varphi_i}$, and $\f_1$, $\f_2$, $h$, and $\bar h$ are
polynomials in the respective logarithms. In particular, the dependence on $\varphi_1$ and
$\varphi_2$ is very restricted.%
\footnote{The form~\eqref{eq:opeform} is valid in the Euclidean region
as well as the $(\p\p\p)$ region. We assume that all other Mandelstam
regions are connected to the Euclidean region through analytic
continuation. Moreover, the remainder function is a function of the
cross ratios. The cross ratios in general
kinematics~\eqref{eq:cr7opevar} depend on the variables $\varphi_1$ and
$\varphi_2$ only through the entire functions $\cos(\varphi_1)$, $\cos(\varphi_2)$, and
$\cos(\varphi_1+\varphi_2)$, and this dependence drops out in the collinear
limit. Hence the
general form~\eqref{eq:opeform} is preserved under the analytic
continuation into the various Mandelstam regions, including the
$(\m\m\m)$ and the $(\m\p\m)$ regions.}
A general combination of multiple
polylogarithms would also produce sine functions of $\varphi_1$,
$\varphi_2$, and $\varphi_1\pm\varphi_2$. It turns out that our parity and
target-projectile symmetric Ansatz is already free of such sine terms,
which is an important cross-check of our result. Moreover, terms where
$\cos(\varphi_1)$ multiplies $\log(\eps_2)$ or $\log(r_2)$ should
be absent, and the same is true for products of $\cos(\varphi_2)$ with
$\log(\eps_1)$ or with $\log(r_1)$.
Imposing these constraints reduces the number of parameters in
$g^{(2)}$ and $g^{(1)}$ to $6$ and $2$ coefficients, respectively, and
moreover sets
\begin{equation}
d_0=0
\,,\qquad
d_1=0
\,.
\end{equation}

When considering the above constraints,
one has to keep in mind that the remainder function in the $(\m\m\m)$
region~\eqref{eq:R7mmmg} consists of the function $g$ as well as two copies of the
six-point $(\m\m)$ region remainder function~\eqref{eq:f}.
In principle, there could be cross-terms between the six-point
functions and the function $g$, such that only their sum vanishes
and satisfies eq.~\eqref{eq:opeform} in the collinear limit. However, the
two instances of the six-point function
separately vanish and satisfy eq.~\eqref{eq:opeform} in the seven-point
Regge collinear limit, for both arguments $v_1=-x$ and $v_2=-y$. Hence
also $g$ has to satisfy these constraints by itself.

\medskip

\noindent
In fact, the combined Regge collinear limit is not unique: By cyclically
rotating the tessellation of the heptagon that defines the OPE
variables~\eqref{eq:TSF} and taking appropriate limits in the
variables $S_i$, we can probe different limits in the space of
multi-Regge kinematics.
Not all collinear limits have an overlap with
the multi-Regge limit: The requirement is that the vanishing of
``small'' cross ratios $u_{j,2}$, $u_{j,3}$ is compatible with the
collinear limit $T_1,T_2\to0$. One further case where this is
satisfied is the cyclic rotation of our polygon variables
by $3$ sites, that is, we use the momentum twistors
$Z\suprm{+3}_{7,i}\equiv Z_{7,i+3}$, where $Z_{7,i}$ are
the momentum twistors~\eqref{eq:ZBSV} used throughout the rest of this work.%
\footnote{Another independent Regge collinear limit is
defined by $Z\suprm{+1}_i\equiv Z_{i+1}$, with $S_1=1/(r_1T_1)$,
$S_2=1/(r_2T_2)$, and $r_i^2=w_i\wb_i$, $F_i^2=w_i/\wb_i$, that
is $x=-F_2(1+F_1r_1)r_2$, $y=-(1+F_2r_2)/(F_1F_2r_1r_2)$. However,
this limit is related to the unshifted kinematics by a combination of
target-projectile symmetry and a permutation of the OPE variables
$\brc{F_i,S_i,T_i}$, and thus does not imply further
independent constraints.}
In this case, the Regge limit is attained by setting
$S_1=1/(r_1T_1)$, $S_2=1/(r_2T_2)$, and letting $T_1,T_2\to0$. The
multi-Regge parameters $w_1$, $w_2$ are then related to the OPE
variables by
\begin{equation}
r_1=\sqrt{w_2\wb_2}
\,,\qquad
r_2=\frac{1}{\sqrt{w_1\wb_1}}
\,,\qquad
F_1=\frac{\sqrt{\wb_2}}{\sqrt{w_2}}
\,,\qquad
F_2=\frac{\sqrt{w_1}}{\sqrt{\wb_1}}
\,,
\label{eq:rFwshifted}
\end{equation}
or equivalently
\begin{alignat}{2}
x&=-\frac{r_1(F_2+r_2)}{F_1r_2}
\,,\qquad
\bar x&=-\frac{F_1r_1(1+F_2r_2)}{F_2r_2}
\,,\nn\\
y&=-\frac{(F_1+r_1)r_2}{F_2r_1}
\,,\qquad
\bar y&=-\frac{F_2(1+F_1r_1)r_2}{F_1r_1}
\,.
\label{eq:coll3}
\end{alignat}
The combined Regge collinear limit is attained from the
multi-Regge limit by letting $r_1,r_2\to\infty$, which implies
\begin{equation}
x\approx-\frac{r_1}{F_1}\to\infty
\,,\qquad
\bar x\approx-r_1F_1\to\infty
\,,\qquad
y\approx-\frac{r_2}{F_2}\to\infty
\,,\qquad
\bar y\approx-r_2F_2\to\infty
\,.
\label{eq:xycoll4}
\end{equation}
In this limit, the basis functions of eq.~\eqref{eq:Gbasis} expand as
follows (see eq.~\eqref{eq:GGcoll3} for the notation):
\begin{alignat}{2}
    \Gsshort{x}_{0} &=
        2 \log\brk{r_1}
        + 2 C_2
        + \order{r^{-2}}
        , &
    \Gsshort{\check{y}}_{0} &=
        -2 \log\brk{r_2}
        - 2 C_1
        + \order{r^{-2}}
        , \nn\\
    \Gsshort{x}_{1} &=
        2 \log\brk{r_1}
        + 2 C_1
        + 2 C_2
        - 2 C_+
        + \order{r^{-3}}
        , &
    \Gsshort{\check{y}}_{1} &=
        2 C_2
        -2 C_+
        + \order{r^{-3}}
        , \nn \\
        &&
    \Gsshort{\check{y}}_{x} &=
        -2 C_+
        + \order{r^{-3}}
        , \nn \\
    \Gsshort{x}_{0,1} &=
        2 \log^2\brk{r_1}
        +2 \log\brk{r_1} \lrbrk{\frac{1}{F_1 r_1}-\frac{1}{F_1 F_2 r_1 r_2}+2 C_2}
        -\frac{F_1}{r_1}
        +\frac{1}{F_1 r_1}
        +\frac{2F_2C_1}{r_2}
        + \order{r^{-3}},
        \mspace{-270mu} && \nn\\
    \Gsshort{\check{y}}_{0,1} &=
        -\frac{2 \log\brk{r_2}}{F_2r_2} \lrbrk{1-\frac{1}{F_1 r_1}}
        +\frac{F_2}{r_2}
        -\frac{1}{F_2 r_2}
        -\frac{2F_1C_2}{r_1}
        + \order{r^{-3}},
        \mspace{-300mu} && \nn\\
    \Gsshort{\check{y}}_{0,x} &=
        \frac{2 \log\brk{r_1}+2 \log\brk{r_2}+1}{F_1 F_2 r_1 r_2}
        -\frac{F_1 F_2}{r_1 r_2}
        + \order{r^{-3}},
        \mspace{-300mu} && \nn\\
    \Gsshort{\check{y}}_{1,x} &=
        \frac{2 \log\brk{r_1}}{F_1 F_2 r_1 r_2}
        +\frac{2C_1}{F_2r_2}
        + \order{r^{-3}},
        && \nn\\
    % \Gsshort{x}_{0,0,1} &=
    %     \frac{4}{3} \bigbrk{\log\brk{r_1}}^3
    %     +2 \bigbrk{\log\brk{r_1}}^2 \Bigsbrk{\frac{1}{F_1 r_1}+2 C_2-\frac{1}{F_1 F_2 r_1 r_2}}
    %     +2 \log\brk{r_1} \Bigsbrk{\frac{1}{F_1 r_1}+\frac{F_2}{F_1 r_1 r_2}}
    %     \nn\\
    %     &\qquad+ 2 C_1
    %     -\frac{F_1 F_2}{r_1 r_2}
    %     +\frac{F_2}{F_1 r_1 r_2}
    %     + \order{r^{-3}},
    %     \nn\\
    % \Gsshort{\check{y}}_{0,0,1} &=
    %     2 \bigbrk{\log\brk{r_2}}^2 \Bigsbrk{\frac{1}{F_2 r_2}-\frac{1}{F_1 F_2 r_1 r_2}}
    %     + 2 \log\brk{r_2} \Bigsbrk{\frac{1}{F_2 r_2}+\frac{F_1}{F_2 r_1 r_2}}
    %     \nn\\
    %     &\qquad+2 C_2
    %     -\frac{F_1 F_2}{r_1 r_2}
    %     +\frac{F_1}{F_2 r_1 r_2}
    %     + \order{r^{-3}},
    %     \nn\\
    \Gsshort{\check{y}}_{0,1,x} &=
        -\frac{2 \log^2\brk{r_1}+2 \log\brk{r_1} \bigbrk{2 \log\brk{r_2} + 1}}{F_1 F_2 r_1 r_2}
        -\frac{2C_1\bigbrk{2 \log\brk{r_2}+1}}{F_2r_2}
        + \order{r^{-3}},
        \mspace{-300mu} && \nn\\
    \Gsshort{\check{y}}_{1,1,x} &=
        -\frac{2 \log\brk{r_1} \bigbrk{\log\brk{r_1} + 1}}{F_1 F_2 r_1 r_2}
        +\frac{F_1}{F_2 r_1 r_2}
        -\frac{1}{F_1 F_2 r_1 r_2}
        + \order{r^{-3}}
        \,. \mspace{-300mu} &&
    \label{eq:GGcoll4}
\end{alignat}
Expanding the Ansatz~\eqref{eq:g2ansatz} for $g$
in the limit~\eqref{eq:xycoll4}, we can again require \emph{(i)} vanishing
in the collinear limit, and \emph{(ii)} agreement with the general
form~\eqref{eq:opeform} of the Wilson loop OPE. Imposing these
constraints further reduces the number of parameters in
$g^{(2)}$ and $g^{(1)}$ to $4$ and $1$ coefficients, respectively,
as shown in \tabref{tab:g2params}.

%%%%%%%%%%%%%%%%%%%%%%%%%%%%%%
\paragraph{The Constrained Ansatz.}

Combining all the above constraints, the full
Ansatz~\eqref{eq:g2ansatz} for the function $g$ reduces to
\begin{align}
g(x,y)&=
-1/2\,\Gsshort{x}_{0}\Gsshort{\cy}_{0}\Gsshort{\cy}_{1}
+1/2\,\Gsshort{x}_{0}\Gsshort{x}_{1}\Gsshort{\cy}_{1}
+1/2\,\Gsshort{\cy}_{0}\Gsshort{x}_{1}\Gsshort{\cy}_{1}
-1/2\,\Gsshort{x}_{0}\Gsshort{x}_{1}\Gsshort{\cy}_{x}
\nn\\&\quad
+1/2\,\Gsshort{\cy}_{0}\Gsshort{\cy}_{1}\Gsshort{\cy}_{x}
-\Gsshort{\cy}_{1}\Gsshort{x}_{0,1}
+\Gsshort{\cy}_{x}\Gsshort{x}_{0,1}
+\Gsshort{x}_{0}\Gsshort{\cy}_{0,1}
-\Gsshort{x}_{1}\Gsshort{\cy}_{0,1}
-\Gsshort{\cy}_{x}\Gsshort{\cy}_{0,1}
\nn\\&\quad
+\Gsshort{x}_{1}\Gsshort{\cy}_{0,x}
-\Gsshort{\cy}_{0}\Gsshort{\cy}_{1,x}
-\Gsshort{x}_{1}\Gsshort{\cy}_{1,x}
+\Gsshort{\cy}_{1}\Gsshort{\cy}_{1,x}
+2\,\Gsshort{\cy}_{0,1,x}
-2\,\Gsshort{\cy}_{1,1,x}
\nn\\&\quad
+\cg0\zeta_2\,\Gsshort{\cy}_{x}
+2\pi i\bigsbrk{
       \cg1\bigbrk{
       \brk{\Gsshort{x}_0-\Gsshort{x}_1}\Gsshort{x}_1
       +\brk{\Gsshort{\cy}_0-\Gsshort{\cy}_1}\Gsshort{\cy}_1
       }
       \nn\\&\quad
       +\cg2\brk{\Gsshort{x}_0-\Gsshort{x}_1}\Gsshort{\cy}_{1}
       +\cg3\brk{\Gsshort{x}_0-\Gsshort{\cy}_0}\Gsshort{\cy}_{x}
       +\cg4(\Gsshort{\cy}_x)^2
}
\,,
\label{eq:g2constrainedapp}
\end{align}
The expansion of this function in the collinear limit~\eqref{eq:xycoll3}
takes the form
\begin{align}
\sbrk{g(x,y)}^\CL={}&
-\frac{\cos(\varphi_1)}{r_1}
 \Bigbrk{8 i \pi \cg3\log(r_1) + 2\cg0\zeta_2}
-\frac{\cos(\varphi_2)}{r_2} 8 i \pi \cg1\log(r_2)
\nn\\ &
-\frac{\cos(\varphi_1+\varphi_2)}{r_1r_2}\,2\Bigbrk{
2\log(r_2)^2+2\log(r_1)\log(r_2)
+4i\pi\brk{\cg1-\cg3}\log(r_1)
\nn\\ & \mspace{150mu}
+\bigbrk{3+4i\pi\brk{2\cg1+\cg2}}\log(r_2)
+2-\cg0\zeta_2
-2 \pi i(\cg1-\cg3)
}
\nn\\ &
+\frac{\cos(\varphi_1-\varphi_2)}{r_1r_2}\,2\Bigbrk{\log(r_2) + 2 \pi i(2\cg1-\cg3)}
\nn\\ &
+\order{r_1^{-2}}
+\order{r_2^{-2}}
\,.
\end{align}
Except for $\cg4$, all remaining coefficients in the constrained Ansatz~\eqref{eq:g2constrainedapp} should be
fixed by the expansion of the true remainder function in this combined
Regge collinear limit.

\medskip

\noindent
All constraints considered above equally apply to the function
$\tilde{g}$, and hence the constrained Ansatz for $g$ equally holds
for $\tilde{g}$.
Of course, the undetermined coefficients may assume
different values for $\tilde{g}$ than for $g$.

%%%%%%%%%%%%%%%%%%%%%%%%%%%%%%
\paragraph{Second Symbol Entry Constraints.}

Next, we will derive further constraints on the functions $g$ and
$\tilde{g}$ by looking at the second entry of the known remainder
function's symbol~\cite{CaronHuot:2011ky}. While the first entry of the symbol encodes the positions of all
branch points on the main sheet, the first and second entries together
determine the positions of all branch points on all sheets
adjacent to the main sheet.
We denote the symbol of the heptagon remainder function by
\begin{equation}
\smbop{R_7^{(2)}}
=\sum_{\mathclap{i\in I,\,j\in J}}\smb{a_i,a_j,X_{ij}}
\,.
\label{eq:smbaaX}
\end{equation}
Letters $a_i$, $i\in I$ in the first entry are drawn from the six cross ratios~\eqref{eq:uj}
as well as the seventh cross ratio $U_{26}$. The second entry includes further letters
that cannot be reduced to cross ratios.
Every pair of first and second
entry is followed by a two-letter symbol $X_{ij}$.
Writing the continuation of the remainder function along some path
$\gen$ as:
\begin{equation}
\gen R_7^{(2)}
=(1+\Delta^\gen)R_7^{(2)}
\,,\qquad
\Delta^\gen R_7^{(2)}=2\pi i h_1 + (2\pi i)^2 h_2 + \dots
\,,
\end{equation}
the coefficients $h_1$ and $h_2$ are functions of weight $3$ and $2$, respectively. Their
symbols have the form
\begin{equation}
\smbop{h_1} = \sum_{\mathclap{i\in I,\,j\in J}}n_i\,\smb{a_j,X_{ij}}
\,,\qquad
\smbop{h_2} = \sum_{\mathclap{i\in I,\,j\in J}}n_{ij} X_{ij}
\,,
\end{equation}
where $n_i$ and $n_{ij}$ are numbers that depend on the chosen path $\gen$
($n_i$ is the winding number of $a_i$ around zero).

We can now
constrain the function $h_2$ by inspecting the known symbol~\eqref{eq:smbaaX}:
We first collect all $X_{ij}$, and compute their multi-Regge limits at
symbol level. In the multi-Regge limit, the entries of the symbols
$X_{ij}$ are rational functions of our Regge variables $T_i$, $w_i$,
and $\wb_i$, $i=5,6$. There are $73$ symbols $X_{ij}$, but not all
of them are independent in the multi-Regge limit. We note that the
six-point function $f$~\eqref{eq:f} is free of terms proportional to $2\pi i$, hence $h_2$
only contributes to the functions $g$ or $\tilde{g}$. We can therefore set the LLA part of $h_2$ to
zero. We do so by first unshuffling the variables $T_1$ and $T_2$ from
the symbol, then identifying the symbol $(T_i)$ with the function
$1/4\brk{\log\eps_i+\log(w_i\wb_i)}$ (using
eqs.~\eqref{eq:FSTtoweps}), and requiring that the coefficients of
$\log\eps_1$ and $\log\eps_2$ must vanish. This imposes $8$
linear constraints on the $73$ numbers $n_{ij}$.

Next, we match the symbol of $h_2$ to
the symbol of a generic weight-two combination of single-valued MPLs
constructed from the basis $\Gs$~\eqref{eq:Gsbasis}. This
implies $28$ further constraints on the $n_{ij}$, and leaves us with a
six-parameter combination of single-valued MPLs. Further matching this combination
against the function multiplying $2\pi i$ in eq.~\eqref{eq:g2constrainedapp}
(\ie imposing parity-invariance, target-projectile symmetry, and
consistency with the Wilson loop OPE in the collinear limit) eliminates $3$
coefficients, and imposes the constraint
\begin{equation}
\cg4=0
\label{eq:c40}
\end{equation}
on our Ansatz~\eqref{eq:g2constrainedapp}. Since the analysis leading
to eq.~\eqref{eq:c40} was independent on the choice of path~$\gen$,
we conclude that eq.~\eqref{eq:c40} has to hold for both functions $g$ and
$\tilde{g}$. We note that the constraints on the numbers $n_{ij}$ are
consistent with
\begin{equation}
n_{ii}=1
\quad\text{where}\quad
a_i=U_{26}
\,,
\end{equation}
which follows if the path $\gen$ winds $U_{26}$ around zero
once (in the negative sense), as is the case for the paths leading
from the $(\p\p\p)$ into the $(\m\m\m)$ and $(\m\p\m)$ regions, see \tabref{table:crsWinding}.

If we require in addition that $u_{1,1}=U_{25}$ and $u_{2,1}=U_{36}$
are held fixed, which is consistent with the continuation into the
$(\m\m\m)$ region, then it follows that
\begin{equation}
n_{ij}=0
\quad\text{if}\quad
a_i\in\brc{U_{25},U_{36}}
\;\vee\;
a_j\in\brc{U_{25},U_{36}}
\,.
\end{equation}
Combining these conditions with the previous constraints enforces that
\begin{equation}
\cg1 = \cg2 = \cg3 = 0
\label{eq:c123zero}
\end{equation}
in our Ansatz~\eqref{eq:g2constrainedapp}, in full agreement
with~\cite{DelDuca:2018hrv}.

%%%%%%%%%%%%%%%%%%%%%%%%%%%%%%%%%%%%%%%%%%%%%%%%%%%%%%%%%%%%
%%%%%%%%%%%%%%%%%%%%%%%%%%%%%%%%%%%%%%%%%%%%%%%%%%%%%%%%%%%%
\section{Discontinuity Tables for Half Windings}
\label{app:mpmDisc}

Here we list the additional contributions $\delta^{\varrho}_I$ to discontinuities $\Delta_I$ listed in eqs.~\eqref{eq:delta}, \eqref{eq:ddelta},
and~\eqref{eq:dddelta}, that arise after appending the half-windings to the path of continuation separately for
each region $\varrho$ according to \tabref{table:crsWinding}:
\begin{equation}
\delta^{\varrho}_I
\defas
\Delta^{\varrho}_I - \Delta_I
\,,\qquad
\varrho\in\brc{(\p\m\m),(\m\m\p),(\m\p\m)}
\,.
\end{equation}
Below, we list these corrections for single $\Delta_i$ and double $\Delta_{i, j}$ discontinuities, while the triple ones
are unchanged $\delta^{\varrho}_{i, j, k} = 0$. We will use the
shorthand notation~\eqref{eq:Ccos}.

%%%%%%%%%%%%%%%%%%%%%%%%%%%%%%%%%%%%%%%%%%%%%%%%%%%%%%%%%%%%
\subsection{The Region \texorpdfstring{$\brk{\p\m\m}$}{(+--)}}

The corrections to the single discontinuities for the $(\p\m\m)$ region read
\begin{align}
\delta^{\p\m\m}_1 &=
    2 \pi i C_1 \log(\eps_1)
    +C_1 (8 \pi i \log(r_1)-4 \pi^2+8 \pi i)
\,,\nn\\
\delta^{\p\m\m}_3 &=
    2 \pi i C_+ \log(\eps_1)
    +C_+ (8 \pi i \log(r_1)+4 \pi i \log(r_2)-4 \pi^2+8 \pi i)
\,.
    \label{eq:deltapmm}
\end{align}
The corrections to the double discontinuities are
\begin{align}
\sfrac13\delta^{\p\m\m}_{1,4} &
= \delta^{\p\m\m}_{4,1}
= 2 \pi i C_1
\,,\nn\\
\sfrac13\delta^{\p\m\m}_{3,4} &
= \delta^{\p\m\m}_{4,3}
= \delta^{\p\m\m}_{3,5}
= 2 \pi i C_+
\,,\nn\\
\delta^{\p\m\m}_{1,2} &
= -6 \pi i C_1+4 \pi i C_+
\,,\nn\\
\delta^{\p\m\m}_{2,1} &
= -2 \pi i C_1+2 \pi i C_+
\,.
\end{align}

%%%%%%%%%%%%%%%%%%%%%%%%%%%%%%%%%%%%%%%%%%%%%%%%%%%%%%%%%%%%
\subsection{The Region \texorpdfstring{$\brk{\m\m\p}$}{(--+)}}

The corrections to the single discontinuities for the $(\m\m\p)$ region read
\begin{align}
\delta^{\m\m\p}_3 ={} &
    -2 \pi i C_+ \log(\eps_1)
    -(2 \pi i C_++2 \pi i C_2) \log(\eps_2)
    -C_2 (8 \pi i \log(r_2)+4 \pi^2+8 \pi i)
    \nn\\ &
    -C_+ (4 \pi i \log(r_1)+8 \pi i \log(r_2)+4 \pi^2+6 \pi i)
    +2 \pi i C_-
\,,\nn\\
\delta^{\m\m\p}_2 ={} &
    -2 \pi i C_2 \log(\eps_2)
    -C_2 (8 \pi i \log(r_2)+4 \pi^2+8 \pi i)
\,.
    \label{eq:deltammp}
\end{align}
The corrections to the double discontinuities are
\begin{align}
\sfrac12\delta^{\m\m\p}_{3,4} &
= \delta^{\m\m\p}_{4,3}
= -\delta^{\m\m\p}_{1,2}
= -2 \pi i C_+
\,,\nn\\
\sfrac13\delta^{\m\m\p}_{3,5} &
= \delta^{\m\m\p}_{5,3}
= -2 \pi i C_+-2 \pi i C_2
\,,\nn\\
\sfrac13\delta^{\m\m\p}_{2,5} &
= \delta^{\m\m\p}_{5,2}
= -2 \pi i C_2
\,.
\end{align}

%%%%%%%%%%%%%%%%%%%%%%%%%%%%%%%%%%%%%%%%%%%%%%%%%%%%%%%%%%%%
\subsection{The Region \texorpdfstring{$\brk{\m\p\m}$}{(-+-)}}

The corrections to the single discontinuities for the $(\m\p\m)$ region read
\begin{align}
\delta^{\m\p\m}_{1} ={}
    &-2 \pi i C_1 \log(\eps_1)
    -C_1 (8 \pi i \log(r_1)+4 \pi^2+8 \pi i)
\,,\nn\\
\delta^{\m\p\m}_{3} ={}
    &(2 \pi i C_++2 \pi i C_2) \log(\eps_2)
    +C_2 (8 \pi i \log(r_2)-4 \pi^2+8 \pi i)
    \nn\\
    &-C_+ (4 \pi i \log(r_1)-4 \pi i \log(r_2)+4 \pi^2+2 \pi i)
    -2 \pi i C_-
\,,\nn\\
\delta^{\m\p\m}_{2} ={}
    &2 \pi i C_2 \log(\eps_2)
    +C_2 (8 \pi i \log(r_2)-4 \pi^2+8 \pi i)
\,.
\end{align}
The corrections to the double discontinuities are
\begin{align}
\sfrac13\delta^{\m\p\m}_{1,2} &
= \delta^{\m\p\m}_{2,1}
= -2 \pi i C_++2 \pi i C_1
\,,\nn\\
\sfrac13\delta^{\m\p\m}_{1,4} &
= \delta^{\m\p\m}_{4,1}
= -2 \pi i C_1
\,,\nn\\
\sfrac13\delta^{\m\p\m}_{2,5} &
= \delta^{\m\p\m}_{5,2}
= 2 \pi i C_2
\,,\nn\\
\delta^{\m\p\m}_{3,4} &
= -2 \pi i C_+
\,,\nn\\
\delta^{\m\p\m}_{3,5} &
= 4 \pi i C_++6 \pi i C_2
\,,\nn\\
\delta^{\m\p\m}_{5,3} &
= 2 \pi i C_++2 \pi i C_2
\,.
\end{align}
%

%%%%%%%%%%%%%%%%%%%%%%%%%%%%%%
% Bibliography
%%%%%%%%%%%%%%%%%%%%%%%%%%%%%%
\bibliographystyle{nb}
\bibliography{references}

\end{document}